\documentclass[sigplan,screen]{acmart}

\AtBeginDocument{%
  \providecommand\BibTeX{{%
    \normalfont B\kern-0.5em{\scshape i\kern-0.25em b}\kern-0.8em\TeX}}}

\renewcommand\footnotetextcopyrightpermission[1]{} % removes footnote with conference information in first column
\acmPrice{15.00}
\acmISBN{978-1-4503-XXXX-X/18/06}

\setcopyright{acmcopyright}
\copyrightyear{2022}
\acmYear{2022}
\acmDOI{XXXXXXX.XXXXXXX}

\acmJournal{JACM}
\acmVolume{37}
\acmNumber{4}
\acmArticle{111}
\acmMonth{8}

\usepackage{breqn}

\usepackage{graphicx}
\usepackage{textcomp}
\usepackage{xcolor}
\usepackage{multirow}
\usepackage{subcaption}
\usepackage{pgf-umlsd}
\usepackage{comment} 
\usepackage{url}
\usepackage{paralist}

\DeclareCaptionLabelFormat{noname}{#2}
\usepackage[linesnumbered,ruled,vlined]{algorithm2e}

\SetCommentSty{mycommfont}
\SetKwInput{KwInput}{Input}
\SetKwInput{KwOutput}{Output}
\SetKwComment{Comment}{/* }{ */}
\begin{document}

\title{Defending against the Label-flipping Attack in Federated Learning}

\author{Najeeb Moharram Jebreel, Josep Domingo-Ferrer, David Sánchez and Alberto Blanco-Justicia}
\email{{najeeb.jebreel, josep.domingo, david.sanchez, alberto.blanco}@urv.cat}
\affiliation{%
  \institution{Universitat Rovira i Virgili}
  \streetaddress{Av. Pa\"{\i}sos Catalans 26}
  \country{Catalonia}
  \city{Tarragona}
  \postcode{E-43007}
}

\renewcommand{\shortauthors}{Najeeb Jebreel, Josep Domingo-Ferrer, David Sánchez, and Alberto Blanco-Justicia.}

\begin{abstract}
Federated learning (FL) provides autonomy and privacy by design to participating peers, who cooperatively build a machine learning (ML) model while keeping their private data in their devices. 
However, that same autonomy opens the door for {\itshape malicious peers} to poison the model by conducting either untargeted or targeted poisoning attacks.  
The {\itshape label-flipping (LF) attack} is a targeted poisoning attack where the attackers poison their training data by flipping the labels of some examples from one class ({\itshape i.e.}, the source class) to another ({\itshape i.e.}, the target class).
Unfortunately, this attack is easy to perform and hard to detect and it negatively impacts on the performance of the global model.
Existing defenses against LF are limited by assumptions on the distribution of the peers' data and/or do not perform well with high-dimensional models. 
In this paper, we deeply investigate the LF attack behavior and find that the contradicting objectives of attackers and honest peers on the source class examples are reflected in the parameter gradients corresponding to the neurons of the source and target classes in the output layer, making those gradients good discriminative features for the attack detection.
Accordingly, we propose a novel defense that first dynamically extracts those gradients from the peers' local updates, and then clusters the extracted gradients, analyzes the resulting clusters and filters out potential bad updates before model aggregation. 
Extensive empirical analysis on three data sets shows the proposed defense's effectiveness against the LF attack regardless of the data distribution or model dimensionality. Also, 
%JOSEP2. Rewritten.
the proposed defense outperforms several state-of-the-art defenses by offering lower test error, higher overall accuracy, higher source class accuracy, lower attack success rate, and higher stability of the source class accuracy. 
\end{abstract}

\begin{CCSXML}
<ccs2012>
 <concept>
  <concept_id>10010520.10010553.10010562</concept_id>
  <concept_desc>Computer systems organization~Embedded systems</concept_desc>
  <concept_significance>500</concept_significance>
 </concept>
 <concept>
  <concept_id>10010520.10010575.10010755</concept_id>
  <concept_desc>Computer systems organization~Redundancy</concept_desc>
  <concept_significance>300</concept_significance>
 </concept>
 <concept>
  <concept_id>10010520.10010553.10010554</concept_id>
  <concept_desc>Computer systems organization~Robotics</concept_desc>
  <concept_significance>100</concept_significance>
 </concept>
 <concept>
  <concept_id>10003033.10003083.10003095</concept_id>
  <concept_desc>Networks~Network reliability</concept_desc>
  <concept_significance>100</concept_significance>
 </concept>
</ccs2012>
\end{CCSXML}

\ccsdesc[500]{Computing methodologies~Distributed machine learning}
\ccsdesc[100]{Security and privacy}

\keywords{Federated learning, Security, Poisoning attacks, Label-flipping attacks}

\maketitle
\pagestyle{plain}

\section{Introduction}
\label{intro}

Federated learning (FL)~\cite{mcmahan2017communication, konevcny2015federated} is an emerging machine learning (ML) paradigm that enables multiple peers to collaboratively train a shared ML model without sharing their private data with a central server. In FL, the peers train a global model received from the server on their local data, and then submit the resulting model updates to the server. The server aggregates the received updates to obtain an updated global model, which it re-distributes among the peers in the next training iteration. Therefore, FL improves privacy and scalability by keeping the peers' local data at their respective premises and by distributing the training load across the peers' devices ({\itshape e.g.}, smartphones)~\cite{bonawitz2019towards}.

Despite these advantages, the distributed nature of FL opens the door for malicious peers to attack the global model~\cite{kairouz2021advances,blanco2021achieving}. Since the server has no control over the peers' behavior, any of them may deviate from the prescribed training protocol to conduct either untargeted poisoning attacks~\cite{blanchard2017machine, wu2020federated} or targeted poisoning attacks~\cite{biggio2012poisoning, fung2020limitations, bagdasaryan2020backdoor}. In the former, the attackers aim to cause model failure or non-convergence; in the latter, they aim to lead the model into misclassifying test examples with specific features into some desired labels.
%JOSEP2. Changed.
Whatever their nature, all these attacks result in bad updates being sent to the server.

The {\itshape label-flipping (LF) attack} ~\cite{biggio2012poisoning,fung2020limitations} is a type of targeted poisoning attack where the attackers poison their training data by flipping the labels of some correct examples from a source class to a target class, {\itshape e.g.}, flipping ''spams'' to ''non-spams'' or ''fraudulent'' activities to ''non-fraudulent''. Although the attack is easy for the attackers to perform, it has a significantly negative impact on the source class accuracy and, sometimes, on the overall accuracy~\cite{fung2020limitations,munoz2019byzantine,tolpegin2020data}. Moreover, the impact of the attack increases as the ratio of attackers and their number of flipped examples increase~\cite{steinhardt2017certified, tolpegin2020data}. 

Several defenses against poisoning attacks (and LF in particular) have been proposed, which we survey in Section \ref{related}.
However, they are either not practical~\cite{nelson2008exploiting, jagielski2018manipulating} or make specific assumptions about the distributions of local training data~\cite{shen2016auror,blanchard2017machine,tolpegin2020data,fung2020limitations, awan2021contra,chen2017distributed,yin2018byzantine}.
For example, \cite{jagielski2018manipulating} assumes the server has some data examples representing the distribution of the peers' data, which is not always a realistic assumption in FL; \cite{shen2016auror,blanchard2017machine,tolpegin2020data,chen2017distributed,yin2018byzantine} assume the data to be independent and identically distributed (iid) among peers, which leads to poor performance when the data are non-iid~\cite{awan2021contra}; \cite{fung2020limitations,awan2021contra} identify peers with a similar objective as attackers, which leads to a high rate of false positives when honest peers have similar local data~\cite{nguyen2021flguard, li2021auto}. Moreover, some methods, such as multi-Krum (MKrum)~\cite{blanchard2017machine} and trimmed mean (TMean)~\cite{yin2018byzantine} assume prior knowledge of the ratio of attackers in the system, which is a strong assumption. 

Besides the assumptions on the distribution of peers' data or their behavior, the dimensionality of the model is an essential factor that impacts on the performance of most of the above methods: high-dimensional models are more vulnerable to poisoning attacks because an attacker can operate small but damaging changes on its local update without being detected~\cite{chang2019cronus}. 
Specifically, in the LF attack, the changes made to a bad update become less evident as the dimensionality of the update increases, because of the relatively small changes the attack causes on the whole update. 

To the best of our knowledge, there is no work that provides an effective defense against LF attacks without being limited by the data distribution and/or model dimensionality.

%JOSEP2. I rewrite.
\textbf{Contributions and plan}. In this paper, we present a novel defense against the LF attack that is effective regardless of the peers' data distribution or the model dimensionality. Specifically, we make the following contributions:
\begin{compactitem} 
    \item We conduct in-depth conceptual and empirical analyses of the attack behavior and we find a useful pattern that helps better discriminate between the attackers' bad updates and the honest peers' good updates. Specifically, we find that the contradictory objectives of attackers and honest peers on the source class' examples are reflected in the parameters' gradients connected to the source and target classes' neurons in the output layer, making those gradients better discriminative features for attack detection. Moreover, we observe that those features stay robust under different data distributions and model sizes. Also, we observe that different types of non-iid data require different strategies to defend against the LF attack.
    
    \item We propose a novel defense that dynamically extracts the potential source and target classes' gradients from the peers' local updates, applies a clustering method on those gradients and analyzes the resulting clusters to filter out potential bad updates before model aggregation.
   
    \item We demonstrate the effectiveness of our defense against the LF attack through an extensive empirical analysis on three data sets with different deep learning model sizes, peers' local data distributions and ratios of attackers up to $50\%$. In addition, we compare our 
    %JOSEP2. Changed below.
    approach with several state-of-the-art defenses and show its superiority at simultaneously delivering low test error, high overall accuracy, high source class accuracy, low attack success rate and stability of the source class accuracy.
    
\end{compactitem}

The rest of this paper is organized as follows. 
Section~\ref{prelim} introduces preliminary notions. Section~\ref{lf_attack_threat_model} formalizes the label-flipping attack and the threat model being considered. Section~\ref{related} discusses countermeasures for poisoning attacks in FL.
Section~\ref{meth} presents the design rationale and the methodology of the proposed defense. 
Section~\ref{analysis} details the experimental setup and reports the obtained results. 
Finally, conclusions and future research lines are gathered in Section~\ref{conclusion}. 

\section{Preliminaries}
\label{prelim}

\subsection{Deep neural network-based classifiers}

A deep neural network (DNN) is a function $F(x)$, obtained by composing $L$ functions
$f^l, l\in [1, L]$, that maps an input $x$ to a predicted output $\hat{y}$. Each $f^l$ is a layer that is parametrized by a weight matrix $w^l$, a bias vector $b^l$ and an activation function $\sigma^l$. $f^l$ takes as input the output of the previous layer $f^{l-1}$. The output of $f^l$ on an input $x$ is computed as $f^l(x) = \sigma^l(w^l \cdot x + b^l)$. Therefore, a DNN can
be formulated as \[    F(x) = \sigma^L(w^L \cdot \sigma^{L-1}(w^{L-1} \dots \sigma^1(w^1 \cdot x + b^1) \dots + b^{L-1} )+ b^L).\]
DNN-based classifiers consist of a feature extraction part and a classification part~\cite{krizhevsky2017imagenet, minaee2021deep}. 
The classification part takes the extracted abstract features and makes the final classification decision. It usually consists of one or more fully connected layers where the output layer contains $|\mathcal{C}|$ neurons, with $\mathcal{C}$ being the set of all possible class values. 
The output layer's vector $o \in \mathbb{R^{|\mathcal{C}|}}$ is usually fed to the softmax function that transforms it to a vector $p$ of probabilities, which are called the confidence scores.
In this paper, we use predictive DNNs as $|\mathcal{C}|$-class classifiers, where the final predicted label $\hat{y}$ is taken to be the index of the highest confidence score in $p$. 
Also, we analyze the output layer of DNNs to filter out updates resulting from the LF attack (called bad updates for short in what follows). 

\subsection{Federated learning}
\label{fl}
In federated learning (FL), $K$ peers and an aggregator server collaboratively build a global model $W$. In each training iteration $t\in[1, T]$, 
the server randomly selects a subset of peers $S$ of size $m = C \cdot K \ge 1$ where $C$ is the fraction of peers that are selected in $t$. After that, the server distributes the current global model $W^{t}$ to all peers in $S$. Besides $W^{t}$, the server sends a set of hyper-parameters to be used to train the local models, which includes the number of local epochs $E$, the local batch size $BS$ and the learning rate $\eta$. After receiving $W^t$, each peer $k \in S$ divides her local data $D_k$ into batches of size $BS$ and performs $E$ SGD training epochs on $D_k$ to compute her update $W^{t+1}_{k}$, which she uploads to the server. 
Typically, the server uses the Federated Averaging (FedAvg) \cite{mcmahan2017communication} method to aggregate the local updates and obtain the updated global model $W^{t+1}$. FedAvg averages the updates proportionally to the number of training samples of each peer.

\section{Label-flipping attack and threat model}
\label{lf_attack_threat_model}

In the label-flipping (LF) attack~\cite{biggio2012poisoning,fung2020limitations, tolpegin2020data}, the attackers poison their local training data by flipping the labels of training examples of a source class $c_{src}$ to a target class $c_{target} \in \mathcal{C}$ while keeping the input data features unchanged. Each attacker poisons her local data set $D_k$ as follows: for all examples in $D_k$ whose class label is $c_{src}$, change their class label to $c_{target}$. 
After poisoning their training data, attackers train their local models using the same hyper-parameters, loss function, optimization algorithm and model architecture sent by the server. Thus, the 
%JOSEP2. Rewritten sentence
attack only requires poisoning the training data, but the learning algorithm remains the same as for honest peers.
Finally, the attackers send their bad updates to the server, so that they are aggregated with other good updates. 

\textbf{Feasibility of the LF attack in FL.} 
Although the LF attack was introduced for centralized ML~\cite{biggio2012poisoning,steinhardt2017certified}, it is more feasible in the FL scenario because the server does not have access to the attackers' local training data. 
Furthermore, this attack can provoke a significant negative impact on the performance of the global model, but it cannot be easily detected because it does not influence non-targeted classes ---it causes minimal changes in the poisoned model~\cite{tolpegin2020data}. Furthermore, LF can be easily performed by non-experts and does not impose much computation overhead on attackers because it is an off-line computation that is done before training. 

\textbf{Assumptions on training data distribution.} 
Since the local data of the peers can come from heterogeneous sources~\cite{bonawitz2019towards,wang2019edge}, they may be either identically distributed (iid) or non-iid. In the iid setting, each peer holds local data representing the whole distribution, which makes the locally computed gradient an
unbiased estimator of the mean of all the peers' gradients. The iid setting requires each peer to have examples of all the classes in a similar proportion as the other peers. In the non-iid setting, the distributions of the peers' local data sets can be different in terms of the classes represented in the data and/or the number of examples each peer has of each class.
We assume that the distributions of the peers' training data may range from extreme non-iid to pure iid.
Consequently, each peer may have local data with i) all the classes being present in a similar proportion as in the other peers' local data (iid setting), ii) some
classes being present in a different proportion (mild non-iid setting), or iii) only one class (extreme non-iid setting, because the class held by a peer is likely to be different from the class held by another peer). The number of peers that have a specific class $c$ in their training data can be denoted as 
$K_c = |\{k \in K|c \in \mbox{Classes}(D_k)\}|$.
 
\textbf{Threat model.}
We consider an attacker or a coalition of $K'_{c_{src}}$ attackers, with $K'_{c_{src}} \leq (K_{c_{src}}/2)$, for the iid and the mild non-iid settings, and $K'_{c_{src}} < K_{c_{target}}$ for the extreme non-iid setting (see Section~\ref{design} for a justification).
The $K'_{c_{src}}$ attackers perform the LF attack by flipping their training examples labeled $c_{src}$ to a chosen target class $c_{target}$ before training their local models. Furthermore, we assume the aggregator to be honest and non-compromised, and the attacker(s) to have no control over the aggregator or the honest peers.
The goal of the attackers is to degrade as much as possible the performance of the global model on the source class examples at test time.

\section{Related work}
\label{related}

The defenses proposed in the literature to counter poisoning attacks (and LF attacks in particular) against FL are based on one of the following principles:

\begin{itemize}
\label{defenses}
    \item \emph{Evaluation metrics.} Approaches under this type exclude or penalize a local
    update if it has a negative impact on an evaluation metric of the global model, {\em e.g.} its accuracy. Specifically, \cite{nelson2008exploiting, jagielski2018manipulating} use a validation data set on the server to compute the loss on a designated metric caused by each local update. Then, updates that negatively impact on the metric are excluded from the global model aggregation. 
    However, realistic validation data require server knowledge on the distribution of the peers' data, which conflicts with the FL idea whereby the server does not see the peers' data.
    
    \item \emph{Clustering updates}. Approaches under this type cluster updates into two groups, where the smaller group is considered potentially malicious and, therefore, disregarded in the model learning process. Auror~\cite{shen2016auror} and multi-Krum (MKrum)~\cite{blanchard2017machine} assume that the peers' data are iid, which results in high false positive and false negative rates when the data are non-iid~\cite{awan2021contra}. Moreover, they require previous knowledge about the characteristics of the training data distribution~\cite{shen2016auror} or the number of expected attackers in the system~\cite{blanchard2017machine}.
    
    \item \emph{Peers' behavior}. This approach assumes that malicious peers behave similarly, which means that their updates will be more similar to each other than to those of honest peers. Consequently, updates are penalized based on their similarity. For example, FoolsGold (FGold)~\cite{fung2020limitations} and CONTRA~\cite{awan2021contra} limit the contribution of potential attackers with similar updates by reducing their learning rates or preventing them from being selected. However, they also tend to incorrectly penalize good updates that are similar, which results in substantial drops in the model performance~\cite{nguyen2021flguard, li2021auto}.
    
    \item \emph{Update aggregation}. This approach uses robust update aggregation methods that are sensitive to outliers at the coordinate level, such as the median~\cite{yin2018byzantine}, the trimmed mean (Tmean)~\cite{yin2018byzantine} or the repeated median (RMedian)~\cite{siegel1982robust}. In this way, bad updates will have little to no influence on the global model after aggregation. Although these methods achieve good performance with updates resulting from iid data for small DL models, their performance deteriorates when updates result from non-iid data, because they discard most of the information in model aggregation. Moreover, their estimation error scales up with the size of the model in a square-root manner~\cite{chang2019cronus}.
    Furthermore, RMedian~\cite{siegel1982robust} involves high computational cost due to the regression process it performs, whereas Tmean~\cite{yin2018byzantine} requires explicit knowledge about the fraction of attackers in the system.
\end{itemize}

Several works focus on analyzing specific parts of the updates to defend against poisoning attacks. \cite{jebreel2020efficient} proposes analyzing the output layer's biases to distinguish bad updates from good ones. However, it only considers the model poisoning attacks in the iid setting. 
%JOSEP2. Rewritten, because FoolsGold has already been discussed.
FGold~\cite{fung2020limitations} analyzes the output layer's weights to counter data poisoning attacks, but 
it has the shortcomings mentioned above. 
\cite{tolpegin2020data} uses PCA to analyze the weights associated with \textit{the possibly attacked source class} and excludes potential bad updates that differ from the majority of updates in those weights. However, the method needs an explicit knowledge about the possibly attacked source class or performs a brute-force search to find it, and is only evaluated under the iid setting with simple DL models. 
CONTRA~\cite{awan2021contra} integrates FGold~\cite{fung2020limitations} with a reputation-based mechanism to penalize potential bad updates and prevent peers with low reputation from being selected. However, the method is only evaluated under mild non-iid settings using different Dirichlet distributions~\cite{minka2000estimating}.
The methods just cited share the shortcomings of (i) making assumptions on the distributions of peers' data and (ii) not providing analytical or empirical evidence of why focusing on specific parts of the updates contributes towards defending against the LF attack.

%JOSEP2. Rewritten.
In contrast, we analytically and empirically justify why focusing on the gradients of the parameters connected to the neurons of the source and target classes in the output layer is more helpful to defend against the attack. 
Also, we propose a novel defense that stays robust under different data distributions and model sizes, and does not require prior knowledge about the number of attackers in the system.

\section{Our defense against LF attacks}
\label{meth}

In this section, we first introduce the rationale of our proposal. Based on that, we present the design of an effective defense against the label-flipping attack.
\subsection{Rationale of our defense}
\label{rational}
The effectiveness of any defense against the LF attack depends on its ability to distinguish good updates sent by honest peers from bad updates sent by attackers. 
In this section, we conduct comprehensive theoretical and empirical analyses of the attack behavior to find a discriminative pattern that better differentiates good updates from bad ones. 

\paragraph{\bf Theoretical analysis of the LF attack}
To understand the behavior of the LF attack from an analytical perspective, let us consider a classification task where each local model is trained with the {\itshape cross-entropy} loss over one-hot encoded labels.
First, the vector $o$ of the output layer neurons ({\em i.e.}, the logits) is fed into the {\itshape softmax} function to compute the vector $p$ of probabilities as 
%\begin{equation}
%    \label{softmax}
 \[        p_k = \frac{e^{o_k}}{\sum_{j = 1}^{|\mathcal{C}|} e^{o_j}},\;\;\; k=1, \ldots, |\mathcal{C}|.\]
%\end{equation}
Then, the loss is computed as
%\begin{equation}
%    \label{loss_cross_entropy}
\[    \mathcal{L}(y, p) = - \sum_{k= 1}^{|\mathcal{C}|} y_k \log(p_k),\]
%\end{equation}
where $y = (y_1, y_2, \ldots, y_{|\mathcal{C}|})$ is the corresponding one-hot encoded true label and $p_k$ denotes the confidence score predicted for the $k^{th}$ class. 
After that, the gradient of the loss w.r.t. the output $o_{i}$ of the $i^{th}$ neuron ({\em i.e.}, the $i^{th}$ neuron error) in the output layer is computed as
\begin{dmath*}
\label{derivative_neuron}
    \delta_i = \frac{\partial \mathcal{L}(y, p)}{\partial o_i}
    =-\sum_{j= 1}^{|\mathcal{C}|} \frac{\partial \mathcal{L}(y, p)}{\partial p_j} \frac{\partial p_j}{\partial o_i} 
   = - \frac{\partial \mathcal{L}(y, p)}{\partial p_i} \frac{\partial p_i}{\partial o_i} - \sum_{j\ne i} \frac{\partial \mathcal{L}(y, p)}{\partial p_j} \frac{\partial p_j}{\partial o_i}
    = p_i - y_i.
\end{dmath*}
Note that $\delta_i$ will always be in the interval $[0, 1]$ when $y_i = 0$ (for the wrong class neuron), while it will always be in the interval $[-1, 0]$ when $y_i = 1$ (for the true class neuron).

The gradient $\nabla b^L_i$ w.r.t. the bias $b^L_i$ of the $i^{th}$ neuron in the output layer can be written as

\begin{equation}
\label{derivative_bias}
    \nabla b^L_i = \frac{\partial \mathcal{L}(y, p)}{\partial b^L_{i}} 
    % = \frac{\partial \mathcal{L}(y, p)}{\partial o_{i}}  \frac{\partial o_i}{\partial b^L_{i}} 
    =  \delta_i \frac{\partial \sigma^L}{\partial (w^L_i \cdot a^{L-1} + b^L_{i})}, 
\end{equation}
where $a^{L-1}$ is the activation output of the previous layer $L-1$.

Likewise, the gradient $\nabla w^L_i$ w.r.t. the weights vector $w^L_i$ connected to the $i^{th}$ neuron in the output layer is
\begin{equation}
\label{derivative_weights}
    \nabla w^L_i = \frac{\partial \mathcal{L}(y, p)}{\partial w^L_i} 
    % = \frac{\partial \mathcal{L}(y, p)}{\partial o_{i}}  \frac{\partial o_i}{\partial w^L_i} 
    =   \delta_i  a^{L-1} \frac{\partial \sigma^L}{\partial (w^L_i \cdot a^{L-1} + b^L_{i})}. 
\end{equation}

From Equations~\eqref{derivative_bias} and~\eqref{derivative_weights}, we can notice that $\delta_i$ directly and highly impacts on the gradients of its connected parameters. 
For example, for the ReLU activation function, which is widely used in DL models, we get 

\begin{dmath*}
%\label{bias_gradient}
    \nabla b^L_i  =
    \left\{
        \begin{array}{ll}
         \delta_i, \enspace \mbox{if $(w^L_i \cdot a^{L-1} + b^L_{i}) > 0$;}\\
           \quad 0, \enspace \enspace \mbox{otherwise;}
        \end{array}
    \right.
\end{dmath*}

and

\begin{dmath*}
%\label{weight_gradient}
    \nabla w^L_i   =
    \left\{
        \begin{array}{ll}
         \delta_i a^{L-1}, \enspace \mbox{if $(w^L_i \cdot a^{L-1} + b^L_{i}) > 0$;}\\
           \quad 0, \enspace \enspace \mbox{otherwise.}
        \end{array}
    \right.
\end{dmath*}

The objective of the attackers is to minimize $p_{c_{src}}$ and maximize $p_{c_{target}}$ for their $c_{src}$ examples, whereas the objective of honest peers is exactly the opposite.
We notice from Expressions \eqref{derivative_neuron}, \eqref{derivative_bias}, and 
\eqref{derivative_weights} that these contradicting objectives will be reflected on the gradients of the parameters connected to the {\itshape relevant} source and target output neurons.
For convenience, in this paper, we use the term {\itshape relevant neurons' gradients} instead of the gradients of the parameters connected to the source and target output neurons. Also, we use the term {\itshape non-relevant neurons' gradients} instead of the gradients of the parameters connected to the neurons different from source and target output neurons.
As a result, as the training evolves, the magnitudes of the relevant neurons' gradients are expected to be larger than those of the non-relevant and non-contradicting neurons. Also, the angle between the relevant neurons' gradients for an honest peer and an attacker is expected to be larger than those of the non-relevant neurons' gradients.
That is because the error of the non-relevant neurons will diminish as the global model training evolves, especially when it starts converging (honest and malicious participants share the same training objectives for non-targeted classes).
On the other hand, the relevant neurons' errors will stay large during model training because of the contradicting objectives.  
Therefore, the relevant neurons' gradients are expected to carry a more valuable and discriminative pattern for an attack detection mechanism than the whole model gradients or the output layer gradients, which carry a lot of information not relevant to the attack.

\paragraph{\bf Empirical analysis of the LF attack}
%\textbf{Empirical analysis of the LF attack}.
To empirically validate the analytical findings discussed above and see how the model size and the data distribution impact on the detection of LF attacks, we used exploratory analysis to visualize the gradients sent by peers in five different FL scenarios under label-flipping attacks: 
% ALBERTO: Should this be described before?
% Najeeb : A possible solution is to bring the experimental setup section before attacks analysis.
MNIST-iid, MNIST-Mild, MNIST-Extreme, CIFAR10-iid and CIFAR10-Mild. 
Besides the whole updates, we visualized the output layer's gradients and the relevant neurons' gradients.
% ALBERTO: If we want to shorten this, this can be omitted
% We intentionally chose source and target classes with similar features to make the attack detection more challenging.
%Najeeb: OK.
%JOSEP2. Slightly rewritten.
The FL attacks in the MNIST benchmarks consisted of flipping class $7$ to class $1$, while in the CIFAR10 benchmarks they consisted of flipping \textit{Dog} to \textit{Cat}. 
For the MNIST benchmarks, we used a simple DL model which contains about $22K$ parameters. For the CIFAR10 benchmarks we used the ResNet18~\cite{he2016deep} architecture, which yields large models containing about $11M$ parameters. 
The details of the experimental setup are given in Section~\ref{setup}. 
In order to visualize the updates, we used Principal Component Analysis (PCA)~\cite{wold1987principal} on the selected gradients and we plotted the first two principal components.
We next report what we observed.

\begin{figure*}[!htbp]
    \centering
    \begin{subfigure}{0.3\textwidth}
      \centering
      \includegraphics[width=1\linewidth]{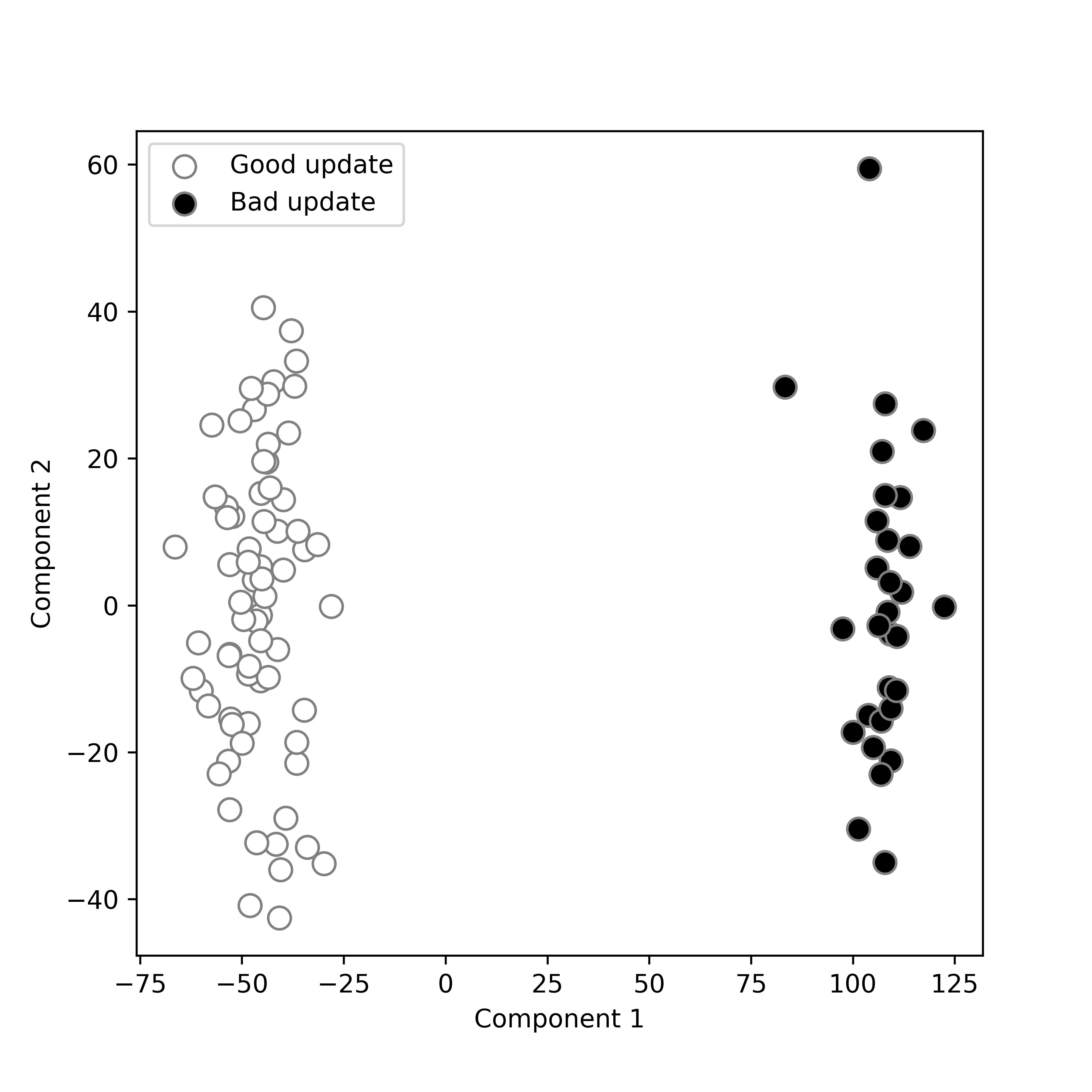}\vspace{-1\baselineskip}
      \caption{Whole}
      \label{fig:pca_mnist_iid_all}
    \end{subfigure}% 
    \begin{subfigure}{0.3\textwidth}
      \centering
      \includegraphics[width=1\linewidth]{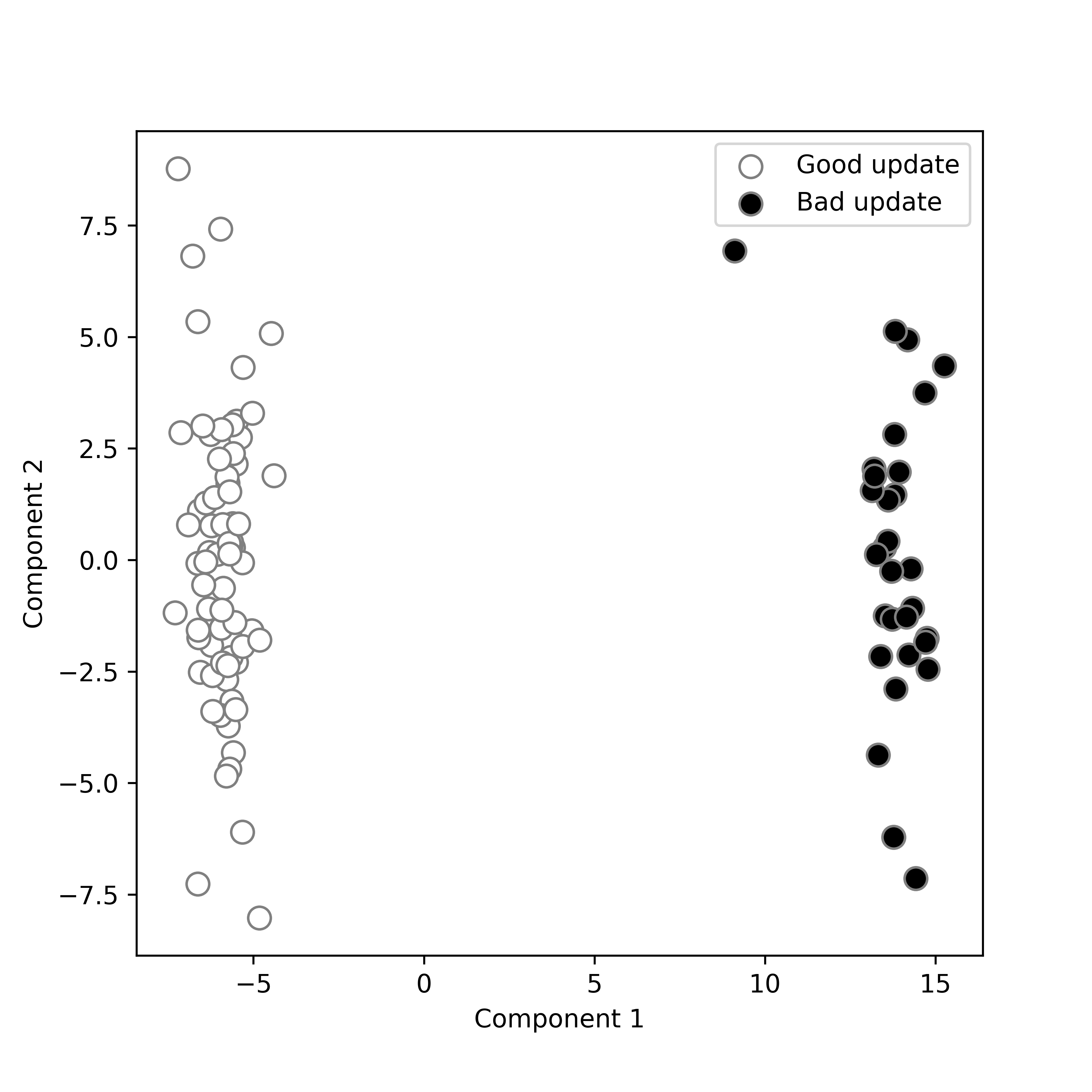}\vspace{-1\baselineskip}
      \caption{Output layer}
      \label{fig:pca_mnist_iid_last}
    \end{subfigure}%
    \begin{subfigure}{0.3\textwidth}
      \centering
      \includegraphics[width=1\linewidth]{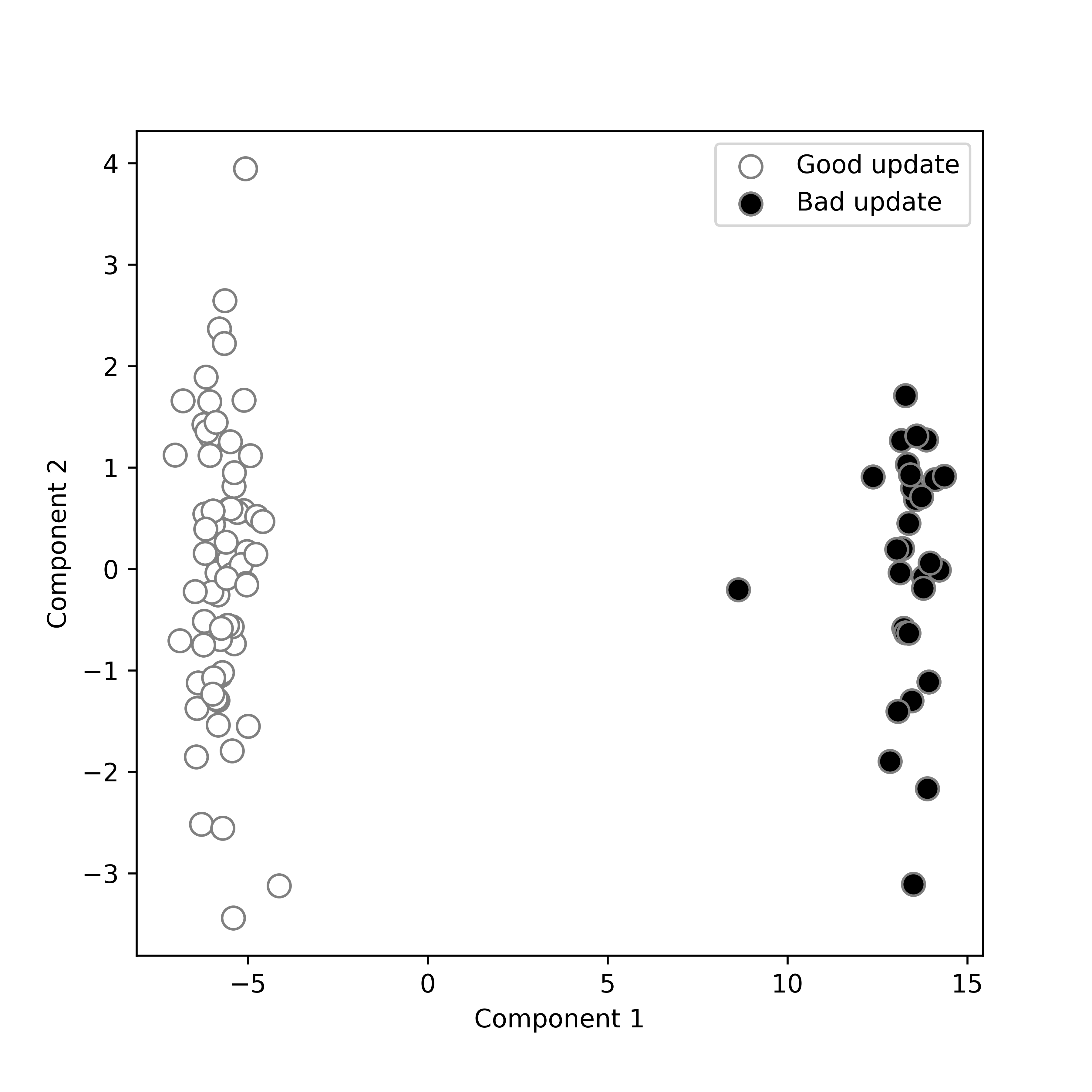}\vspace{-1\baselineskip}
      \caption{Relevant neurons}
      \label{fig:pca_mnist_iid_st}
    \end{subfigure}%
\caption{First two PCs of the MNIST-iid benchmark gradients}
\label{fig:pca_mnist_iid}
\end{figure*}

\begin{figure*}[!htbp]
    \centering
    \begin{subfigure}{0.3\textwidth}
      \centering
      \includegraphics[width=1\linewidth]{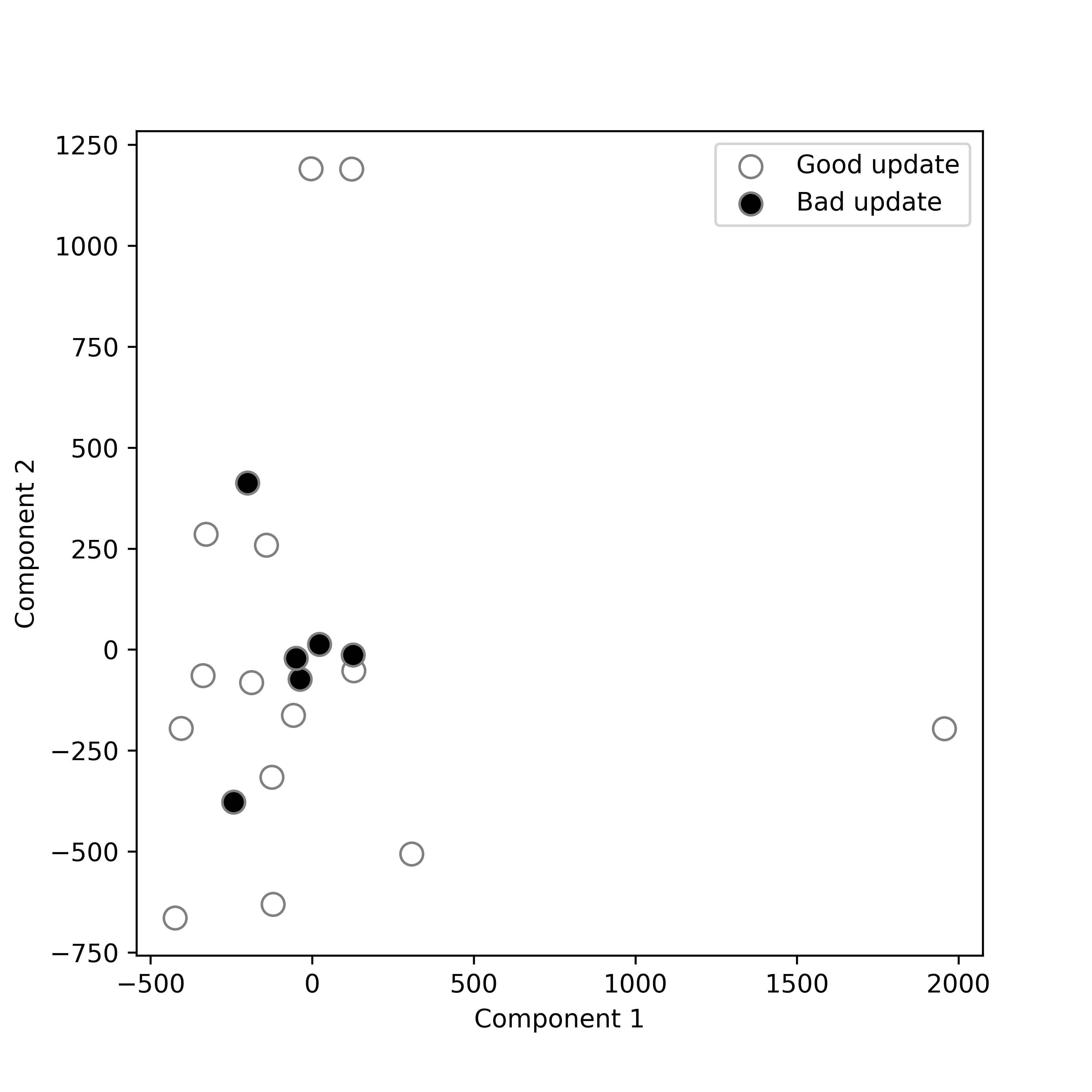}\vspace{-1\baselineskip}
      \caption{Whole}
      \label{fig:pca_cifar10_iid_all}
    \end{subfigure}% 
    \begin{subfigure}{0.3\textwidth}
      \centering
      \includegraphics[width=1\linewidth]{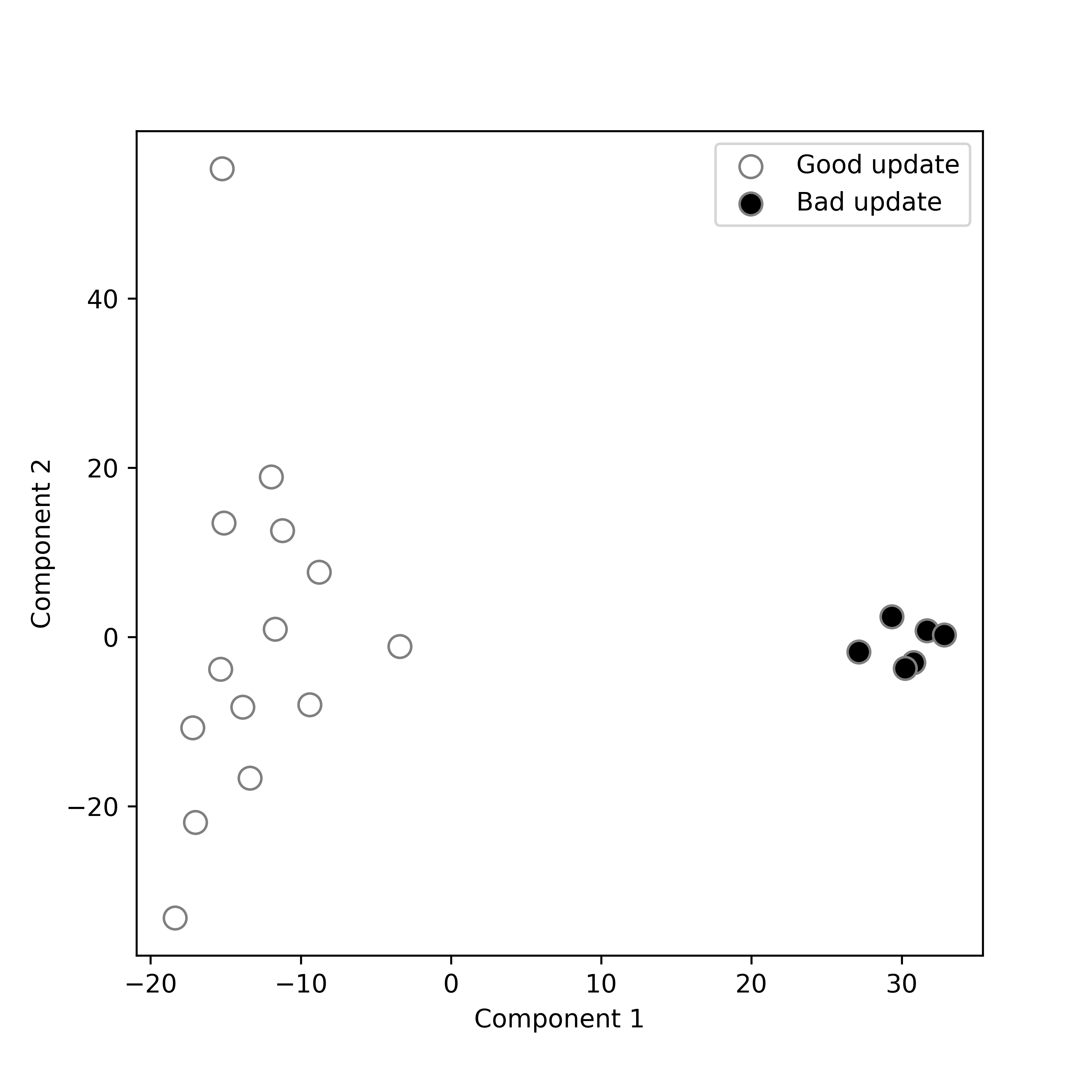}\vspace{-1\baselineskip}
      \caption{Output layer}
      \label{fig:pca_cifar10_iid_last}
    \end{subfigure}%
    \begin{subfigure}{0.3\textwidth}
      \centering
      \includegraphics[width=1\linewidth]{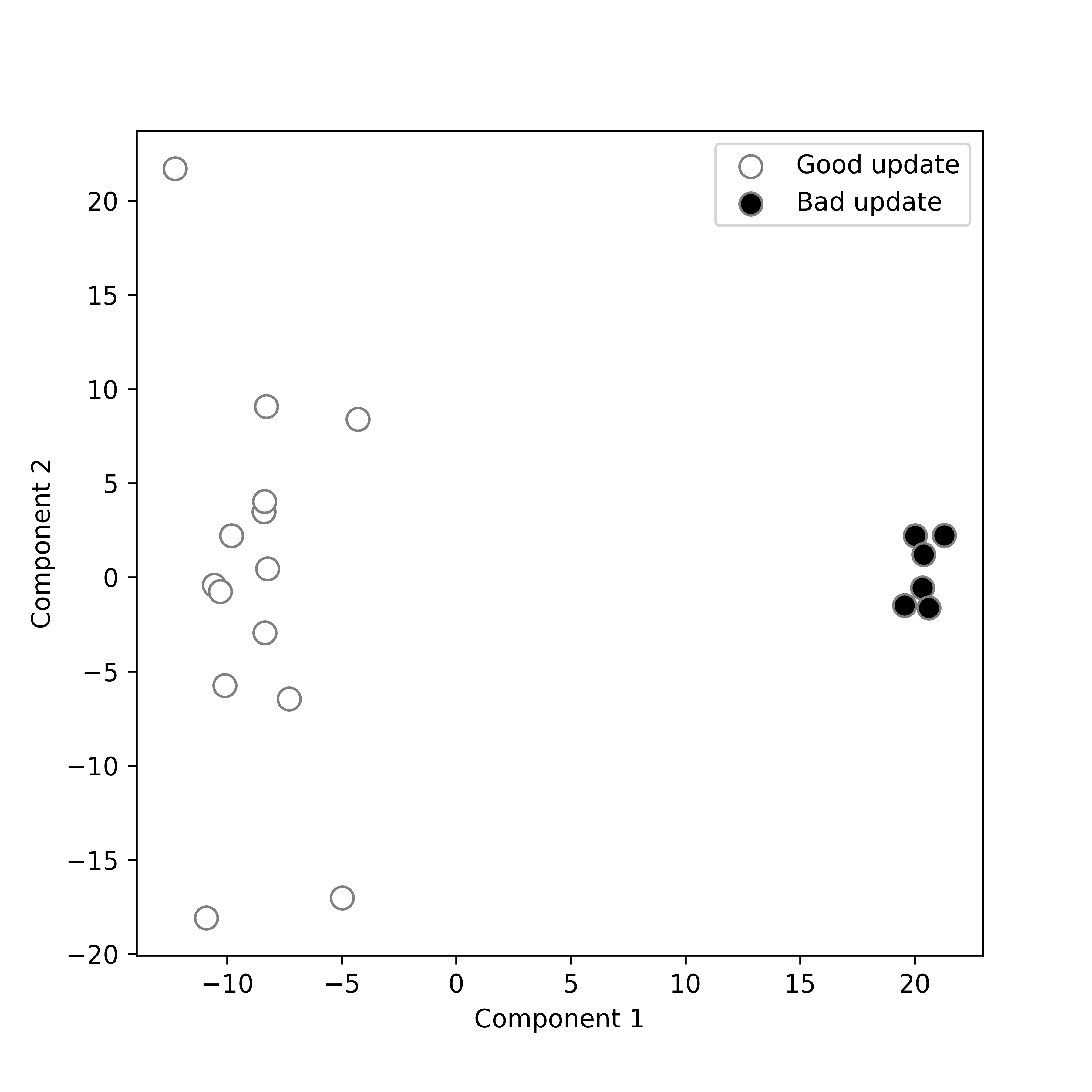}\vspace{-1\baselineskip}
      \caption{Relevant neurons}
      \label{fig:pca_cifar10_iid_st}
    \end{subfigure}%
\caption{First two PCs of the CIFAR10-iid benchmark gradients}
\label{fig:pca_cifar10_iid}
\end{figure*}

\noindent 1) \textbf{Impact of model dimensionality}. Figures~\ref{fig:pca_mnist_iid} and~\ref{fig:pca_cifar10_iid} show the gradients of whole local updates, gradients corresponding to the output layers, and relevant gradients corresponding to the source and target neurons from the MNIST-iid ($30$ bad updates out of $100$) and the CIFAR10-iid ($6$ bad updates out of $20$) benchmarks, respectively.
In these two benchmarks, the training data were iid among peers.

%JOSEP2. I recall which model is small and which is large.
The figures show that, when the model size is small (MNIST-iid), good and bad updates can be easily separated,
% whether we use the whole gradients, the output layer gradients, or gradients of the parameters connected to the source and target classes' neurons.
whichever set of gradients is considered.
On the other hand, when the model size is large (CIFAR10-iid), the attack's influence does not seem to be enough to distinguish good from bad updates
%JOSEP2. Rewritten.
when using whole update gradients;
% analyzing the whole update leads to bad updates getting close to good ones, which makes it challenging to tell them apart. 
yet, the gradients of the output layer or those of the relevant neurons still allow for a crisp differentiation between good and bad updates.

In fact, several factors make it challenging to detect LF attacks by analyzing an entire high-dimensional update.
First, the computed errors for the neurons in a certain layer depend on all the errors for the neurons in the subsequent layers and their connected weights~\cite{rumelhart1986learning}. Thus, as the model size gets larger,
% as we backpropagate the errors
the impact of the attack is mixed with that of the non-relevant neurons.
Second, the early layers of DL models usually extract common features that are not class-specific~\cite{nasr2019comprehensive}.
Third, in general, most parameters in DL models are redundant~\cite{denil2013predicting}. 
These factors cause the magnitudes of the whole gradients of good and bad updates and the angles between them to be similar, making DL models with large dimensions an ideal environment for a successful label-flipping attack.

To confirm these observations, we performed the following experiment with the CIFAR10-iid benchmark. 
First, a chosen peer trained her local model on her data honestly, which yielded a good update. 
Then, the same peer flipped the labels of the source class \textit{Cat} to the target class \textit{Dog} and then trained her local model on the poisoned training data, which yielded a bad update.
After that, we computed the magnitudes of and the angle between 
i) the whole updates, 
ii) the output layer gradients, 
iii)  the relevant gradients related to $c_{src}$ and $c_{target}$.
% Also, we computed the angle between each of them for the whole gradients, the output layer's gradients and the gradients corresponding to the neurons' weights and biases of the source and target classes.
Table~\ref{tab:norm_angle} shows the obtained results, which confirm our analytical and empirical findings. 
It is clear that both whole gradients had approximately the same magnitude, and the angle between them was close to zero. 
On the other hand, the difference between the output layer gradients was large and even more significant in the case of the relevant neurons' gradients.
As for non-relevant neurons, their gradients' magnitude and angle were not significantly affected because in such neurons there was no contradiction between the objectives of the good and the bad updates.

\begin{table*}[!ht]
\centering
\caption{Comparison of the magnitudes and the angle of the gradients of a good and a bad update for the whole update, the output layer 
%JOSEP2. Rewritten.
parameters, the parameters of the relevant source and target neurons, and the parameters of the non-relevant neurons.}
\label{tab:norm_angle}
\resizebox{0.7\textwidth}{!}{%
\begin{tabular}{|cc|c|c|c|c|}
\hline
\multicolumn{2}{|c|}{Gradients}               & Whole & Output layer & Relevant neurons & Non-relevant neurons \\ \hline
\multicolumn{1}{|c|}{\multirow{2}{*}{Magnitude}} & Good & 351123         & 72.94                 & 23.38                                & 55.30                   \\ \cline{2-6} 
\multicolumn{1}{|c|}{}                         & Bad  & 351107         & 100.23                & 64.43                                & 65.95                   \\ \hline
\multicolumn{2}{|c|}{Angle}                           & 0.41           & 69.19                 & 115                         & 18                      \\ \hline
\end{tabular}%
}
\end{table*}

\begin{figure*}[!htbp]
    \centering
    \begin{subfigure}{0.4\textwidth}
      \centering
      \includegraphics[width=1\linewidth]{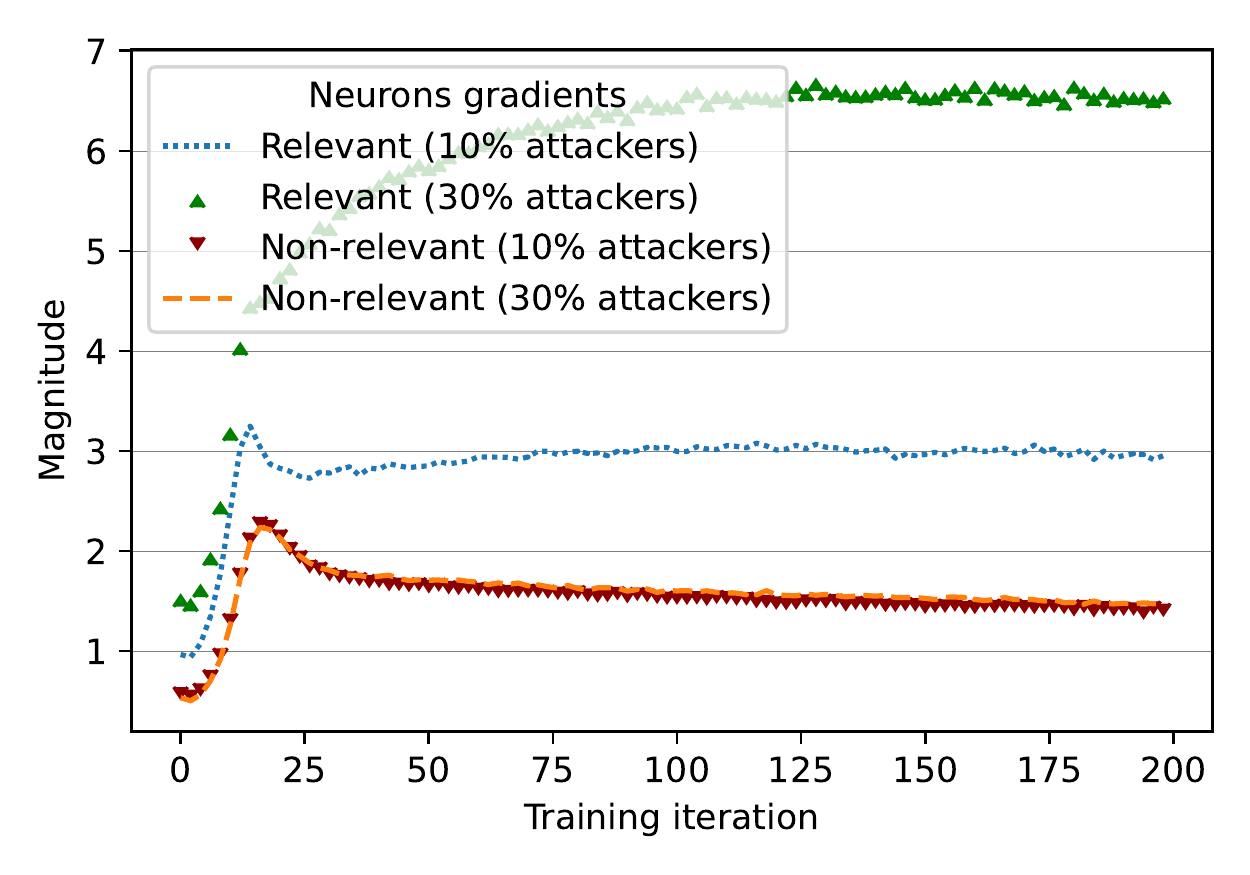}
      \caption{MNIST-iid}
      \label{fig:mnist_iid_magnitude}
    \end{subfigure}% 
    \begin{subfigure}{0.4\textwidth}
      \centering
      \includegraphics[width=1\linewidth]{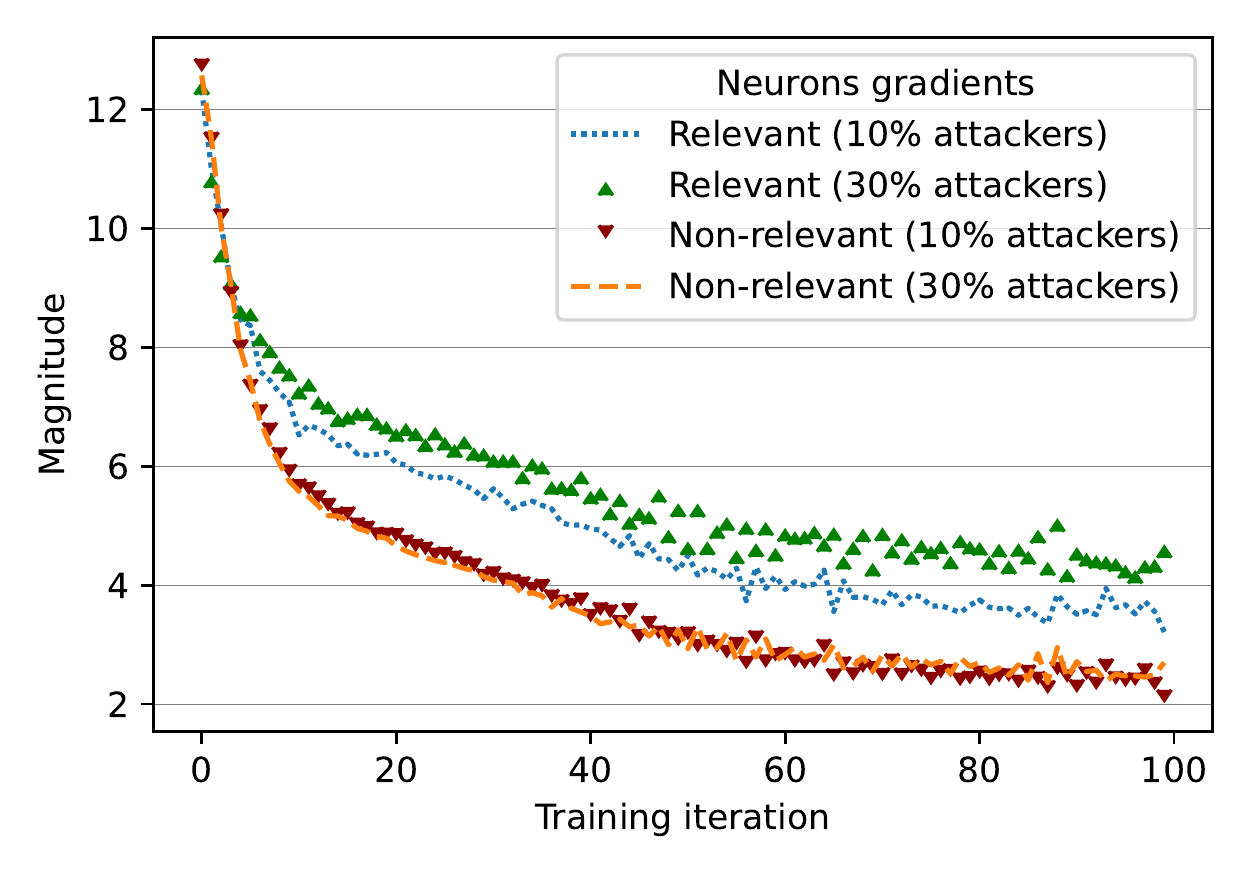}
      \caption{CIFAR10-Mild}
      \label{fig:cifar10_mild_magnitude}
    \end{subfigure}
%JOSEP2. Rewritten caption.
\caption{Gradient magnitudes during training for relevant and non-relevant neurons}
\label{fig:gradients_magnitude}
\end{figure*}

To underscore this point and see how the gradients of the non-relevant neurons vanish as the training evolves, while the gradients of the relevant neurons remain larger, we show the gradients' magnitudes during training in Figure~\ref{fig:gradients_magnitude}.
The magnitudes of those gradients for the MNIST-iid and the CIFAR10-Mild benchmarks are shown for ratios of attackers $10\%$ and $30\%$. 
We can see that, although the attackers' ratio and the data distribution had an impact on the magnitudes of those gradients, 
%JOSEP2. A bit rewritten.
the gradients' magnitudes for the relevant source and target class neurons always remained larger.

\begin{figure*}[!htbp]
    \centering
    \begin{subfigure}{0.3\textwidth}
      \centering
      \includegraphics[width=1\linewidth]{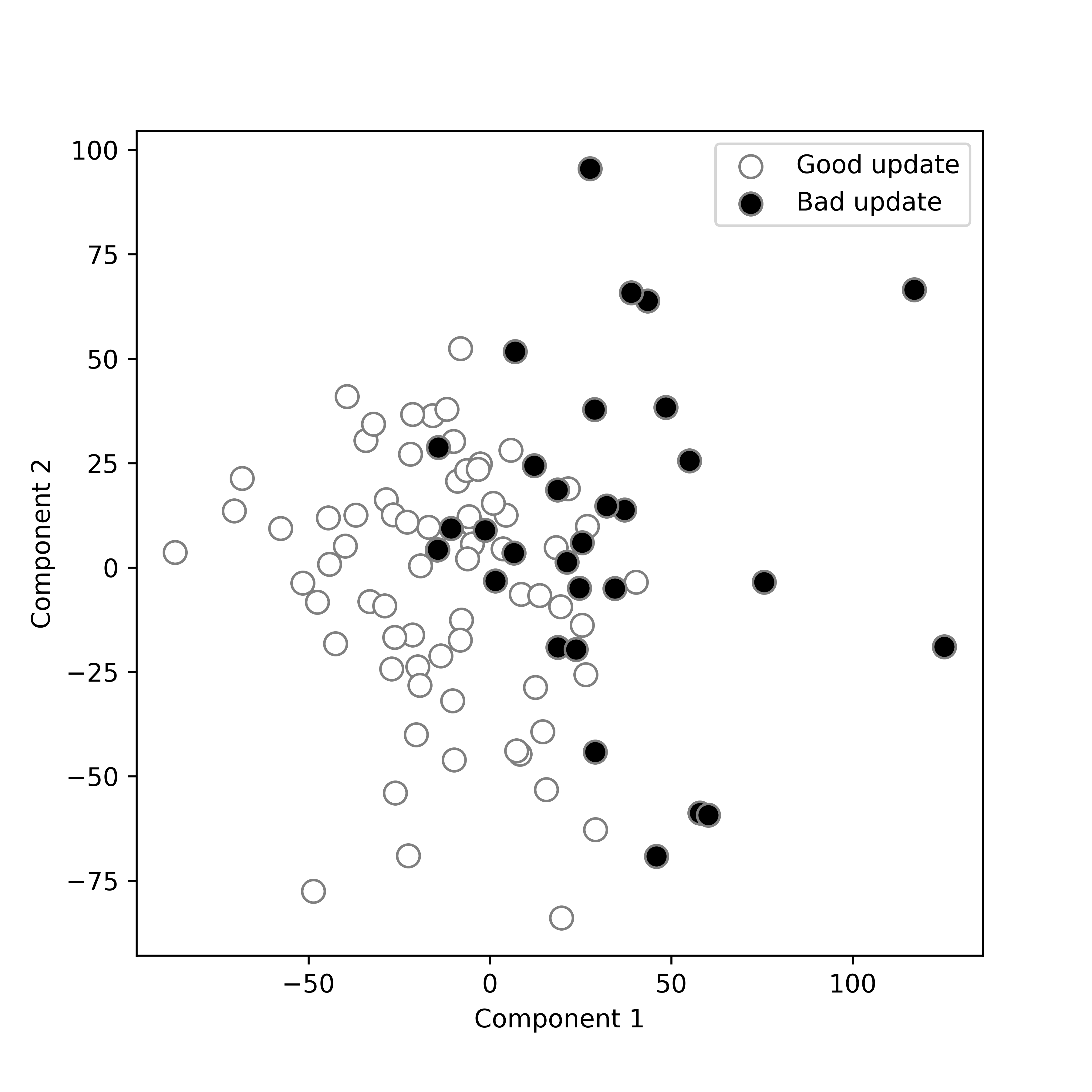}\vspace{-1\baselineskip}
      \caption{Whole}
      \label{fig:pca_mnist_mild_all}
    \end{subfigure}% 
    \begin{subfigure}{0.3\textwidth}
      \centering
      \includegraphics[width=1\linewidth]{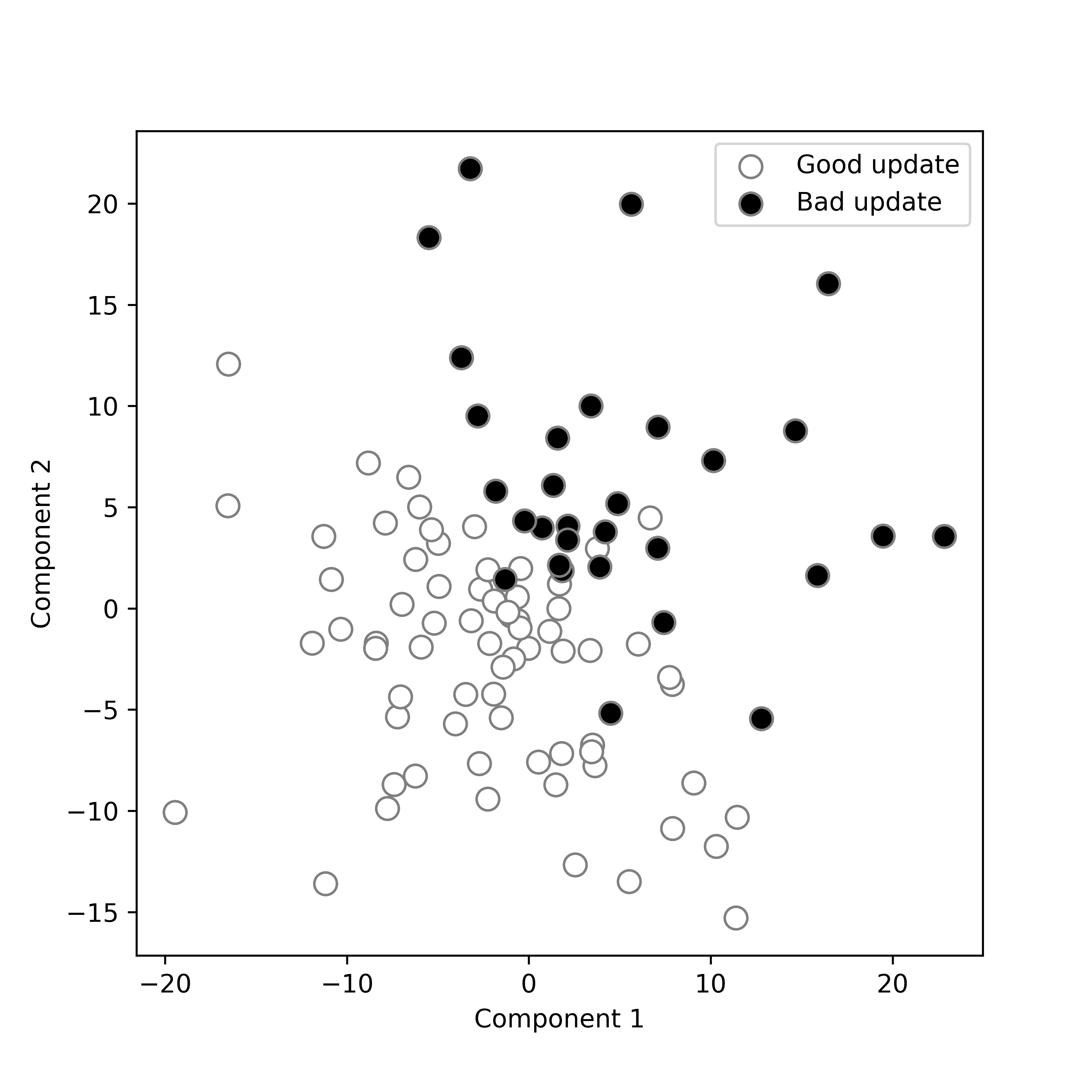}\vspace{-1\baselineskip}
      \caption{Output layer}
      \label{fig:pca_mnist_mild_last}
    \end{subfigure}%
    \begin{subfigure}{0.3\textwidth}
      \centering
      \includegraphics[width=1\linewidth]{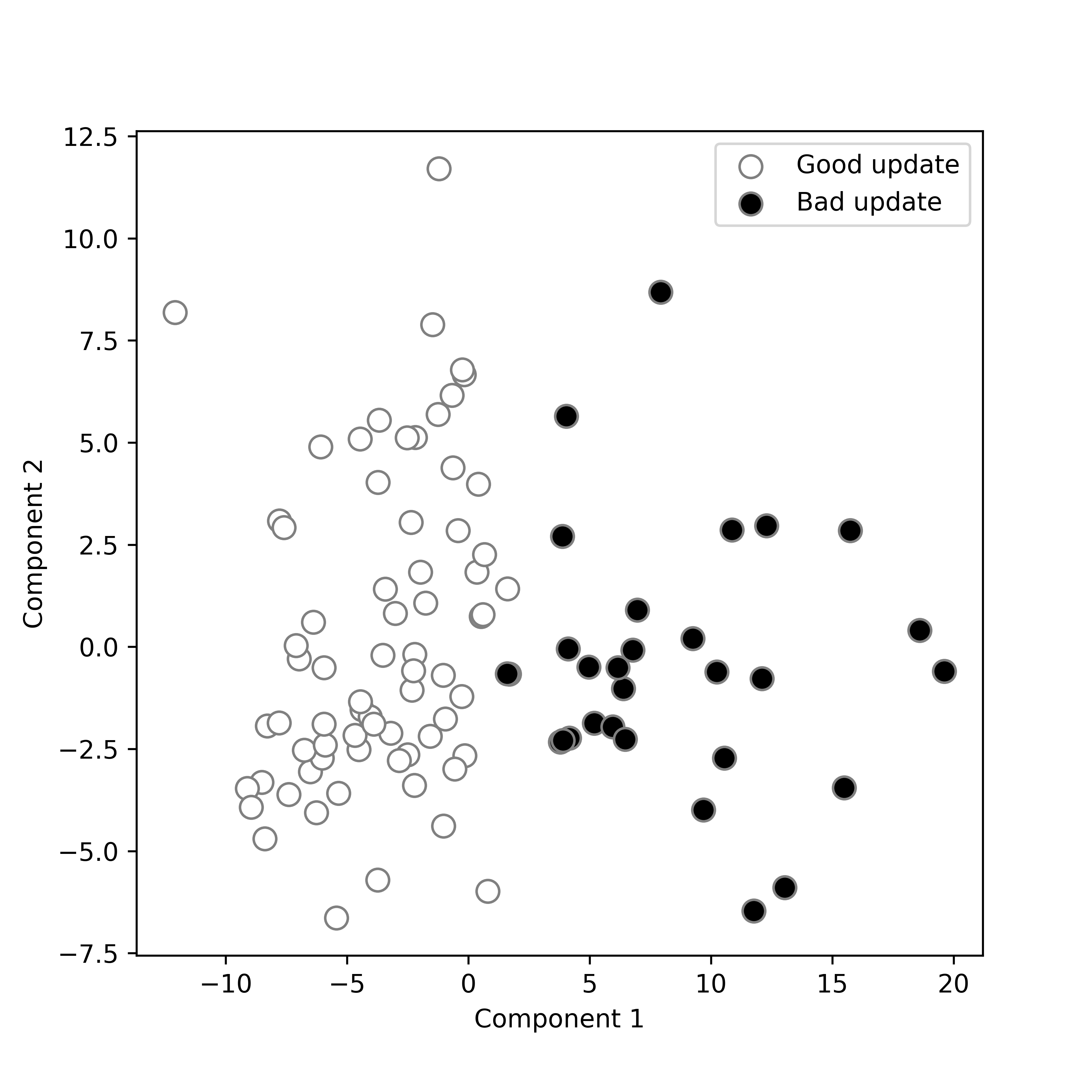}\vspace{-1\baselineskip}
      \caption{Relevant neurons}
      \label{fig:pca_mnist_mild_st}
    \end{subfigure}%
\caption{First two PCs of the MNIST-Mild benchmark gradients}
\label{fig:pca_mnist_mild}
\end{figure*}

\begin{figure*}[!htbp]
    \centering
    \begin{subfigure}{0.3\textwidth}
      \centering
      \includegraphics[width=1\linewidth]{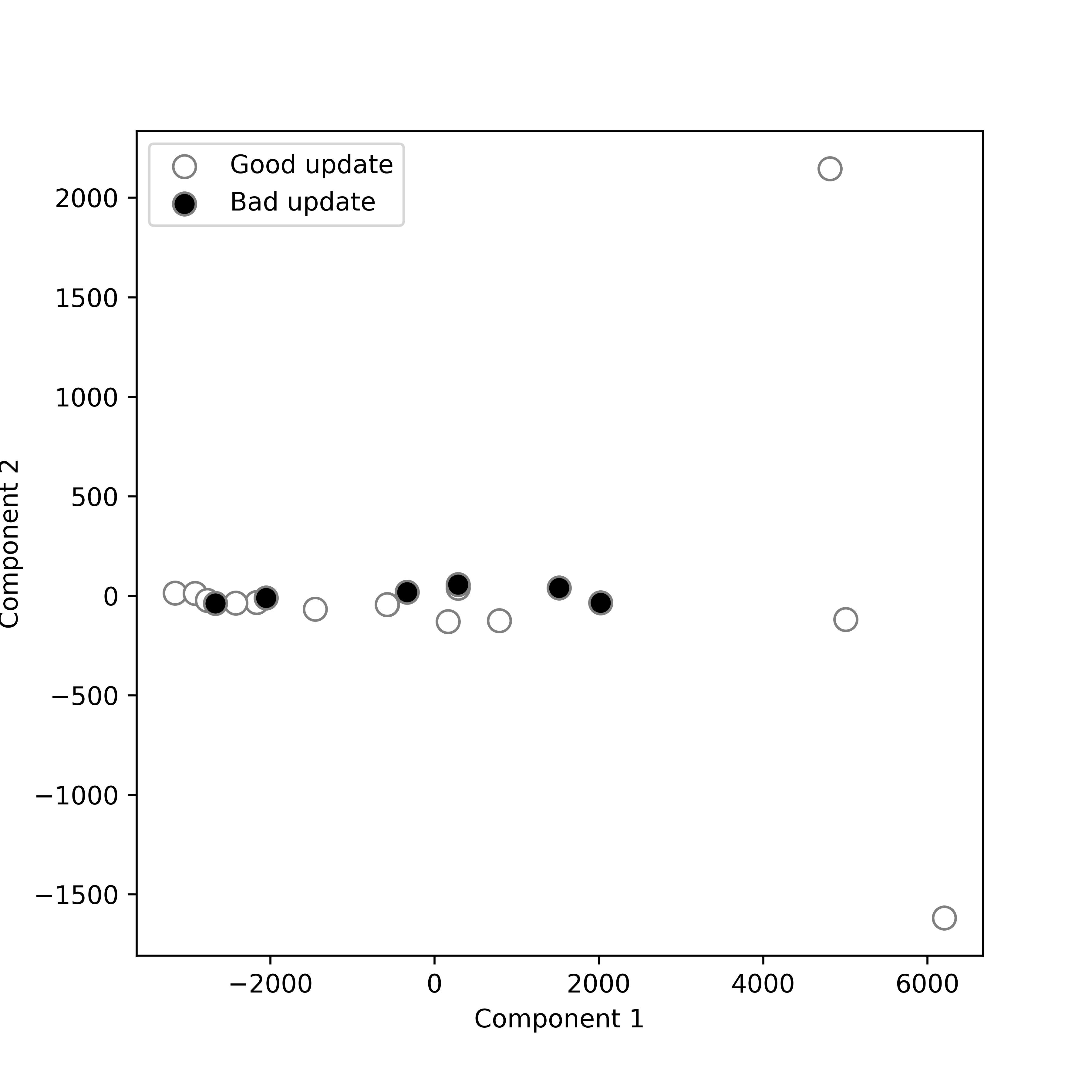}\vspace{-1\baselineskip}
      \caption{Whole}
      \label{fig:pca_cifar10_mild_all}
    \end{subfigure}% 
    \begin{subfigure}{0.3\textwidth}
      \centering
      \includegraphics[width=1\linewidth]{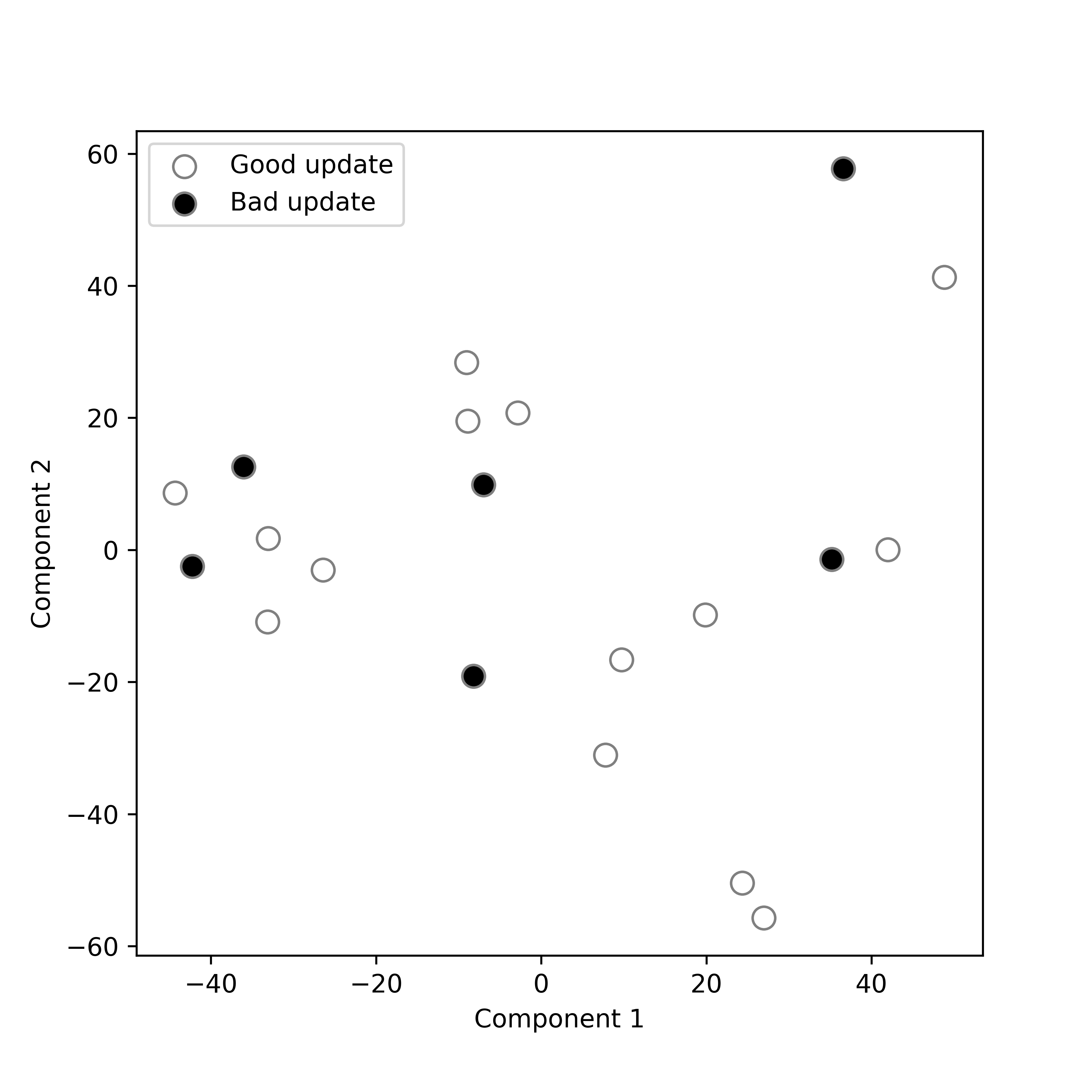}\vspace{-1\baselineskip}
      \caption{Output layer}
      \label{fig:pca_cifar10_mild_last}
    \end{subfigure}%
    \begin{subfigure}{0.3\textwidth}
      \centering
      \includegraphics[width=1\linewidth]{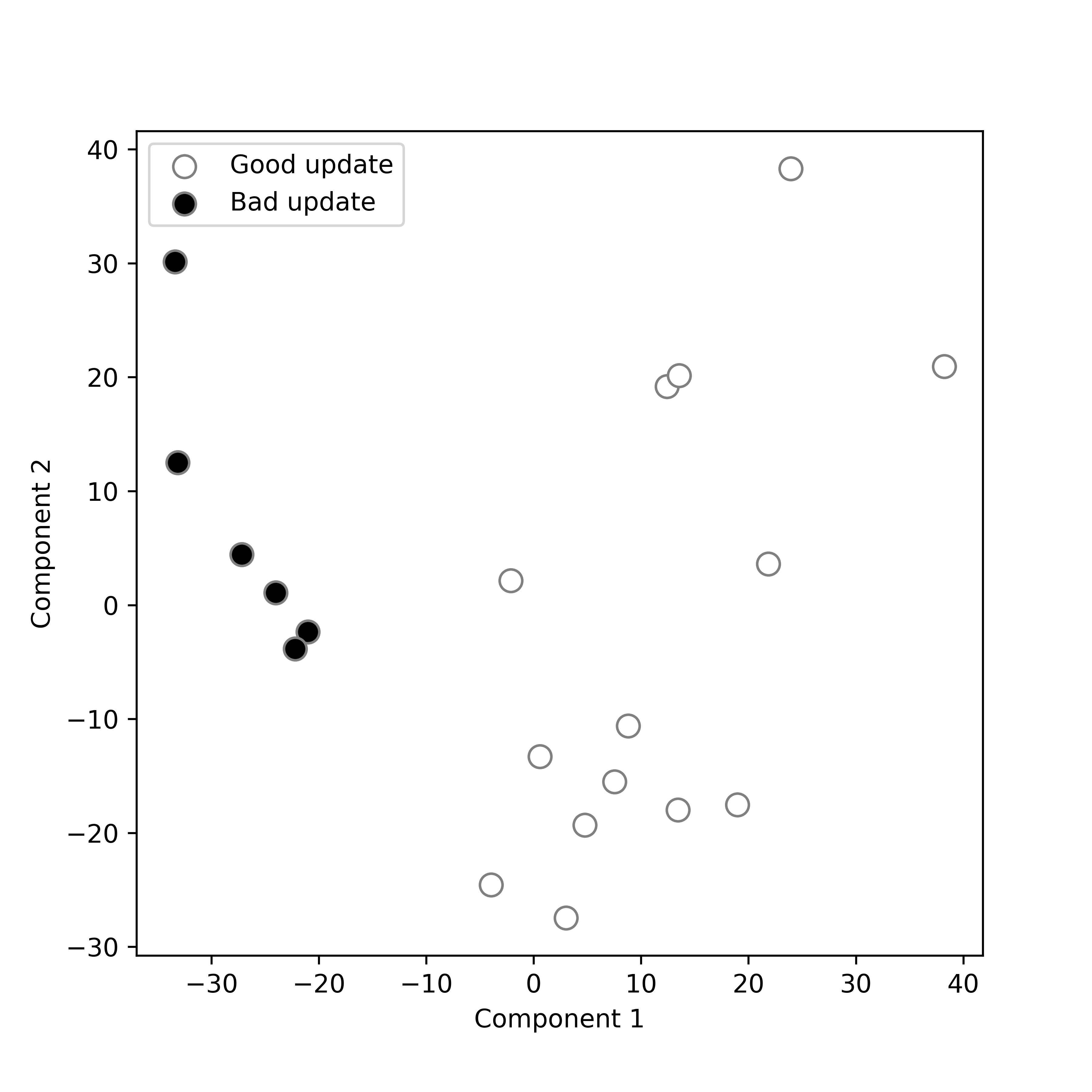}\vspace{-1\baselineskip}
      \caption{Relevant neurons}
      \label{fig:pca_cifar10_mild_st}
    \end{subfigure}%
\caption{First two PCs of the CIFAR10-Mild benchmark gradients}
\label{fig:pca_cifar10_mild}
\end{figure*}

\noindent 2) \textbf{Impact of data distribution}. Figure~\ref{fig:pca_mnist_mild} shows the gradients of $100$ local updates from the MNIST-Mild benchmark and their corresponding output layer and relevant neurons' gradients, where $30$ updates were bad. 
Figure~\ref{fig:pca_cifar10_mild} shows the same for the CIFAR10-Mild benchmark, where $6$ out of $20$ local updates were bad.
In these two benchmarks, the training data were distributed following a mild non-iid Dirichlet distribution among peers~\cite{minka2000estimating} with $\alpha = 1$.
%JOSEP2. Rewritten below.
Figure~\ref{fig:pca_mnist_mild} shows that, despite the model used for the MNIST-Mild benchmark being small, distinguishing between good and bad updates was harder than in the iid setting shown in Figure~\ref{fig:pca_mnist_iid}. 
It also shows that the use of the relevant neurons' gradients provided the best separation compared to whole update gradients or output layer gradients.

Figure~\ref{fig:pca_cifar10_mild} shows that the combined impact of model size and data distribution in the CIFAR-Mild benchmark made it very challenging to separate bad updates from good ones using  %JOSEP2. Rewritten.
whole update gradients or even using the output layer gradients. 
On the other hand, the relevant neurons' gradients gave a clearer separation.
%JOSEP2. Suppressed because it was not very clear.
%The reason for this is that the relevant neurons' gradients do not make a big difference in the output layer's gradients due to the high diversity of the local data in this benchmark.

From the previous analyses, we can observe that analyzing the gradients of the parameters connected to the source and target class neurons led to better discrimination between good updates and bad ones for both the iid and the mild non-iid settings. 
We can also observe that, in general, those gradients formed two clusters: one cluster for the good updates and another cluster for the bad updates.
Moreover, the attackers' gradients were more similar among them and caused their clusters to be denser than the honest peers' clusters. 
% ALBERTO: I would take this out if we need to save space, 
% since this argument is used already in the background section.
% That is because the attackers always minimize $p_{c_{src}}$, whereas the honest peers maximize $p_{c_{src}}$ when their examples belong to $c{src}$, and they also minimize it for the examples different from $c{src}$. That will cause the pairwise angles between attackers' gradients to be smaller than those of honest peers, which was also observed in~\cite{fung2020limitations}, where they show that attackers contribute updates that appear more similar to each other than those among honest peers.
%Najeeb: OK. In case the reviewers asked for explaining that point we could use the omitted text.

\begin{figure*}[!htbp]
    \centering
      \includegraphics[width=0.4\linewidth]{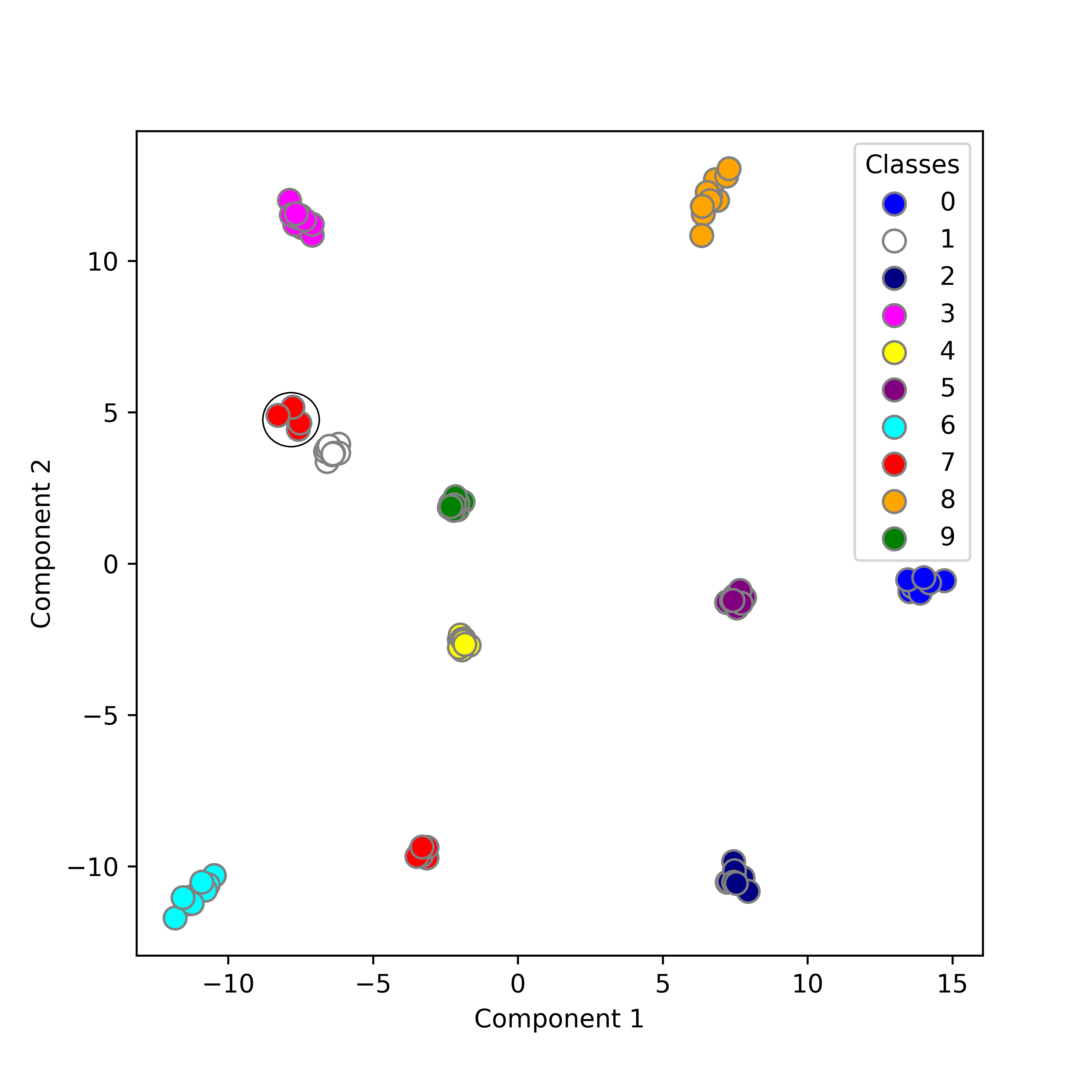}
\caption{Extreme non-iid setting. First two PCs of the MNIST-Extreme source and target classes' neurons gradients. Circled updates are bad.}
\label{fig:pca_mnist_extreme}
\end{figure*}

However, what would be the case when the data are extremely non-iid, that is, when each peer has local training data of a single class?
Figure~\ref{fig:pca_mnist_extreme} shows the gradients of the relevant neurons' gradients of $100$ local updates from the MNIST-Extreme benchmark, where each peer provided examples of a single class. 
In this experiment, $4$ attackers out of the $10$ peers who had examples of the class $7$, flipped the labels of their training examples from the source class $7$ to the target class $1$. 
The figure shows that the gradients of the updates of each class form an individual cluster, and the $4$ bad updates form a cluster that is very close to the cluster of the target class $1$ updates. 
The explanation is that, in the extreme non-iid setting, most peers have classes different from $c_{src}$ in their data, and hence, the honest peers have less influence on $\delta_{c_{src}}$ than in the iid or the mild non-iid settings.
Therefore, the alteration of $\delta_{c_{src}}$ via decrease of $p_{c_{src}}$ is less detectable than the alteration of $\delta_{c_{target}}$ via increase of $p_{c_{target}}$, because averaging local updates should decrease both $p_{c_{src}}$ and $p_{c_{target}}$ 
%JOSEP2. Added parenthetical remark.
(in this extreme non-iid setting, each class is absent from the local data of most peers).
In this benchmark, bad updates got close to the target class cluster because both the attackers and class $1$ honest peers shared a common objective, which is maximizing $p_{c_{1}}$.
On the other hand, what made them different is the attackers' aim to minimize $p_{c_{7}}$ to mitigate the remaining impact of the honest peers in $K_{c_{7}}$ after the global model aggregation.

Based on the analyses and observations presented so far, we concluded that {\em an effective defense against the label-flipping attack needs to consider the following aspects:}
\begin{itemize}
%JOSEP2. Rewritten.
    \item Only the gradients of the parameters connected to the source and target class neurons in the output layer must
    be extracted and analyzed.
    
    \item If the data are iid or mild non-iid, the extracted gradients need to be separated into two clusters that are compared to identify which of them contains the bad updates. 
    
    \item If the data are extremely non-iid, the extracted gradients need to be dynamically clustered so that the gradients of the peers that have data belonging to the same class fall in the same individual cluster. Then, the bad updates cluster must 
    be compared with the target class's cluster.
\end{itemize}

\subsection{Design of our defense}
\label{design}

Considering the observations and conclusions discussed in the previous section, we present our proposed defense against the label-flipping attack in federated learning systems.
 
Unlike other defenses, our proposal does not require a prior assumption on the peers' data distribution, is not affected by model dimensionality, and does not require prior knowledge about the proportion of attackers. 
In each training iteration, our defense first separates the 
%JOSEP2. Rewritten
gradients of the output layer parameters
%output layer's parameters' gradients 
from the local updates. Then it dynamically identifies the two neurons with the highest gradient magnitudes as the potential source and target class neurons, and extracts the gradients of the parameters connected to them.
Next, it applies a proper clustering method on those extracted gradients based on the peers' data distribution.
Unlike existing approaches, we do not use a fixed strategy to address all types of local data distributions. 
Instead, we cluster the extracted gradients into two clusters using k-means~\cite{hartigan1979algorithm} for the iid and mild non-iid settings, while we cluster them into multiple clusters using HDBSCAN~\cite{campello2013density} for the extreme non-iid setting. 
% ALBERTO: This was said before
% That is because those gradients usually form two clusters in the iid and the mild non-iid settings, while they form multiple clusters in the extreme non-iid setting.
%Najeeb: OK.
Thereafter, we further analyze the resulting clusters to identify the bad cluster. 
In the iid and the mild non-iid settings, we consider the size and the density of clusters. The smaller and/or the denser cluster is identified as a potentially bad cluster.
In the extreme non-iid setting, we compare the two clusters with the same highest neuron gradients' magnitudes. 
The smaller cluster is identified as a potentially bad cluster.
Finally, we exclude the updates corresponding to the potentially bad cluster from the aggregation phase. 
Note that discovering whether the data of peers are iid, mild non-iid, or extreme non-iid can be achieved by either i) projecting the extracted gradients into two dimensions and seeing the shape of the formed clusters, ii) asking each peer what classes she holds, or iii) using the sign of the bias gradient of each class output neuron (as mentioned in the previous section, the error of a peer's output neuron lies within $[-1, 0]$ for the classes she holds, while it lies within $[0, 1]$ for the classes she does not hold).
In any case, assuming knowledge on the peers' class distribution is a much weaker requirement than assuming the peers' data follow certain distributions ({\em i.e.} many related works directly assume the iid setting). 

We formalize our method in Algorithm~\ref{algorithm1}. 
The aggregator server $A$ starts a federated learning task by selecting a random set $S$ of $m$ peers, initializes the global model $W^0$ and sends it to the $m$ selected peers. 
Then, each peer $k\in S$ locally trains $W^{t}$ on her data $D_k$ and sends her local update $W_{k}^{t+1}$ back to $A$. 
Once $A$ receives the $m$ local updates, it computes their corresponding gradients as $\{\nabla W_{k}^{t} = (W^{t} - W_{k}^{t+1})/\eta |k\in S\}$. 
After that, $A$ separates the gradients connected to the output layer neurons to obtain the set $\{\nabla^{L,t}_k|k \in S\}$.

\begin{algorithm}[!ht]
\SetKwProg{Fn}{Function}{}{end}
\caption{Defending against the label-flipping attack}
\label{algorithm1}
\SetAlgoLined
\KwInput{$K, C, BS, E, \eta, T$}
\KwOutput{$W^T$, the global model after $T$ training rounds}

$A$ initializes $W^{0}$

\For{each round $t \in [0, T-1]$}{

    $m \leftarrow \max(C \cdot K, 1)$
    
    $S \leftarrow$ random set of $m$ peers
    
    $A$ sends $W^t$ to all peers in $S$
    
    \For{each peer $k \in S$ \textbf{in parallel }}{
        
            $W_{k}^{t+1} \leftarrow$\FuncSty{PEER\_UPDATE($k, W^{t}$)}\tcp{\small $A$ sends
            $W^t$ to each peer $k$ who trains $W^t$ using her data $D_k$ locally, and sends her local update  $W_{k}^{t+1}$ back to the aggregator}  
    }
    
    Let $\{\nabla^{L,t}_k|k \in S\}$ be the peers' output layer gradients at iteration $t$

    \For{each peer $k \in S$}{
        
         \For{each neuron $i \in [1, \mathcal{|C|}]$}{
         
        Let $||\nabla_{i, k}^{L, t}||$ be the magnitude of the gradients connected to the output layer neuron $i$ of the peer $k$ at iteration $t$
        }
          
        Let $imax_{1, k}, imax_{2, k}$ be the neurons with the highest two magnitudes in peer $k$'s output layer
    }

     \uIf {data are iid or mild non-iid}{
     
     Let $||\nabla_{i, S}^{L, t}|| =  \mathlarger{\sum\limits}_{k \in S} ||\nabla_{i, k}^{L, t}||$ \tcp{\small Neuron-wise magnitude aggregation}
     
     Let $imax_{1, S}, imax_{1, S}$ be the neurons with the highest two magnitudes in $(||\nabla_{1, S}^{L, t}||, .., ||\nabla_{i, S}^{L, t}||, .., ||\nabla_{|\mathcal{C}|, S}^{L, t}||)$ \tcp{\small Identifying potential source and target classes}  
     
     $bad\_peers \leftarrow$\FuncSty{FILTER\_MILD($\{\nabla^{L,t}_k|k \in S\},imax_{1, S}, imax_{1, S}$)}
     
     }
     
    \Else{
        
        $bad\_peers\leftarrow$\FuncSty{FILTER\_EXTREME($\{\nabla^{L,t}_{imax_{1, k}}, \nabla^{L,t}_{imax_{2, k}}|k \in S\}$)}
        }

    $A$ aggregates $W^{t+1} \leftarrow$ \FuncSty{FedAvg$(\{W_{k}^{t+1}|k \notin bad\_peers\})$}.}
\end{algorithm}

\textbf{Identifying potential source and target classes.} 
After separating the gradients of the output layer, we need to identify the potential source and target classes, which is key to our defense. 
% ALBERTO: Interesting, but we already know the approach and its justification,
% so I would remove this.
% A possible solution is to perform a brute-force search by taking two classes at a time and identifying the two classes that give the best separation between gradients as the source and target classes. 
% However, this is impractical as its cost will increase with the number of classes as the number of the required searches will be $\binom{\mathcal{C}}{2}$. For example, in the MNIST or the CIFAR10 benchmarks used in our work, we need to perform $\binom{10}{2} = 45$ searches.
% Besides, this would not be feasible if the data distribution is extreme non-iid because they do not form two clusters. 
%Alternatively, 
%Najeeb: OK.
As we have shown in the previous section, the magnitudes of the gradients connected to the source and target class neurons for the attackers and honest peers are expected to be larger than the magnitudes of the other non-relevant classes.
Thus, we can dynamically identify the potential source and target class neurons by analyzing the magnitudes of the gradients connected to the output layer neurons.
To do so, for each peer $k \in S$, we compute the neuron-wise magnitude of each output layer neuron's gradients $||\nabla_{i, k}^{L, t}||$ and identify the two neurons with the highest two magnitudes $imax_{1, k}$ and $imax_{2, k}$ as potential source and target class for that peer under the extreme non-iid setting.
For the iid or mild non-iid settings, after computing the output layer neuron magnitudes for all peers in $S$, we aggregate their neuron wise gradient magnitudes into the vector $(||\nabla_{1, S}^{L, t}||, .., ||\nabla_{i, S}^{L, t}||, .., ||\nabla_{\mathcal{|C|}, S}^{L, t}||)$. 
We then identify the potential source and target class neurons $imax_{1, S}$ and $imax_{2, S}$ as the two neurons with the highest two magnitudes in the aggregated vector.

\textbf{Filtering in case of iid and mild non-iid updates.} 
For the local updates resulting from the iid and mild non-iid settings, we filter out bad updates by using the \FuncSty{FILTER\_MILD} procedure detailed in Procedure~\ref{proc1}. 
First, we start by extracting the gradients connected to the identified potential source and target classes $imax_{1, S}$ and $imax_{2, S}$ from the output layer gradients of each peer.
Then, we use the k-means~\cite{hartigan1979algorithm} method with two clusters to group the extracted gradients into two clusters $cl_1$ and $cl_2$. 
Once the two clusters are formed, we need to decide which of them contains the potential bad updates. To make this critical decision, we consider two factors: the size and the density of clusters. We mark the smaller and/or denser cluster as potentially bad. 
This makes sense: when the two clusters have similar densities, the smaller is probably the bad one, but if the two clusters are close in size, the denser and more homogeneous cluster is probably the bad one. The higher similarity between the attackers makes their cluster usually denser, as shown in the previous section. 
To compute the density of a cluster, we compute the pairwise angle $\theta_{ij}$ between each pair of gradient vectors $i$ and $j$ in the cluster.
Then, for each gradient vector $i$ in the cluster, we find $\theta_{max, i}$ as the maximum pairwise angle for that vector.
That is because no matter how far apart two attackers' gradients are, they will be closer to each other due to the larger similarity of their directions compared to that of two honest peers.
After that, we compute the average of the maximum pairwise angles for the cluster to obtain the 
%JOSEP2. dns is in fact the inverse of the density. I add inverse.
inverse density value $dns$ of the cluster.
This way, the denser the cluster, the lower $dns$ will be. 
After computing $dns_1$ and $dns_2$ for $cl_1$ and $cl_2$, we compute $score_1$ and $score_2$ by re-weighting the computed clusters' inverse densities proportionally to their sizes.
If both clusters have similar inverse densities, the smaller cluster will probably have the lower score, or if they have similar sizes, the denser cluster will probably have the lower score.  
Finally, we use $score_1$ and $score_2$ to decide which cluster contains the potential bad updates.
We compute the set $bad\_peers$ as the peers in the cluster with the minimum score.

% ALBERTO: Najeeb, I see that FILTER_MILD and FILTER_EXTREME have different
% input parameters, but that internally, data seems to represent the same
% thing. Is it like that? In that case, the input parameters for both methods
% could be the same.
%Najeeb: Actually, they are different. The one for the MILD first take the neuron-wise gradients mean among all the peers to \nable^L, S. Then, the potential source and target classes (imax1,S and imax2, S) are identified as the two neurons with the two largest magnitudes in \nable^L. Then the gradients corresponding to imax1,S and imax2, S are extracted from each peer's last layer gradients.
%The thing for the EXTREME case is to identify the potential source and target classes for each peer individually (imax1,k and imax2,k). Then those ones are extracted from k's last layer gradients.
\setcounter{algocf}{0}
\renewcommand*{\algorithmcfname}{Procedure}
\begin{algorithm}[!ht]
  \DontPrintSemicolon
  \SetKwFunction{FMain}{FILTER_MILD}
  \caption{Filtering iid and mild non-iid updates}
  \label{proc1}
  \SetKwProg{Fn}{}{:}{}
  \Fn{\FMain{$\{\nabla^{L,t}_k|k \in S\},imax_{1, S}, imax_{1, S}$}}{
     
     $data \leftarrow \{\nabla^{L,t}_{i, k}|(k \in S, i \in \{imax_{1, S}, imax_{1, S}\})\}$
     
     $cl_1, cl_2 \leftarrow$ \FuncSty{kmeans$(data, num\_clusters = 2)$}

%JOSEP2. I add inverse to density
     \tcp{\small Computing cluster inverse densities}
     
     $dns_1 \leftarrow$ \FuncSty{CLUSTER_INVERSE_DENSITY$(cl_1)$}
     
     $dns_2 \leftarrow$ \FuncSty{CLUSTER_INVERSE_DENSITY$(cl_2)$}
     
     \tcp{\small Re-weighting clusters inverse densities}
     
      $score_1 = |cl_1|/(|cl_1| + |cl_2|)*dns_1$
      
      $score_2 = |cl_2|/(|cl_1| + |cl_2|)*dns_2$
     
     \uIf {$score_1 < score_2$}{
            
            $bad\_peers \leftarrow \{k|k \in cl_1\}$
            
     }
     
    \Else{
        
            $bad\_peers \leftarrow \{k|k \in cl_2\}$
       
        }

    \KwRet $bad\_peers$
}

\SetKwFunction{FMain}{CLUSTER_INVERSE_DENSITY}
  \SetKwProg{Fn}{}{:}{}
  \Fn{\FMain{$\{\nabla_i\}_{i = 1}^{n}$}}{
  
     \For{each $\nabla_i $}{
        
        \For{each $\nabla_j$}{
            
            Let $\theta_{ij}$ be the angle between $\nabla_i$ and $\nabla_j$
        }
        
        Let $\theta_{max, i} = max_j(\theta_i)$
        
     }
     
        $dns = \frac{1}{n} \sum_i \theta_{max, i}$
        
      \KwRet $dns$
     
}

\end{algorithm}

\textbf{Filtering in case of extreme non-iid updates.} For the local updates resulting from extreme non-iid data, we filter out potential bad updates by using the \FuncSty{FILTER\_EXTREME} procedure described in Procedure~\ref{proc2}. 
First, from each peer's output layer gradients, we extract the gradients connected to the potential source and target class neurons of that peer ($imax_{1, k}$ and $imax_{2, k}$).
After that, we use HDBSCAN~\cite{campello2013density}, that clusters its inputs based on their density and dynamically determines the required number of clusters. 
%JOSEP2. HDBSCAN is newer than microaggregation, but it is similar: form clusters of at least a given size, without setting in advance the number of clusters. The difference is that microaggregation *always* places a data point in some cluster, while DBSCAN leaves %outliers out of any cluster.
This method fits our need: we do not know how many classes there are, but we need to separate the gradients resulting from each class training data into an individual cluster.
Another interesting feature of HDBSCAN is that it requires only one main parameter to build clusters: the minimum cluster size, which must be greater than or equal to $2$. 
We use a minimum cluster size of $2$ because, in this way, a cluster can be formed if there are at least two gradients vectors with similar input features. 
Otherwise, a gradient vector will be marked as an outlier and assigned the label $-1$ if it does not belong to any formed cluster.
After clustering the extracted gradients, we compute the neuron-wise mean of the output layer gradients for the peers in each cluster. 
Then, for each mean, we compute the magnitudes of the gradient vectors corresponding to the parameters of the mean's output neurons.
That is, for the mean $\mu_j$ corresponding to the $j$-th cluster, we compute $||\nabla{1, j}||,.., ||\nabla{i, j}||, ..,||\nabla{|\mathcal{C}|, j}||$ where $||\nabla{i, j}||$ is the magnitude of the gradient vector of the $i$-th neuron in $\mu_j$. 
After that, for each mean $\mu_j$, we identify the index of the neuron that has the maximum gradient vector magnitude as $imax{1, j}$. 
In the extreme non-iid setting, each cluster corresponds to a specific class and its maximum magnitude corresponds to the neuron of that specific class, as we have discussed in the previous section. 
As a result, when the means of two clusters have the same $imax_{1}$, one of them could be a potential attackers' cluster. 
We assume that the attackers' cluster must have a smaller size than the target class cluster, and therefore we identify the smaller cluster as a potential bad cluster. 
Note that even if honest peers who hold {\em source} class examples are a minority compared to the attackers, our defense will preserve the contributions of that minority provided that the number of attackers is less than the number of peers of the {\em target} class.
Also, we identify gradient vectors that do not belong to any cluster (labeled -1 by HDBSCAN) as potential bad gradients.
This ensures that even if there is only one attacker in the system, we can also detect him/her.  
Finally, we compute the set $bad\_peers$ as the peers with gradients in the cluster identified as bad or labeled $-1$.

\begin{algorithm}[!ht]
  \DontPrintSemicolon
  \SetKwFunction{FMain}{FILTER_EXTREME}
  \caption{Filtering extreme non-iid updates}
  \label{proc2}
  \SetKwProg{Fn}{}{:}{}
  \Fn{\FMain{$\{\nabla^{L,t}_{imax_{1, k}}, \nabla^{L,t}_{imax_{2, k}}|k \in S\}$}}{

         $data \leftarrow \{\nabla^{L,t}_{imax_{1, k}}, \nabla^{L,t}_{imax_{2, k}}|k \in S\}$

          $\{cl_j\}_{j = 1}^{Z}\leftarrow$\FuncSty{HDBSCAN($data, min\_cluster\_size = 2$)}\tcp*{\small $Z$ is the number of clusters formed}
         
        $\{\mu_j\}_{j=1}^{Z}\leftarrow$(\FuncSty{MEAN($cl_1$)},..,
        \FuncSty{MEAN($cl_j$)},..,
        \FuncSty{MEAN($cl_Z$)}) \tcp*{\small $\mu_j$ is the neuron-wise mean of the output layer magnitudes of the $j^{th}$ cluster}
        
        Let $imax_{1, 1}, .., imax_{1, j}, imax_{1, Z}$ be the indices of the neurons with the highest magnitude for the computed means 
        \tcp*{\small $imax_{1, j}$ is the index of the neuron with the highest magnitude in $\mu_j$}
        
        \For{$i \in [1, Z]$}{
            \For{$j \in [i, Z]$}{
                \uIf {$(imax_{1, i} = imax_{1, j}) \hspace{2mm} and \hspace{2mm} (|cl_i| > |cl_j|)$}{
                    $bad\_cluster \leftarrow cl_j$    
                
                }
            }
    }
        
        $bad\_peers \leftarrow \{k|k \in S, k \in \{bad\_cluster, -1\}\}$
    
        \KwRet $bad\_peers$.
  }
\end{algorithm}

\textbf{Aggregating potential good updates}. After identifying the potentially bad peers, the server $A$ computes
%JOSEP. Fixed brackets.
\FuncSty{FedAvg$(\{W_{k}^{t+1}|k \notin bad\_peers\})$} to obtain the updated global model $W^{t+1}$.

\section{Empirical analysis}
\label{analysis}

%JOSEP. Rewritten.
In this section we compare the performance of our method with that of several state-of-the-art countermeasures against poisoning attacks. Our code and data are available for reproducibility purposes at \url{https://github.com/anonymized1/LF-Fighter}. 

\subsection{Experimental setup}
\label{setup}
We used the PyTorch framework to implement the experiments on an AMD Ryzen 5 3600 6-core CPU with 32 GB RAM, an NVIDIA GTX 1660 GPU, and Windows 10 OS. 

%\begin{itemize}
%\item 
\textbf{Data sets and models.}
\label{data_models}
We tested the proposed method on three data sets (see Table~\ref{tab:datasets_models}):
\begin{compactitem}
\item MNIST. It contains $70K$ handwritten digit images from $0$ to $9$~\cite{lecun1999object}. The images are divided into a training set ($60K$ examples) and a testing set ($10K$ examples). We used a two-layer convolutional neural network (CNN) with two fully connected layers on this data set. 
\item CIFAR10. It consists of $60K$ colored images of $10$ different classes \cite{krizhevsky2009learning}. The data set is divided into $50K$ training examples and $10K$ testing examples. We used the ResNet18 CNN model with one fully connected layer \cite{he2016deep} on this data set. 
\item IMDB. Specifically, we used the IMDB Large Movie Review data set
\cite{maas2011learning} for binary sentiment classification. The data set is a collection of $50K$ movie reviews and their corresponding sentiment binary labels (either positive or negative). We divided the data set into $40K$ training examples and $10K$ testing examples. We used a Bidirectional Long/Short-Term Memory (BiLSTM) model with an embedding layer that maps each word to a 100-dimensional vector.
%David2 to Najeeb: 100 what?
%Najeeb: Done!
The model ends with a fully connected layer followed by a sigmoid function to produce the final predicted sentiment for an input review. 
\end{compactitem}

\begin{table}[ht]
\centering
\caption{Data sets and models used in the experiments}
\label{tab:datasets_models}
\resizebox{0.5\textwidth}{!}{%
\begin{tabular}{|l|c|c|c|c|}
\hline
\multicolumn{1}{|c|}{Task}            & Data set & \# Examples & Model  & \# Parameters \\ \hline
%JOSEP2. We do not need to shorten task names.
\multirow{2}{*}{Image classification} & MNIST    & 70K         & CNN    & $\sim$22K     \\ \cline{2-5} 
                                      & CIFAR10  & 60K         & ResNet18  & $\sim$11M     \\ \hline
Sentiment analysis                    & IMDB     & 50K         & BiLSTM & $\sim$12M     \\ \hline
\end{tabular}%
}
\end{table}

\textbf{Data distribution and training.}
\label{data_training}
We defined the following benchmarks by distributing the data from the data sets above among the participating peers in the following way:

\begin{compactitem}

\item MNIST-iid. We randomly and uniformly divided the MNIST training data among $100$ peers. The CNN model was trained for $200$ iterations.
In each iteration, the FL server asked the peers to train their models for $3$ local epochs and a local batch size of $64$. 
The participants used the cross-entropy loss function and the stochastic gradient descent (SGD) optimizer with learning rate = $0.001$ and momentum = $0.9$ to train their models. 

\item MNIST-Mild. We adopted a Dirichlet distribution~\cite{minka2000estimating} with a hyperparameter $\alpha = 1$ to generate \textit{mild non-iid} data for $100$ participating peers. The training settings were the same as for  MNIST-iid.

\item MNIST-Extreme. We simulated an \textit{extreme non-iid} setting with $100$ peers where each peer was randomly assigned examples of a single class out of the MNIST data set. 
Out of the $100$ peers, only $10$ had examples of the source class $7$ and $10$ had examples of the target class $1$. 
The training settings were the same as for MNIST-iid.

\item CIFAR10-iid. We randomly and uniformly divided the CIFAR10 training data among $20$ peers.
The ResNet18 model was trained during $100$ iterations. In each iteration, the FL server asked the $20$ peers to train the model for $3$ local epochs and a local batch size $32$. The peers used the cross-entropy loss function and the SGD optimizer with learning rate = $0.01$ and momentum = $0.9$. 

\item CIFAR10-Mild. We adopted a Dirichlet distribution~\cite{minka2000estimating} with a hyperparameter $\alpha = 1$ to generate \textit{mild non-iid} data for $20$ participating peers. The training settings were the same as for CIFAR10-iid.

\item IMDB. We randomly and uniformly split the $40K$ training examples among $20$ peers to simulate an \emph{iid} setting. 
The BiLSTM was trained during $50$ iterations. In each iteration, the FL server asked the $20$ peers to train the model for $1$ local epoch and a local batch size of $32$. The peers used the binary cross-entropy with logit loss function and the \textit{Adam} optimizer with learning rate = $0.001$.
\end{compactitem}

\textbf{Attack scenarios.}
\label{attack_set} 
%JOSEP. Rearranged.
In all the experiments with MNIST, the attackers flipped the examples with the source class $7$ to the target class $1$.
In the CIFAR10 experiments, the attackers flipped the examples with the label \emph{Dog} to \emph{Cat} before training their local models, whereas for IMDB, the attackers flipped the examples with the label \emph{positive} to \emph{negative}.

In all benchmarks, the ratio of attackers ranged in $\{0\%, 10\%, 20\%, 30\%, 40\%, 50\% \}$. 
Note that in all the benchmarks, the $40\%$ ratio of attackers corresponds to $m' = (m/2) - 2$, which is the theoretical upper bound of the number of attackers MKrum~\cite{blanchard2017machine} can defend against.

\textbf{Evaluation metrics.}
\label{metrics} We used the following evaluation metrics on the test set examples for each benchmark to assess the impact of the LF attack on the learned model and the performance of the proposed method w.r.t. the state-of-the-art methods:
\begin{compactitem}
%JOSEP2. Written more compactly.
\item \emph{Test error (TE)}. Error resulting from the loss function used in training. The lower TE, the better.
\item \emph{Overall accuracy (All-Acc)}. Number of correct predictions divided by the total number of predictions for all the examples. The greater All-Acc, the better.
\item \emph{Source class accuracy (Src-Acc)}. Number of the source class examples correctly predicted divided by the total number of the source class examples. The greater Src-Acc, the better.
\item \emph{Attack success rate (ASR)}. Proportion of the source class examples incorrectly classified as the target class. The lower ASR, the better.
\item \emph{Coefficient of variation (CV)}. Ratio of the standard deviation $\sigma$ to the mean $\mu$, that is, $CV = \frac{\sigma}{\mu}$. The lower CV, the better. 
\end{compactitem}
While TE, All-Acc, Src-Acc and ASR are used in previous works to evaluate robustness against poisoning attacks~\cite{blanchard2017machine, tolpegin2020data, fung2020limitations}, we also use the CV metric to assess the stability of Src-Acc during training. We justify our choice of this metric in Section~\ref{stability}.
%JOSEP2. protect all of -> perform well
An effective defense needs to simultaneously perform well in terms of TE, All-Acc, Src-Acc, ASR and CV.

\subsection{Results}
\label{evaluation}

%JOSEP2. ...in terms of...
First, we report the robustness against the attack in terms of TE, All-Acc, Src-Acc and ASR for different ratios of attackers.
Then, we report the stability of the source class accuracy under the LF attack.
%JOSEP2. Rewritten.
Finally, we report the runtime of our defense. 
In all the experiments, along with the results of our method, we also give the results of several countermeasures discussed in Section~\ref{related}, including the standard FedAvg~\cite{mcmahan2017communication} aggregation method (not meant to counter poisoning attacks), the median~\cite{yin2018byzantine}, the repeated median (RMedian)~\cite{siegel1982robust}, the trimmed mean (TMean)~\cite{yin2018byzantine}, multi-Krum (MKrum)~\cite{blanchard2017machine}, and FoolsGold (FGold)~\cite{fung2020limitations}.

Note that for a $50\%$ ratio of attackers, we employ the median instead of the trimmed mean (both are equivalent in this case). Also, due to space restrictions and because the results of the repeated median were almost identical to those of the median, we do not include the former results in this section. We refer the reader to the paper repository on GitHub\footnote{\url{https://github.com/anonymized1/LF-Fighter}} for more detailed numerical results of all the benchmarks and defenses.

\begin{figure*}[htbp]
    \centering
    \begin{subfigure}{0.34\textwidth}
      \centering
      \includegraphics[width=1\linewidth]{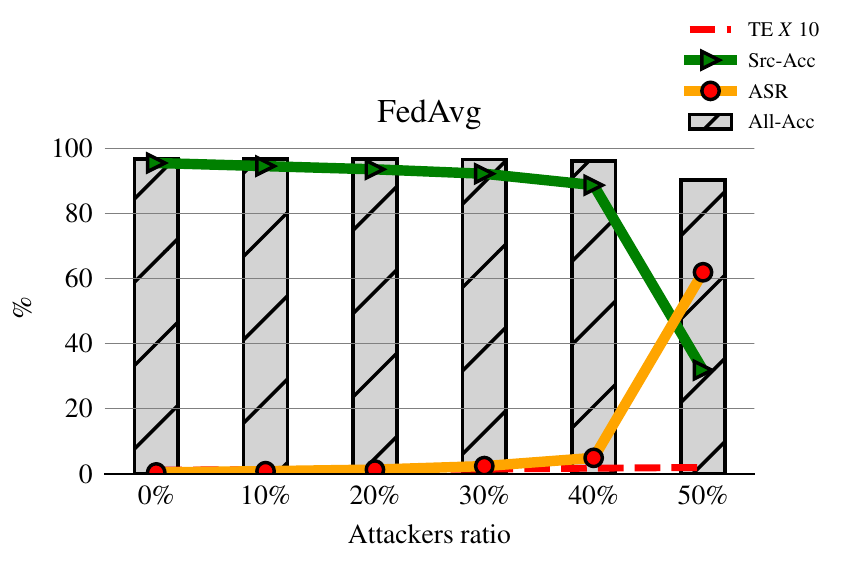}\vspace{-1.5\baselineskip}
    \end{subfigure}% 
    \begin{subfigure}{0.34\textwidth}
      \centering
      \includegraphics[width=1\linewidth]{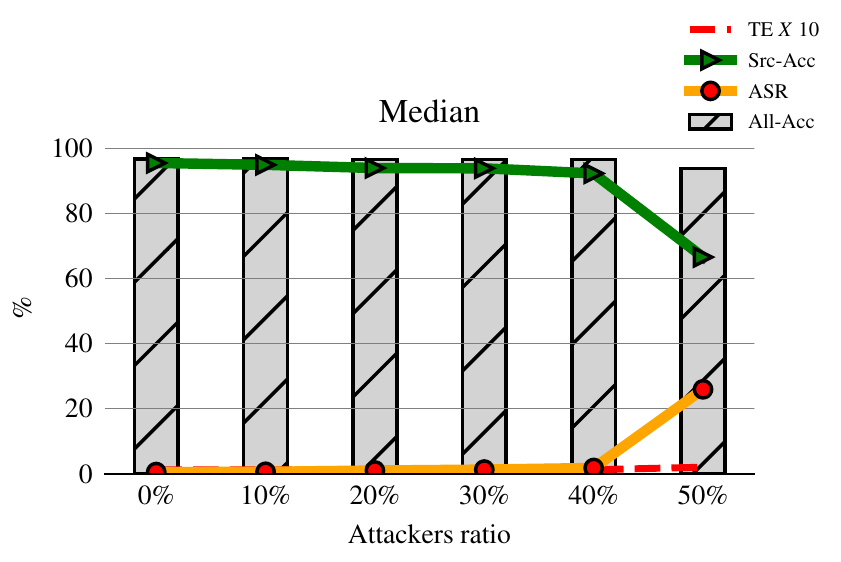}\vspace{-1.5\baselineskip}
    \end{subfigure}%
     \begin{subfigure}{0.34\textwidth}
      \centering
      \includegraphics[width=1\linewidth]{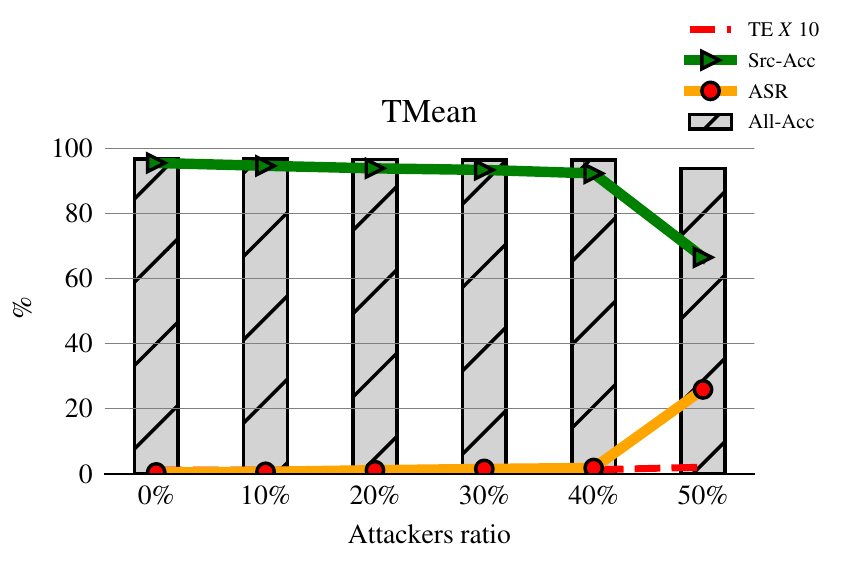}\vspace{-1.5\baselineskip}
    \end{subfigure}%
    \vspace{0.5\baselineskip}
    
     \begin{subfigure}{0.34\textwidth}
      \centering
      \includegraphics[width=1\linewidth]{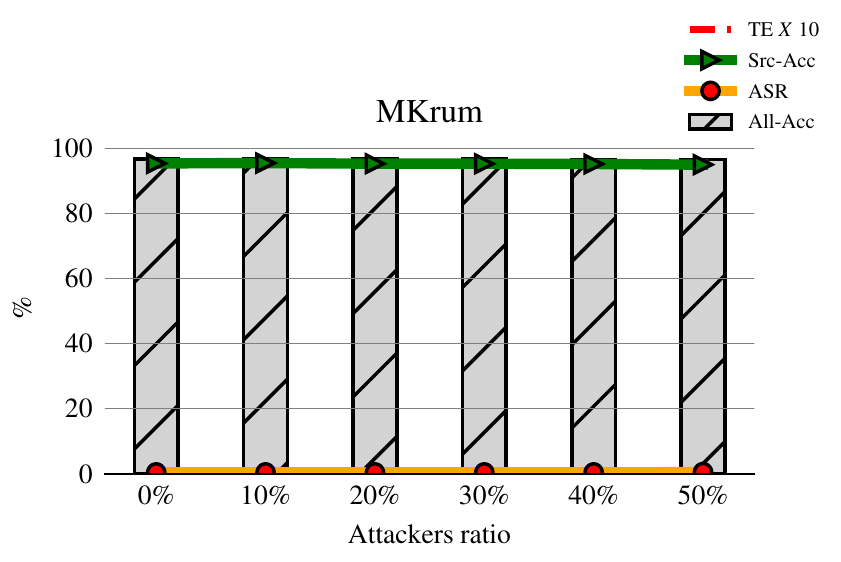}\vspace{-1.5\baselineskip}
    \end{subfigure}%
     \begin{subfigure}{0.34\textwidth}
      \centering
      \includegraphics[width=1\linewidth]{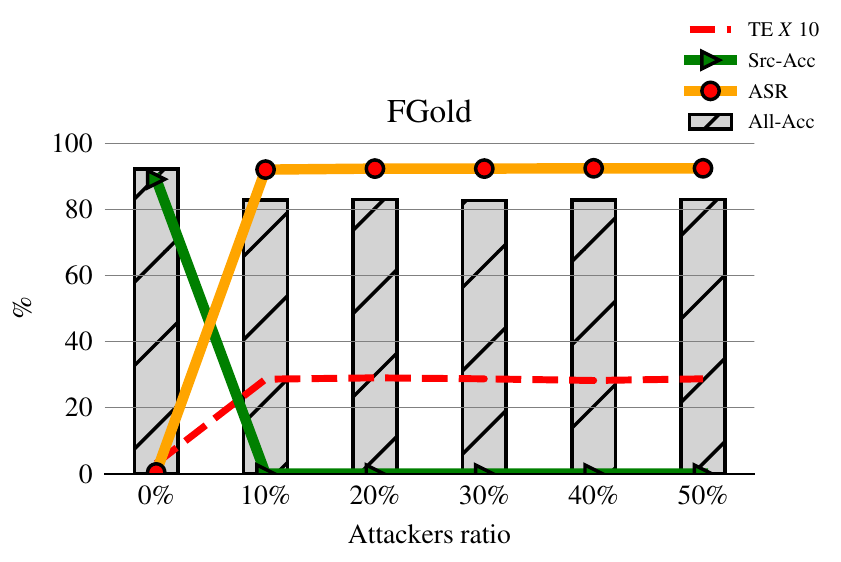}\vspace{-1.5\baselineskip}
    \end{subfigure}%
     \begin{subfigure}{0.34\textwidth}
      \centering
      \includegraphics[width=1\linewidth]{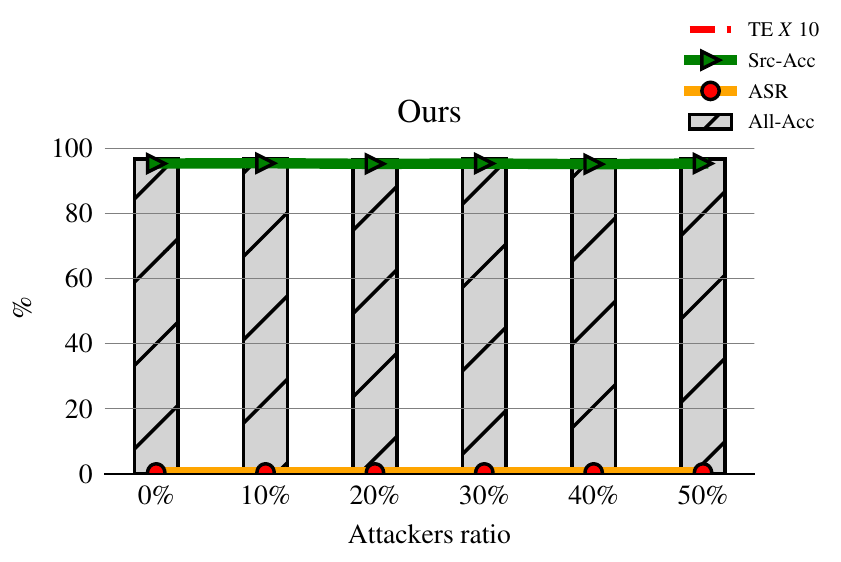}\vspace{-1.5\baselineskip}
    \end{subfigure}%
    \vspace{0.5\baselineskip}
\caption{Robustness against the label-flipping attack with the MNIST-iid benchmark}
\label{fig:mnist_iid}
\end{figure*}

\textbf{Robustness against the label-flipping attack.}
\label{defending}
We evaluated the robustness of our method against the LF attack on the used benchmarks using the attack scenarios described in Section~\ref{attack_set}. We report the average results of the last $10$ training rounds to ensure a fair comparison among methods. 
Note that we scaled the TE by $10$ to make its results fit in the figures.
Figure~\ref{fig:mnist_iid} shows the results obtained with the MNIST-iid benchmark. 
We can see all the defenses, except FoolsGold, perfectly defended against the attack with ratios of attackers up to $40\%$. However, when the attackers' ratio was $50\%$, most failed to counter the attack, except MKrum and our defense, which stayed robust against the attack with all ratios.
On the other hand, FoolsGold achieved the worst performance in the presence of attackers in all the metrics. Once the attackers appeared in the system, the accuracy of the source class plummeted, while the attackers had the highest success rate (close to $100\%$) compared to the other defenses.
That happened because of the high similarity between the honest peers' output layer gradients which FoolsGold takes into account. That led to wrongly penalizing honest peers' updates and wrongly including some of the attackers' bad updates in the global model.
Note that, due to the small variability of the MNIST data set, each local update of an honest peer was an unbiased estimator of the mean of all the good local updates. Therefore, the coordinate-wise aggregation methods, like the median and the trimmed mean, or the update-wise aggregation methods, like MKrum, achieved such good performance in this benchmark.
Another interesting note is that, although FedAvg is not meant to mitigate poisoning attacks, it achieved a good performance in this benchmark. That was also observed in~\cite{shejwalkar2021back}, where the authors argued that, in some cases, FedAvg is more robust against poisoning attacks than many of the state-of-the-art countermeasures.

\begin{figure*}[htbp]
    \centering
    \begin{subfigure}{0.34\textwidth}
      \centering
      \includegraphics[width=1\linewidth]{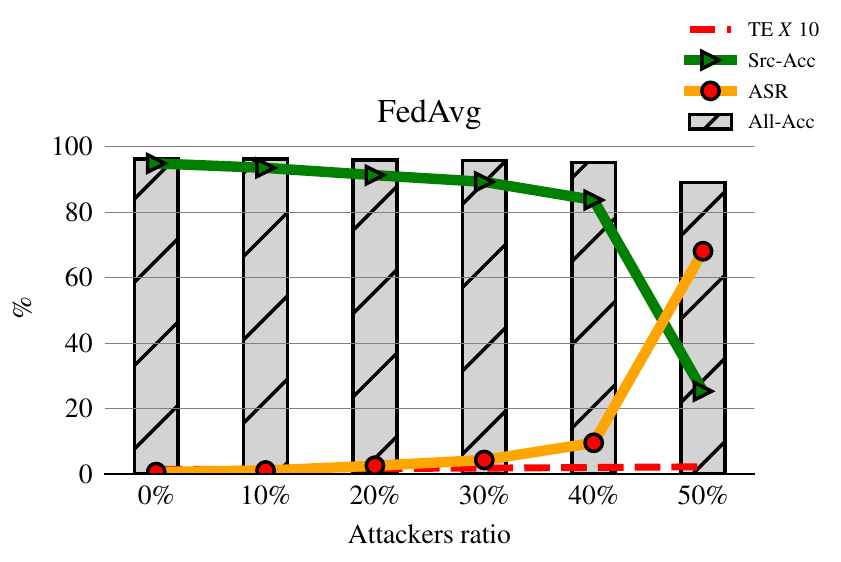}\vspace{-1.5\baselineskip}
    \end{subfigure}% 
    \begin{subfigure}{0.34\textwidth}
      \centering
      \includegraphics[width=1\linewidth]{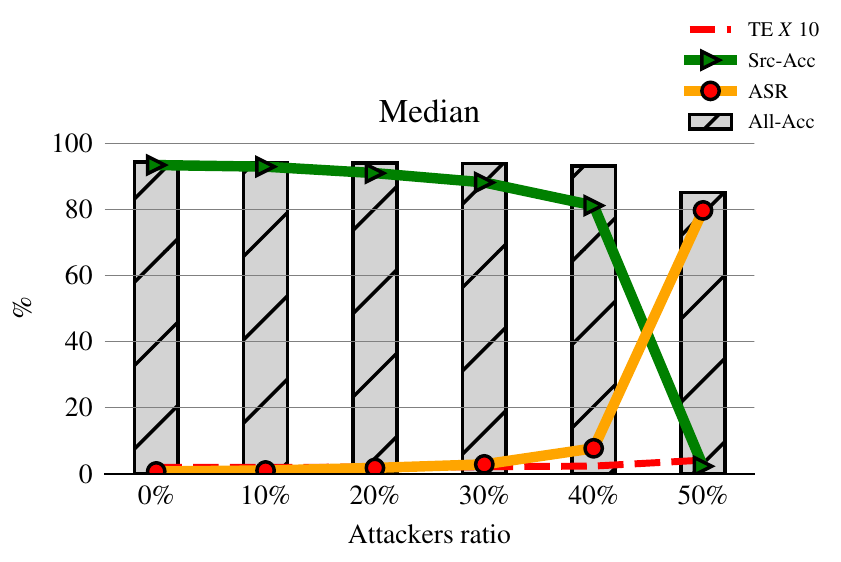}\vspace{-1.5\baselineskip}
    \end{subfigure}%
     \begin{subfigure}{0.34\textwidth}
      \centering
      \includegraphics[width=1\linewidth]{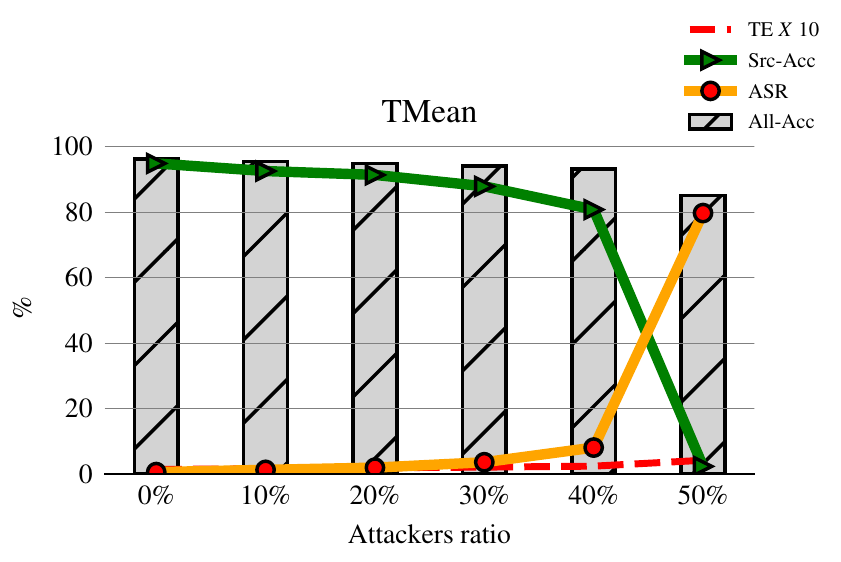}\vspace{-1.5\baselineskip}
    \end{subfigure}%
    \vspace{0.5\baselineskip}
    
     \begin{subfigure}{0.34\textwidth}
      \centering
      \includegraphics[width=1\linewidth]{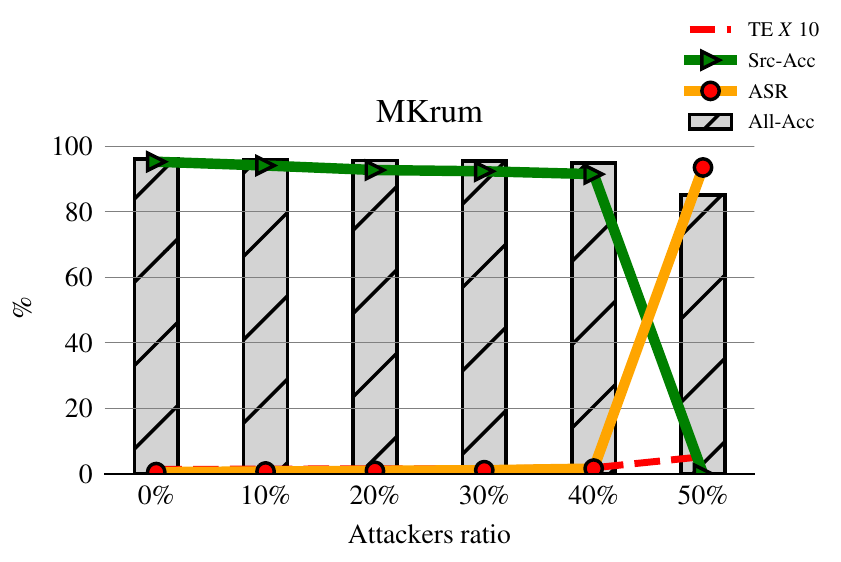}\vspace{-1.5\baselineskip}
    \end{subfigure}%
     \begin{subfigure}{0.34\textwidth}
      \centering
      \includegraphics[width=1\linewidth]{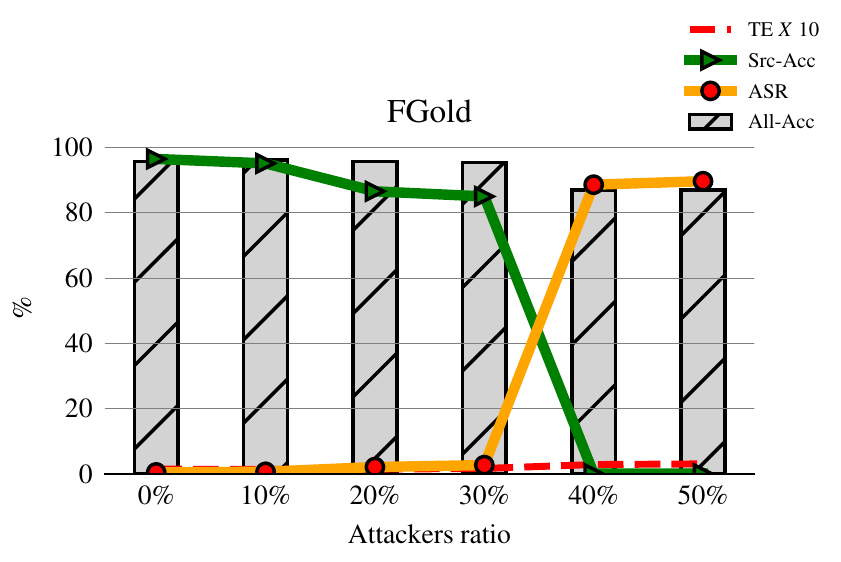}\vspace{-1.5\baselineskip}
    \end{subfigure}%
     \begin{subfigure}{0.34\textwidth}
      \centering
      \includegraphics[width=1\linewidth]{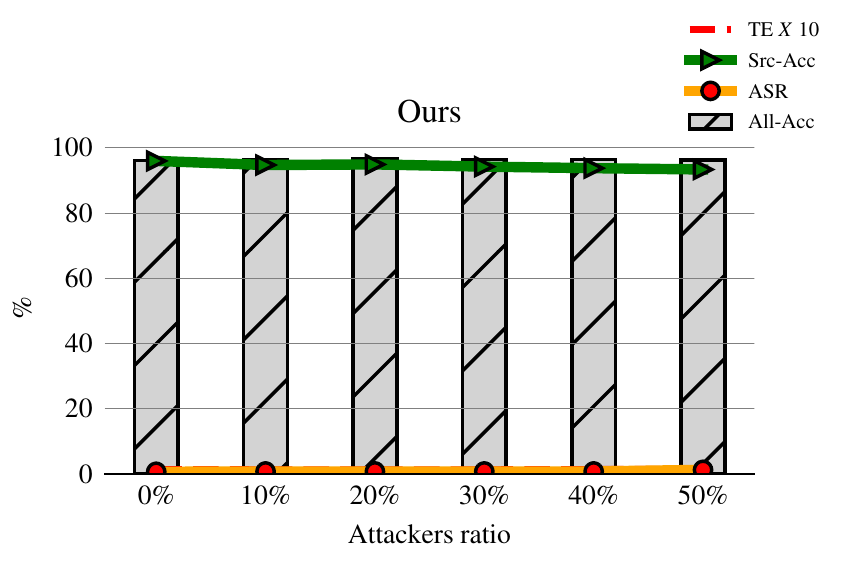}\vspace{-1.5\baselineskip}
    \end{subfigure}%
     \vspace{0.5\baselineskip}
\caption{Robustness against the label-flipping attack with the MNIST-Mild benchmark}
\label{fig:mnist_mild}
\end{figure*}

Figure~\ref{fig:mnist_mild} shows the results obtained with the MNIST-Mild benchmark. 
Although the data were non-iid, the performance was close to that in MNIST-iid.
The reasons are the simplicity of the MNIST dataset and the small size of the model. 
It is also worth noting that FoolsGold performed better in this benchmark than in MNIST-iid. The honest peers' gradients were more diverse in this benchmark due to their data distribution. 
On the other hand, our defense outperformed all the defenses and stayed robust even when the attackers' ratio reached $50\%$.

\begin{figure*}[htbp]
    \centering
    \begin{subfigure}{0.34\textwidth}
      \centering
      \includegraphics[width=1\linewidth]{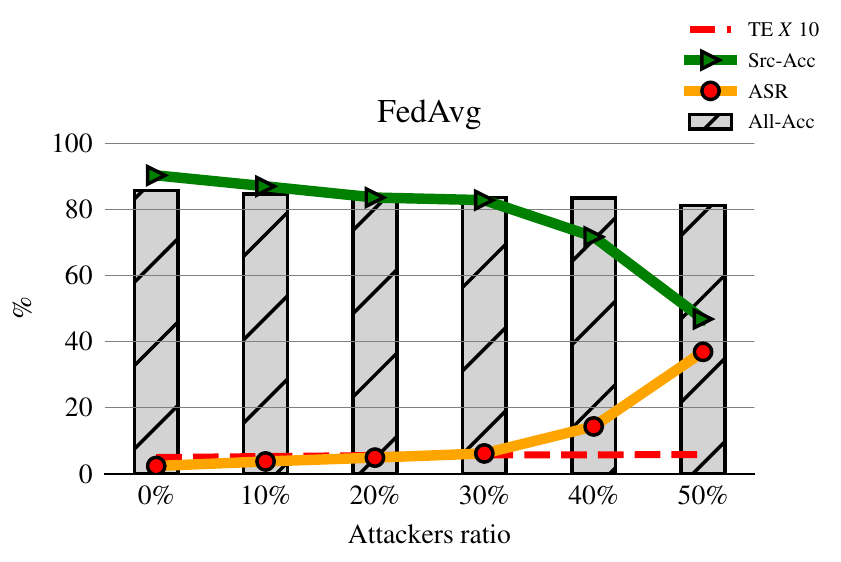}\vspace{-1.5\baselineskip}
    \end{subfigure}% 
    \begin{subfigure}{0.34\textwidth}
      \centering
      \includegraphics[width=1\linewidth]{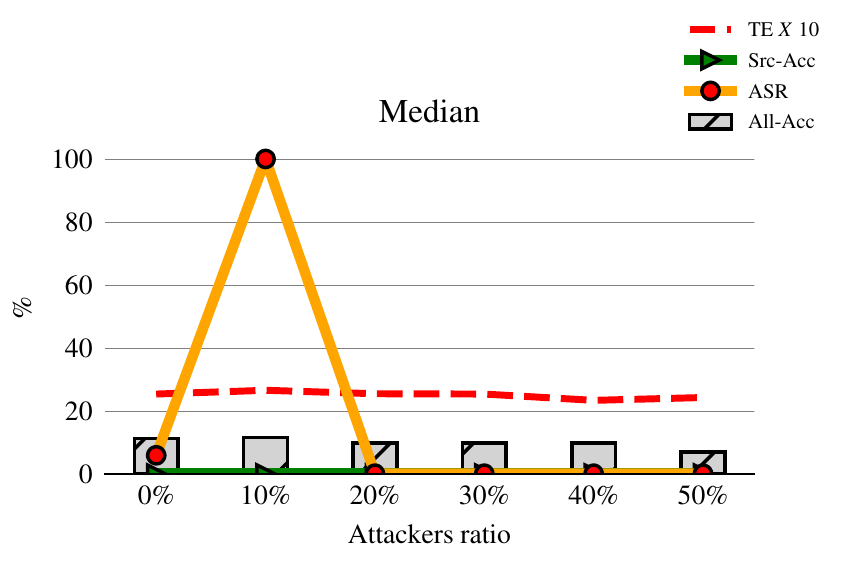}\vspace{-1.5\baselineskip}
    \end{subfigure}%
     \begin{subfigure}{0.34\textwidth}
      \centering
      \includegraphics[width=1\linewidth]{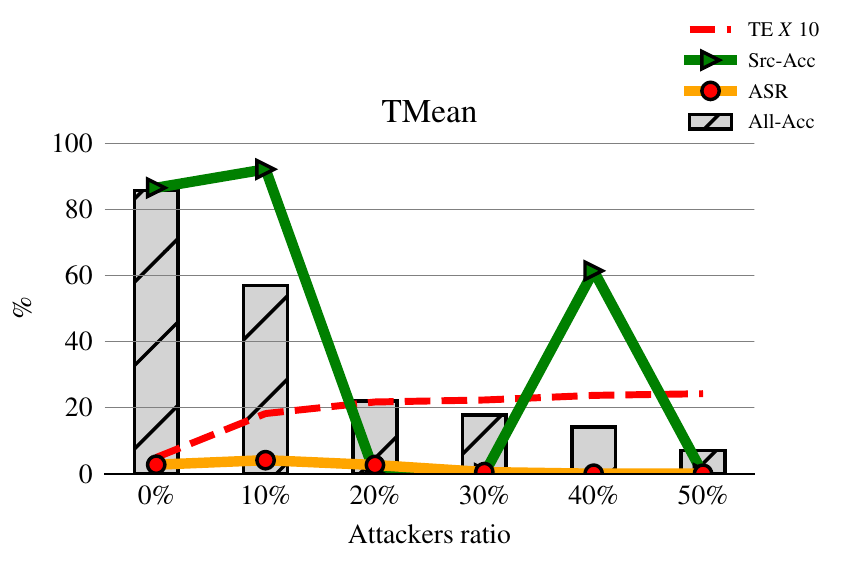}\vspace{-1.5\baselineskip}
    \end{subfigure}%
    \vspace{0.5\baselineskip}
    
     \begin{subfigure}{0.34\textwidth}
      \centering
      \includegraphics[width=1\linewidth]{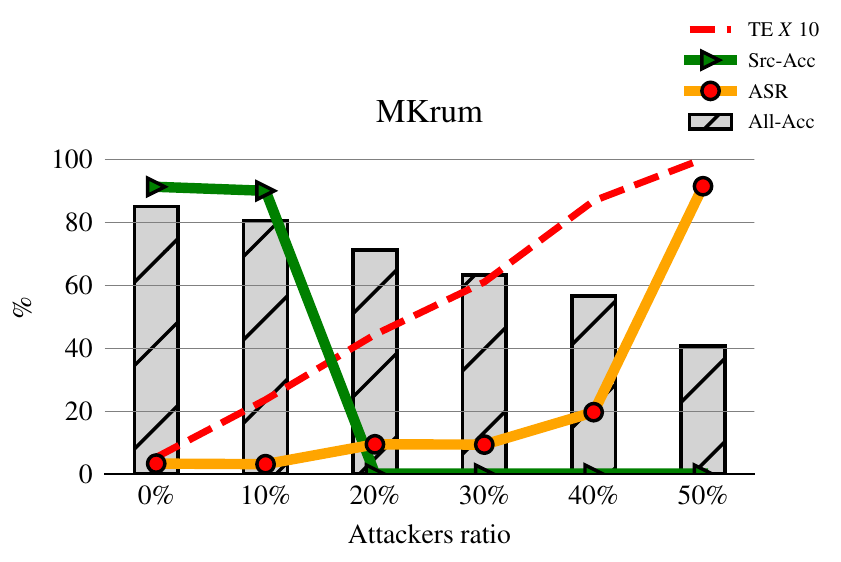}\vspace{-1.5\baselineskip}
    \end{subfigure}%
     \begin{subfigure}{0.34\textwidth}
      \centering
      \includegraphics[width=1\linewidth]{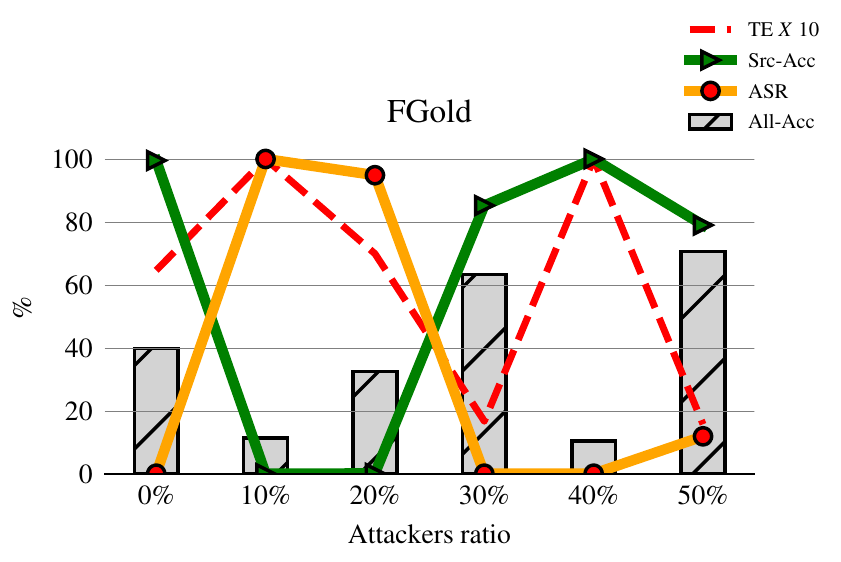}\vspace{-1.5\baselineskip}
    \end{subfigure}%
     \begin{subfigure}{0.34\textwidth}
      \centering
      \includegraphics[width=1\linewidth]{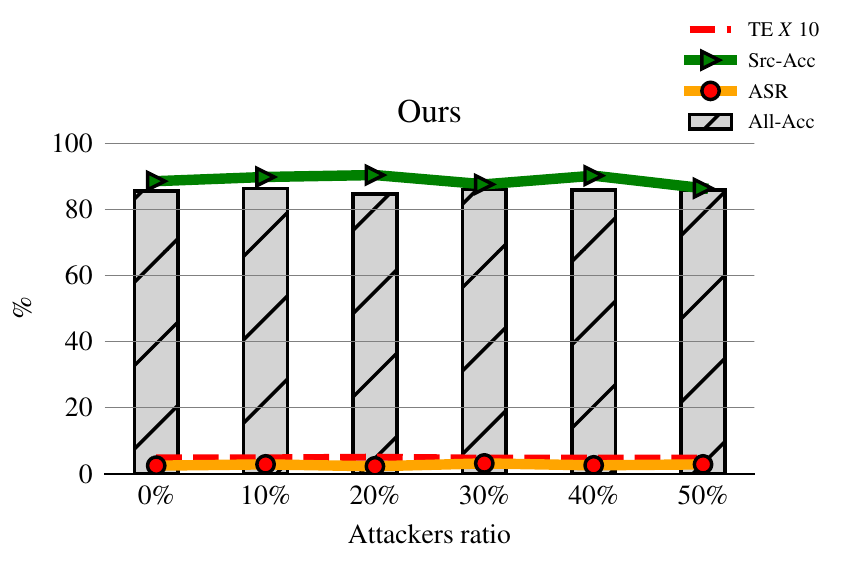}\vspace{-1.5\baselineskip}
    \end{subfigure}%
     \vspace{0.5\baselineskip}
\caption{Robustness against the label-flipping attack with the MNIST-Extreme benchmark}
\label{fig:mnist_extreme}
\end{figure*}

Figure~\ref{fig:mnist_extreme} shows the results obtained on MNIST-Extreme. 
%JOSEP2. Rewritten.
Our defense was effective for all considered attacker ratios.  
In comparison to other methods, our defense achieved simultaneous perfect performance in terms of the test error, overall accuracy, the source class (class 7) accuracy and the attack success rate. Thanks to comparing the attackers with the target class cluster, we achieved $0$ false positives, and we preserved the contributions of all the other good clusters, including the source class cluster. It is also worth noting that the trimmed mean and MKrum were less affected by the attack than the other methods. This is because they considered a larger number of peer updates. On the other hand, the median and the repeated median had poor performance because the data were extremely non-iid. Thus, these methods discarded a lot of information in the global model aggregation. Also, FoolsGold performed poorly because the clusters of good updates were penalized due to their high similarity.
%JOSEP. protect -> perform. Rewritten below.
Note that some defenses may sometimes perform well on a subset of metrics but perform poorly on the rest. For example, the median achieved an ASR of $0\%$ in most cases, but did poorly regarding TE, All-Acc 
or Src-Acc.
Also, FoolsGold did well for Src-Acc and ASR in some cases, but failed for TE and All-Acc.
As we have mentioned, it is essential to provide good performance for all metrics, which our defense did.

\begin{figure*}[htbp]
    \centering
    \begin{subfigure}{0.34\textwidth}
      \centering
      \includegraphics[width=1\linewidth]{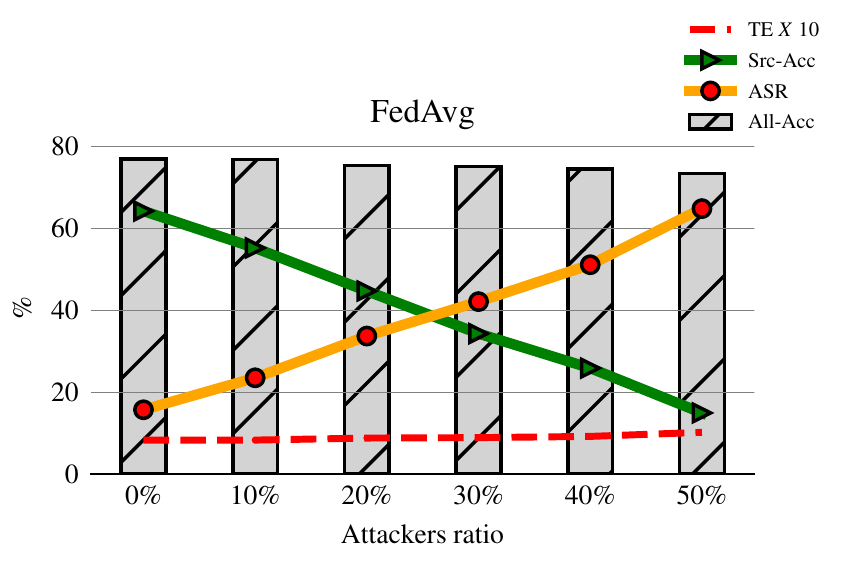}\vspace{-1.5\baselineskip}
    \end{subfigure}% 
    \begin{subfigure}{0.34\textwidth}
      \centering
      \includegraphics[width=1\linewidth]{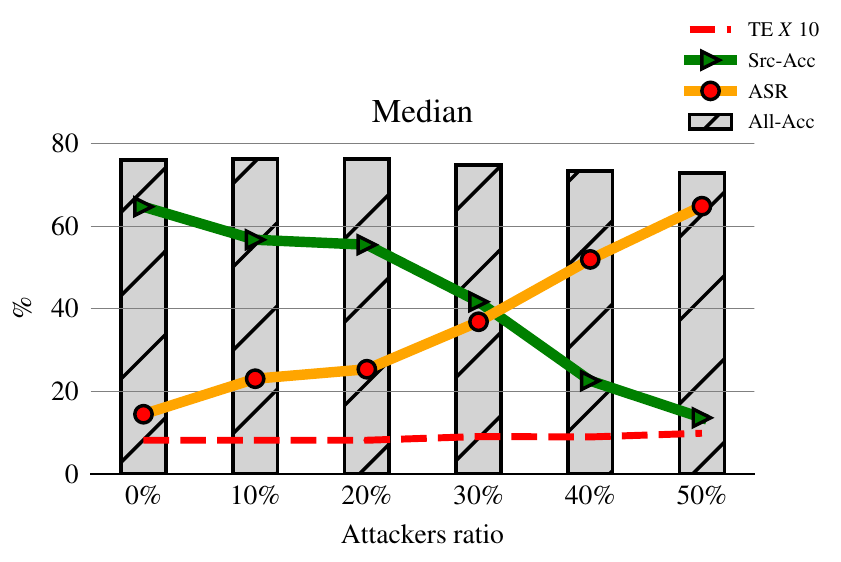}\vspace{-1.5\baselineskip}
    \end{subfigure}%
     \begin{subfigure}{0.34\textwidth}
      \centering
      \includegraphics[width=1\linewidth]{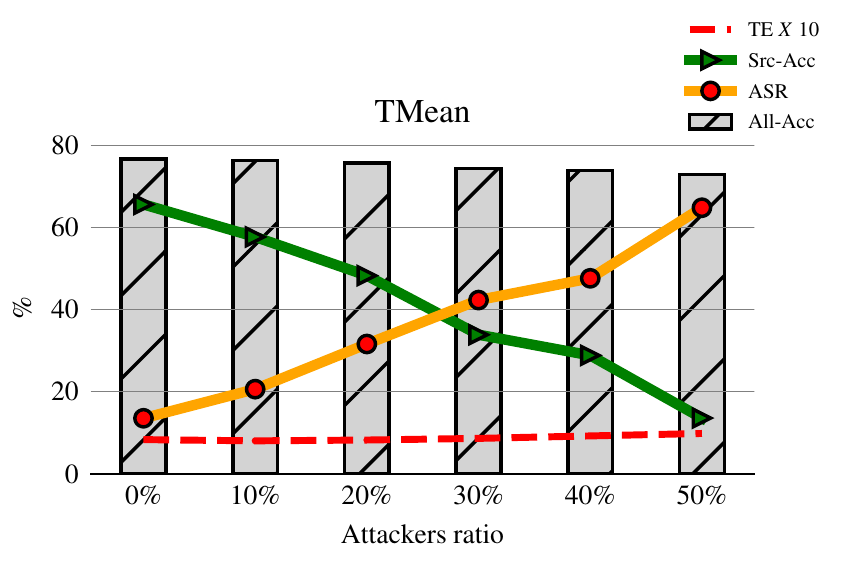}\vspace{-1.5\baselineskip}
    \end{subfigure}%
    \vspace{0.5\baselineskip}
    
     \begin{subfigure}{0.34\textwidth}
      \centering
      \includegraphics[width=1\linewidth]{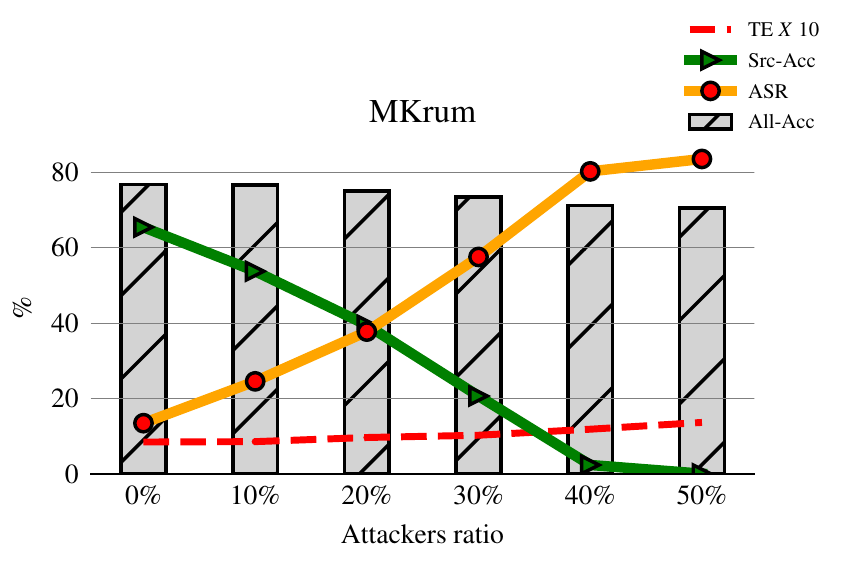}\vspace{-1.5\baselineskip}
    \end{subfigure}%
     \begin{subfigure}{0.34\textwidth}
      \centering
      \includegraphics[width=1\linewidth]{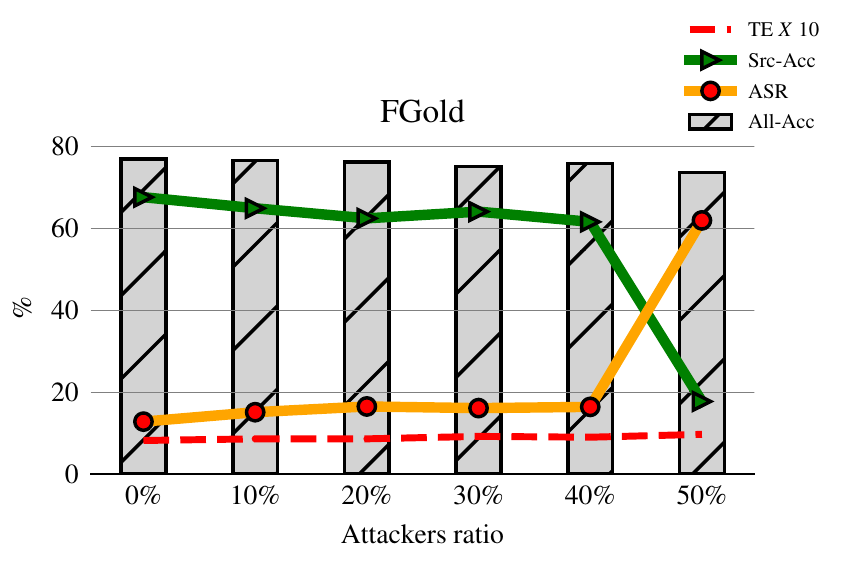}\vspace{-1.5\baselineskip}
    \end{subfigure}%
     \begin{subfigure}{0.34\textwidth}
      \centering
      \includegraphics[width=1\linewidth]{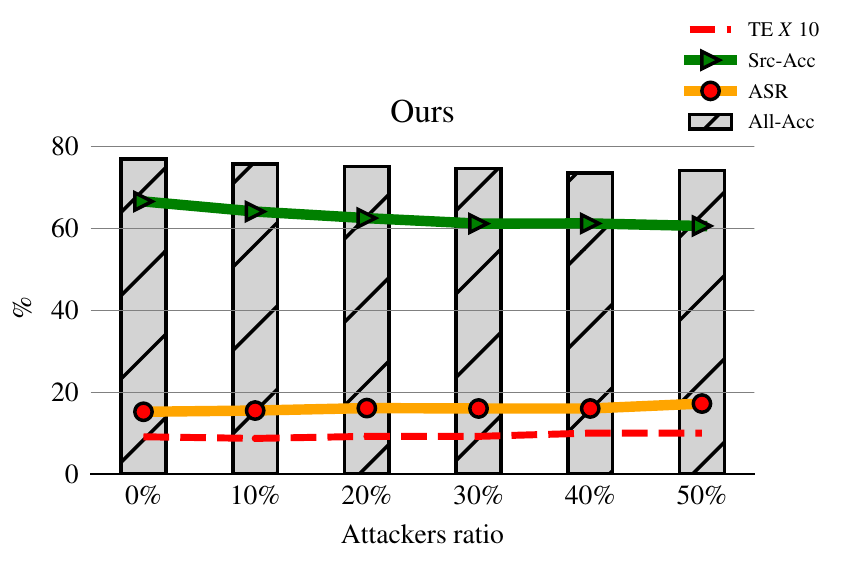}\vspace{-1.5\baselineskip}
    \end{subfigure}%
     \vspace{0.5\baselineskip}
\caption{Robustness against the label-flipping attack with the CIFAR10-iid benchmark}
\label{fig:cifar10_iid}
\end{figure*}

Figure~\ref{fig:cifar10_iid} shows the results on the CIFAR10-iid benchmark. We can see that the performance of all methods, except FoolsGold and ours, was highly affected and degraded as the ratio of the attackers increased. This is because these methods considered the whole local updates, and the size of the used model was large (about $11M$ parameters). 
%JOSEP2. changed.
In this vast amount of information they could not properly distinguish the good updates from the bad ones.
FoolsGold effectively defended against the attack in general because it only analyzed the output layer gradients. Since the data were iid, and the CIFAR10 data set is more varied than MNIST, the attackers' output layer gradients were more similar than the honest peers' ones, and therefore, FoolsGold penalized the attackers and kept the honest peers contributions.
%JOSEP2. Rewritten.
Our defense stayed robust against the attack and achieved the best simultaneous performance for all the metrics. 
This is because it was able to perfectly maintain the honest peers' contributions and exclude all the attackers even when the attackers' ratio was $50\%$. 

\begin{figure*}[htbp]
    \centering
    \begin{subfigure}{0.34\textwidth}
      \centering
      \includegraphics[width=1\linewidth]{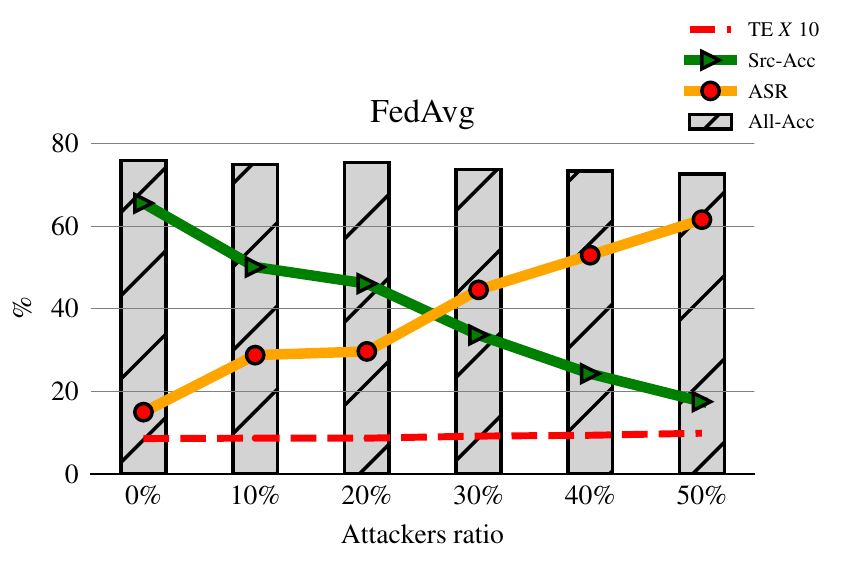}\vspace{-1.5\baselineskip}
    \end{subfigure}% 
    \begin{subfigure}{0.34\textwidth}
      \centering
      \includegraphics[width=1\linewidth]{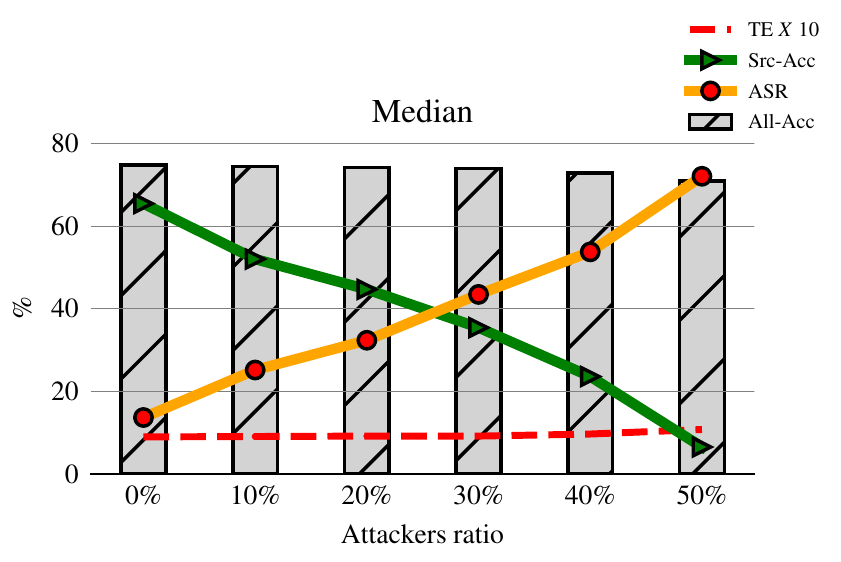}\vspace{-1.5\baselineskip}
    \end{subfigure}%
     \begin{subfigure}{0.34\textwidth}
      \centering
      \includegraphics[width=1\linewidth]{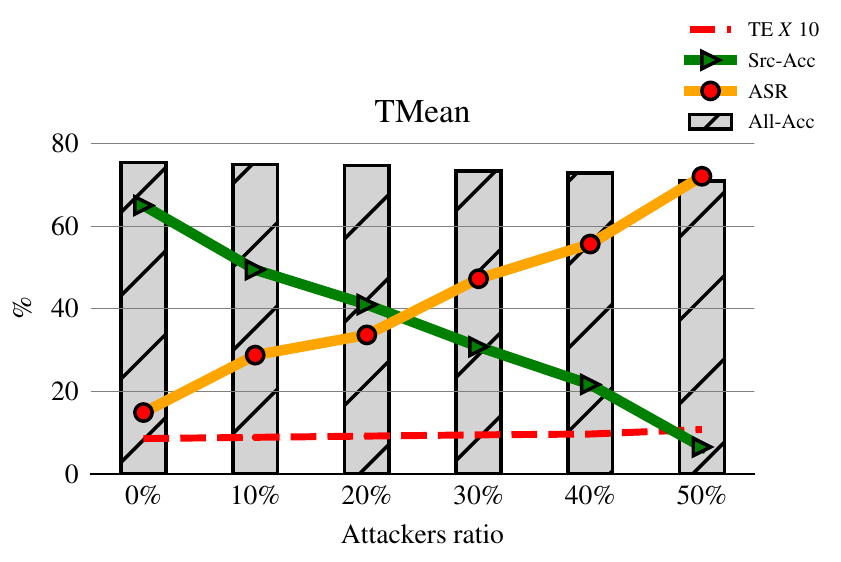}\vspace{-1.5\baselineskip}
    \end{subfigure}%
    \vspace{0.5\baselineskip}
    
     \begin{subfigure}{0.34\textwidth}
      \centering
      \includegraphics[width=1\linewidth]{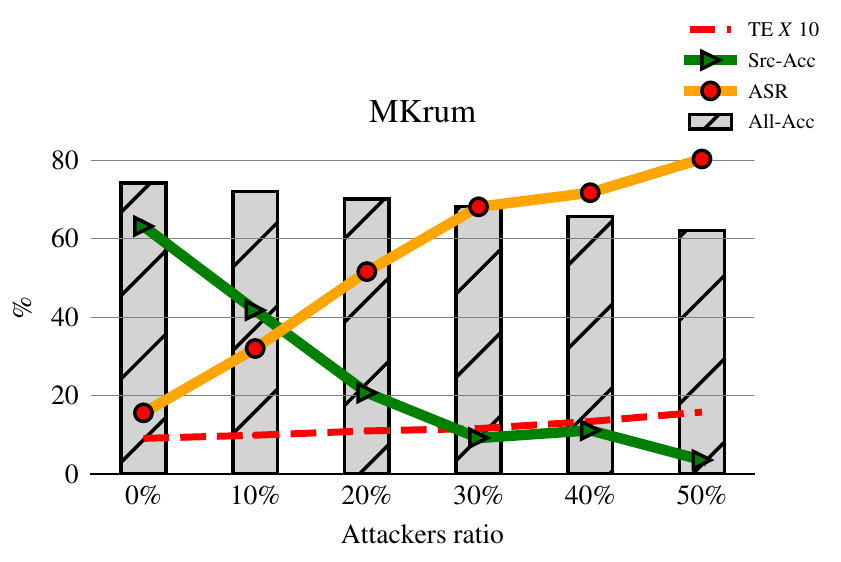}\vspace{-1.5\baselineskip}
    \end{subfigure}%
     \begin{subfigure}{0.34\textwidth}
      \centering
      \includegraphics[width=1\linewidth]{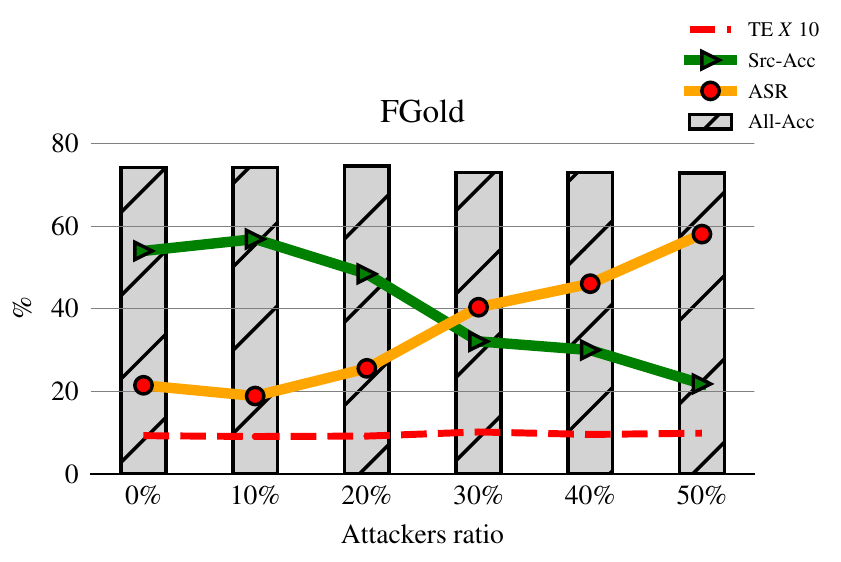}\vspace{-1.5\baselineskip}
    \end{subfigure}%
     \begin{subfigure}{0.34\textwidth}
      \centering
      \includegraphics[width=1\linewidth]{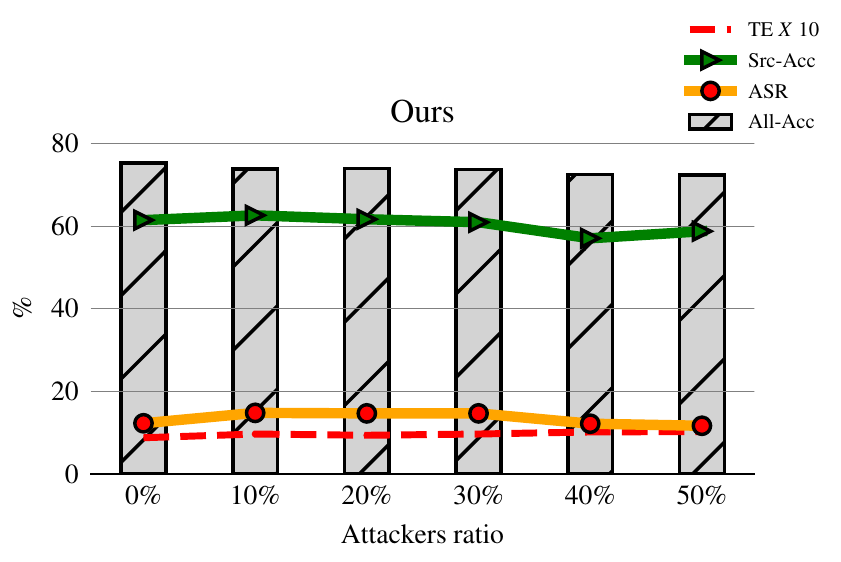}\vspace{-1.5\baselineskip}
    \end{subfigure}%
     \vspace{0.5\baselineskip}
\caption{Robustness against the label-flipping attack with the CIFAR10-Mild benchmark}
\label{fig:cifar10_mild}
\end{figure*}

Figure~\ref{fig:cifar10_mild} shows the results on the CIFAR10-Mild benchmark. 
We can see that, in this benchmark, the performance of all the methods except ours was worse due to the combined impact of the data distribution and the model dimensionality on the differentiation between the good updates and the bad ones.
%JOSEP2. Rewritten.
Although FoolsGold performed well in CIFAR10-iid, its performance substantially degraded in this benchmark. 
The reason for this is the combined impact on the output layer gradients of the CIFAR10 data set variability and the non-iid distribution of the data among peers, as shown in Figure~\ref{fig:pca_cifar10_mild_last}. This made all the output layer gradients highly diverse, and thus the gradients of the source and the target classes did not make a big difference in the distribution of the output layer gradients. 
On the other hand, thanks to the robust discriminative pattern we used to distinguish between updates, our defense stayed robust against the attack for all  attacker ratios. In fact, it offered best simultaneous performance on all the metrics.
Since our method considered only the source and target class neuron gradients (the gradients relevant to the attack) and excluded the non-relevant gradients, it was able to adequately differentiate between the good updates and the bad ones.

\begin{figure*}[htbp]
    \centering
    \begin{subfigure}{0.34\textwidth}
      \centering
      \includegraphics[width=1\linewidth]{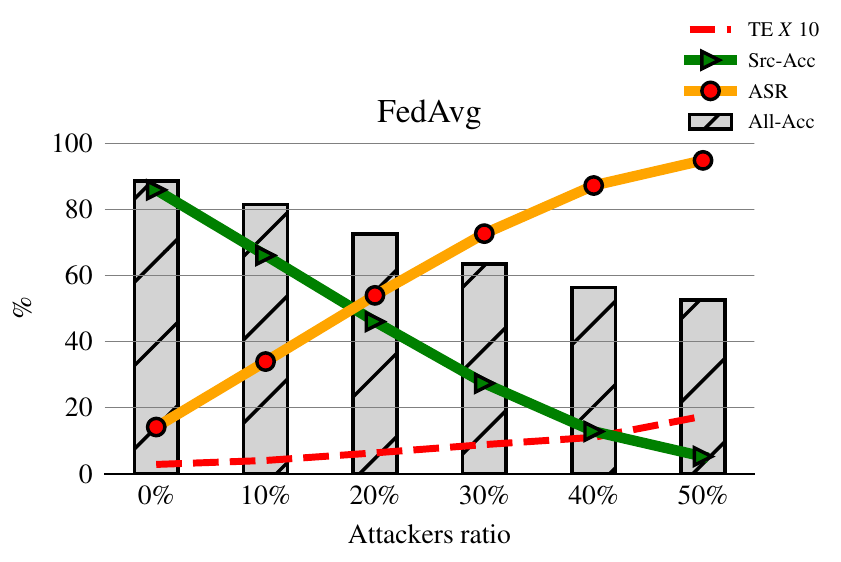}\vspace{-1.5\baselineskip}
    \end{subfigure}% 
    \begin{subfigure}{0.34\textwidth}
      \centering
      \includegraphics[width=1\linewidth]{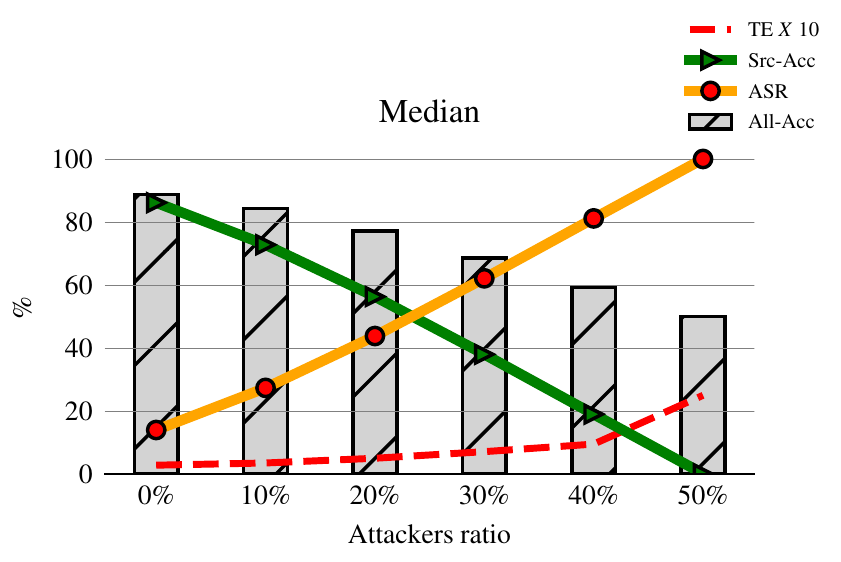}\vspace{-1.5\baselineskip}
    \end{subfigure}%
     \begin{subfigure}{0.36\textwidth}
      \centering
      \includegraphics[width=1\linewidth]{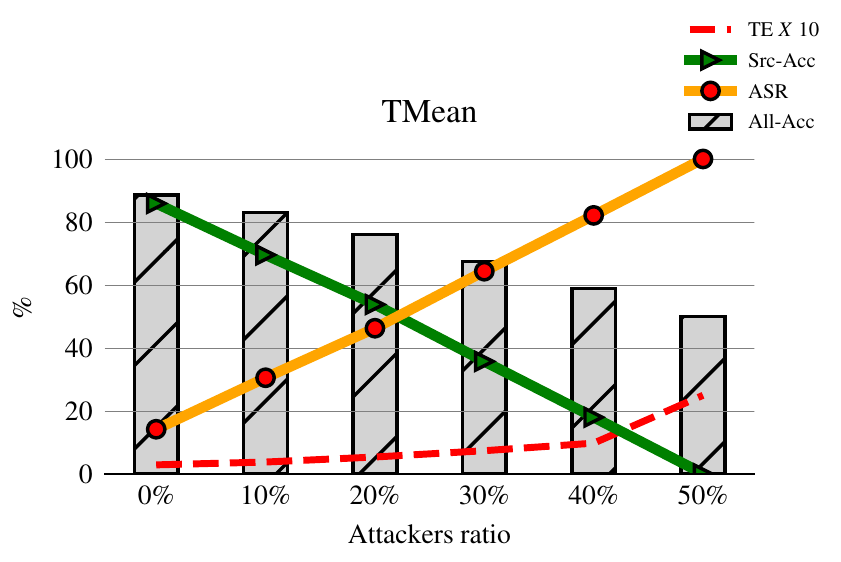}\vspace{-1.5\baselineskip}
    \end{subfigure}%
    \vspace{0.5\baselineskip}
    
     \begin{subfigure}{0.34\textwidth}
      \centering
      \includegraphics[width=1\linewidth]{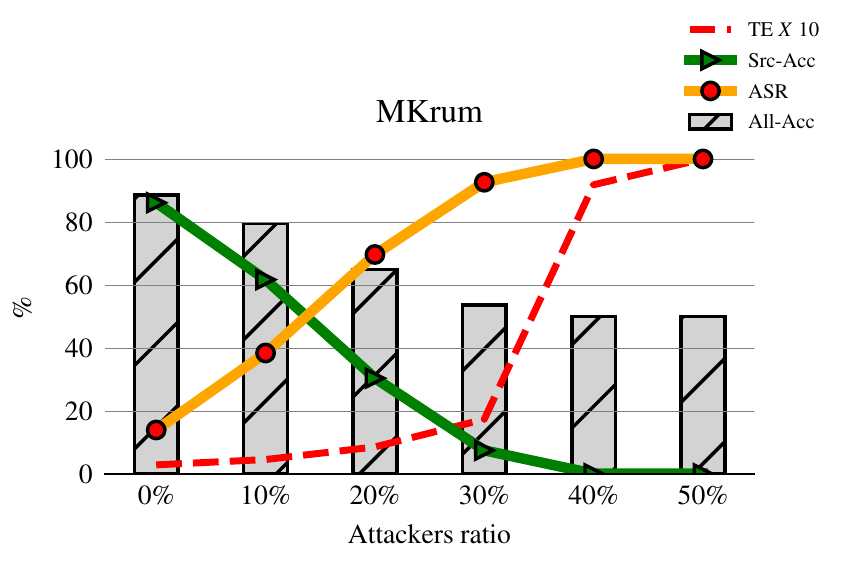}\vspace{-1.5\baselineskip}
    \end{subfigure}%
     \begin{subfigure}{0.34\textwidth}
      \centering
      \includegraphics[width=1\linewidth]{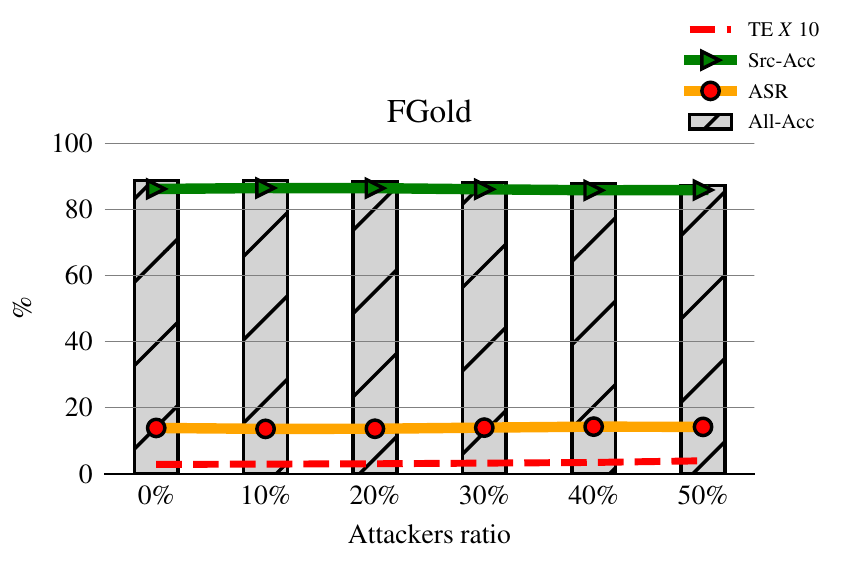}\vspace{-1.5\baselineskip}
    \end{subfigure}%
     \begin{subfigure}{0.34\textwidth}
      \centering
      \includegraphics[width=1\linewidth]{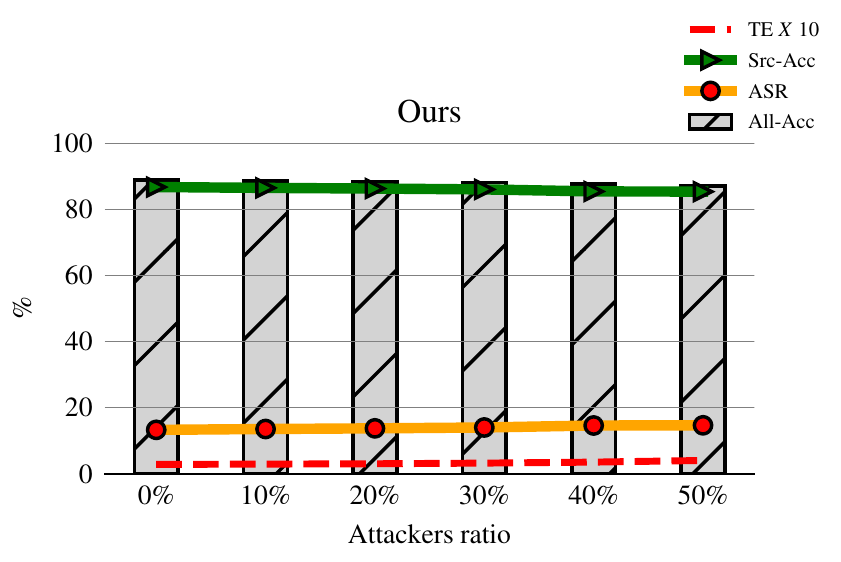}\vspace{-1.5\baselineskip}
    \end{subfigure}%
     \vspace{0.5\baselineskip}
\caption{Robustness against the label-flipping attack with the IMDB benchmark}
\label{fig:imdb}
\end{figure*}

Figure~\ref{fig:imdb} shows the results on the IMDB benchmark. 
%JOSEP2. Rewritten.
Our defense and FoolsGold had almost the same performance and outperformed the other methods for all the metrics. 
FoolsGold performed well in this benchmark because it is its ideal setting: updates for honest peers were somewhat different due to the different reviews they gave, while updates for attackers became very close to each other because they shared the same objective. Another reason was that the number of classes in the output layer was only two. Hence, all the parameters' gradients in the output layer were relevant to the attack.

\textbf{Accuracy stability.}
\label{stability}
The stability of the global model convergence (and its accuracy in particular) during training is a problem in FL, especially when training data are non-iid~\cite{li2018federated, karimireddy2020scaffold}. Furthermore, with an LF attack targeting a particular source class, the evolution of the accuracy of the source class becomes more unstable. 
Since an updated global model may be used after some intermediate training rounds, as in \cite{hard2018federated}, this may entail degradation of the accuracy of the source class at inference time. 
Keeping the accuracy stable is needed to prevent such consequences.
For this reason, we decided to use the CV metric to measure the stability of the source class accuracy.
Table~\ref{tab:stability} shows the CV of the accuracy of the source class in the used benchmarks for the different defense mechanisms. We can see that our proposal outperformed the other methods in most cases, and achieved a stability very close to that of FedAvg when the attackers' ratio was $0\%$ ({\em i.e.}, absence of attack). This is thanks to the perfect protection our method provided for the source class. NaN values in the table resulted from zero values of the source class accuracy in all the training rounds for those cases.

\begin{table*}[!ht]
\centering
\caption{Coefficient of variation (CV) of the source class accuracy during training for the considered benchmarks with different attacker ratios. The best figure in each column is shown in boldface.}
\label{tab:stability}
\resizebox{\textwidth}{!}{%
\begin{tabular}{l|ccccccc|ccccccc|}
\hline
\multicolumn{1}{|c|}{\multirow{2}{*}{\textbf{\begin{tabular}[c]{@{}c@{}}Attackers ratio/ \\ Method\end{tabular}}}} & \multicolumn{1}{c|}{FedAvg} & \multicolumn{1}{c|}{Median} & \multicolumn{1}{c|}{RMedian} & \multicolumn{1}{c|}{TMean} & \multicolumn{1}{c|}{MKrum} & \multicolumn{1}{c|}{FGold} & Ours          & \multicolumn{1}{c|}{FedAvg} & \multicolumn{1}{c|}{Median} & \multicolumn{1}{c|}{RMedian} & \multicolumn{1}{c|}{TMean} & \multicolumn{1}{c|}{MKrum} & \multicolumn{1}{c|}{FGold} & Ours          \\ \cline{2-15} 
\multicolumn{1}{|c|}{}                                                                                             & \multicolumn{7}{c|}{\textbf{MNIST-iid}}                                                                                                                                                         & \multicolumn{7}{c|}{\textbf{MNIST-Mild}}                                                                                                                                                         \\ \cline{1-1}
\multicolumn{1}{|l|}{0\%}                                                                                          & \textbf{0.11}               & \textbf{0.11}               & \textbf{0.11}                & \textbf{0.11}              & \textbf{0.11}              & 0.31                       & \textbf{0.11} & \textbf{0.08}               & 0.12                        & 0.12                         & \textbf{0.08}              & 0.09                       & 0.09                       & 0.10          \\ \cline{1-1}
\multicolumn{1}{|l|}{10\%}                                                                                         & 0.14                        & 0.12                        & 0.12                         & 0.12                       & \textbf{0.11}              & 7.31                       & \textbf{0.11} & 0.12                        & 0.20                        & 0.20                         & 0.16                       & 0.15                       & \textbf{0.10}              & 0.15          \\ \cline{1-1}
\multicolumn{1}{|l|}{20\%}                                                                                         & 0.19                        & 0.14                        & 0.14                         & 0.14                       & \textbf{0.11}              & 6.46                       & \textbf{0.11} & 0.19                        & 0.28                        & 0.28                         & 0.25                       & 0.24                       & \textbf{0.17}              & \textbf{0.17} \\ \cline{1-1}
\multicolumn{1}{|l|}{30\%}                                                                                         & 0.24                        & 0.16                        & 0.16                         & 0.17                       & \textbf{0.11}              & 6.89                       & \textbf{0.11} & 0.24                        & 0.33                        & 0.33                         & 0.32                       & 0.36                       & \textbf{0.19}              & \textbf{0.19} \\ \cline{1-1}
\multicolumn{1}{|l|}{40\%}                                                                                         & 0.31                        & 0.20                        & 0.20                         & 0.20                       & \textbf{0.11}              & 6.39                       & \textbf{0.11} & 0.27                        & 0.43                        & 0.43                         & 0.43                       & 0.44                       & 1.35                       & \textbf{0.24} \\ \cline{1-1}
\multicolumn{1}{|l|}{50\%}                                                                                         & 0.42                        & 0.56                        & 0.56                         & 0.56                       & 0.29                       & 6.34                       & \textbf{0.21} & 0.36                        & 0.63                        & 0.63                         & 0.63                       & 3.14                       & 2.75                       & \textbf{0.27} \\ \hline
                                                                                                                   & \multicolumn{7}{c|}{\textbf{MNIST-Extreme}}                                                                                                                                                     & \multicolumn{7}{c|}{\textbf{IMDB}}                                                                                                                                                              \\ \cline{1-1}
\multicolumn{1}{|l|}{0\%}                                                                                          & \textbf{0.27}               & 2.50                        & 2.49                         & \textbf{0.27}              & 0.32                       & 0.38                       & 0.29          & 0.10                        & 0.10                        & 0.10                         & 0.10                       & 0.10                       & 0.10                       & \textbf{0.09} \\ \cline{1-1}
\multicolumn{1}{|l|}{10\%}                                                                                         & \textbf{0.33}               & 2.70                        & 3.14                         & 1.15                       & 0.41                       & NaN                        & 0.34          & 0.15                        & 0.11                        & 0.11                         & 0.12                       & 0.16                       & \textbf{0.09}              & \textbf{0.09} \\ \cline{1-1}
\multicolumn{1}{|l|}{20\%}                                                                                         & 0.44                        & 2.90                        & 2.96                         & 1.86                       & 14.14                      & 1.65                       & \textbf{0.29} & 0.22                        & 0.15                        & 0.15                         & 0.16                       & 0.28                       & \textbf{0.10}              & \textbf{0.10} \\ \cline{1-1}
\multicolumn{1}{|l|}{30\%}                                                                                         & 0.53                        & 3.14                        & 2.82                         & 2.28                       & NaN                        & 0.58                       & \textbf{0.26} & 0.29                        & 0.17                        & 0.17                         & 0.18                       & 0.43                       & \textbf{0.10}              & \textbf{0.10} \\ \cline{1-1}
\multicolumn{1}{|l|}{40\%}                                                                                         & 0.77                        & 3.07                        & 2.88                         & 3.43                       & NaN                        & 0.43                       & \textbf{0.31} & 0.35                        & 0.27                        & 0.27                         & 0.27                       & NaN                        & \textbf{0.09}              & \textbf{0.09} \\ \cline{1-1}
\multicolumn{1}{|l|}{50\%}                                                                                         & 1.25                        & 3.43                        & 3.23                         & 3.43                       & NaN                        & 2.74                       & \textbf{0.37} & 0.42                        & 0.84                        & 0.75                         & 0.84                       & NaN                        & \textbf{0.10}              & \textbf{0.10} \\ \hline
                                                                                                                   & \multicolumn{7}{c|}{\textbf{CIFAR10-iid}}                                                                                                                                                       & \multicolumn{7}{c|}{\textbf{CIFAR10-Mild}}                                                                                                                                                      \\ \cline{1-1}
\multicolumn{1}{|l|}{0\%}                                                                                          & 0.13                        & \textbf{0.12}               & \textbf{0.12}                & 0.13                       & \textbf{0.12}              & 0.13                       & 0.13          & 0.14                        & 0.14                        & 0.14                         & 0.14                       & 0.14                       & \textbf{0.13}              & 0.14          \\ \cline{1-1}
\multicolumn{1}{|l|}{10\%}                                                                                         & 0.16                        & 0.16                        & 0.16                         & 0.16                       & 0.17                       & 0.16                       & \textbf{0.14} & 0.19                        & 0.18                        & 0.19                         & 0.17              & 0.20                       & 0.17              & \textbf{0.14}        \\ \cline{1-1}
\multicolumn{1}{|l|}{20\%}                                                                                         & 0.19                        & 0.21                        & 0.20                         & 0.18                       & 0.23                       & 0.20                       & \textbf{0.13} & 0.21                        & 0.21                        & 0.20                         & 0.21                       & 0.26                       & 0.19                       & \textbf{0.14} \\ \cline{1-1}
\multicolumn{1}{|l|}{30\%}                                                                                         & 0.23                        & 0.23                        & 0.24                         & 0.23                       & 0.32                       & 0.27                       & \textbf{0.14} & 0.23                        & 0.22                        & 0.23                         & 0.25                       & 0.43                       & 0.22                       & \textbf{0.15} \\ \cline{1-1}
\multicolumn{1}{|l|}{40\%}                                                                                         & 0.26                        & 0.31                        & 0.26                         & 0.29                       & 1.25                       & 0.40                       & \textbf{0.14} & 0.28                        & 0.28                        & 0.27                         & 0.28                       & 0.45                       & 0.26                       & \textbf{0.15} \\ \cline{1-1}
\multicolumn{1}{|l|}{50\%}                                                                                         & 0.36                        & 0.45                        & 0.45                         & 0.45                       & 6.33                       & 0.35                       & \textbf{0.24} & 0.38                        & 0.42                        & 0.43                         & 0.42                       & 0.54                       & 0.40                       & \textbf{0.17} \\ \hline
\end{tabular}%
}
\end{table*}

\begin{figure*}[!htbp]
    \centering
%JOSEP2. I make the top figure the same size as the others.
    \begin{subfigure}{0.5\textwidth}
      \centering
      \includegraphics[width=1\linewidth]{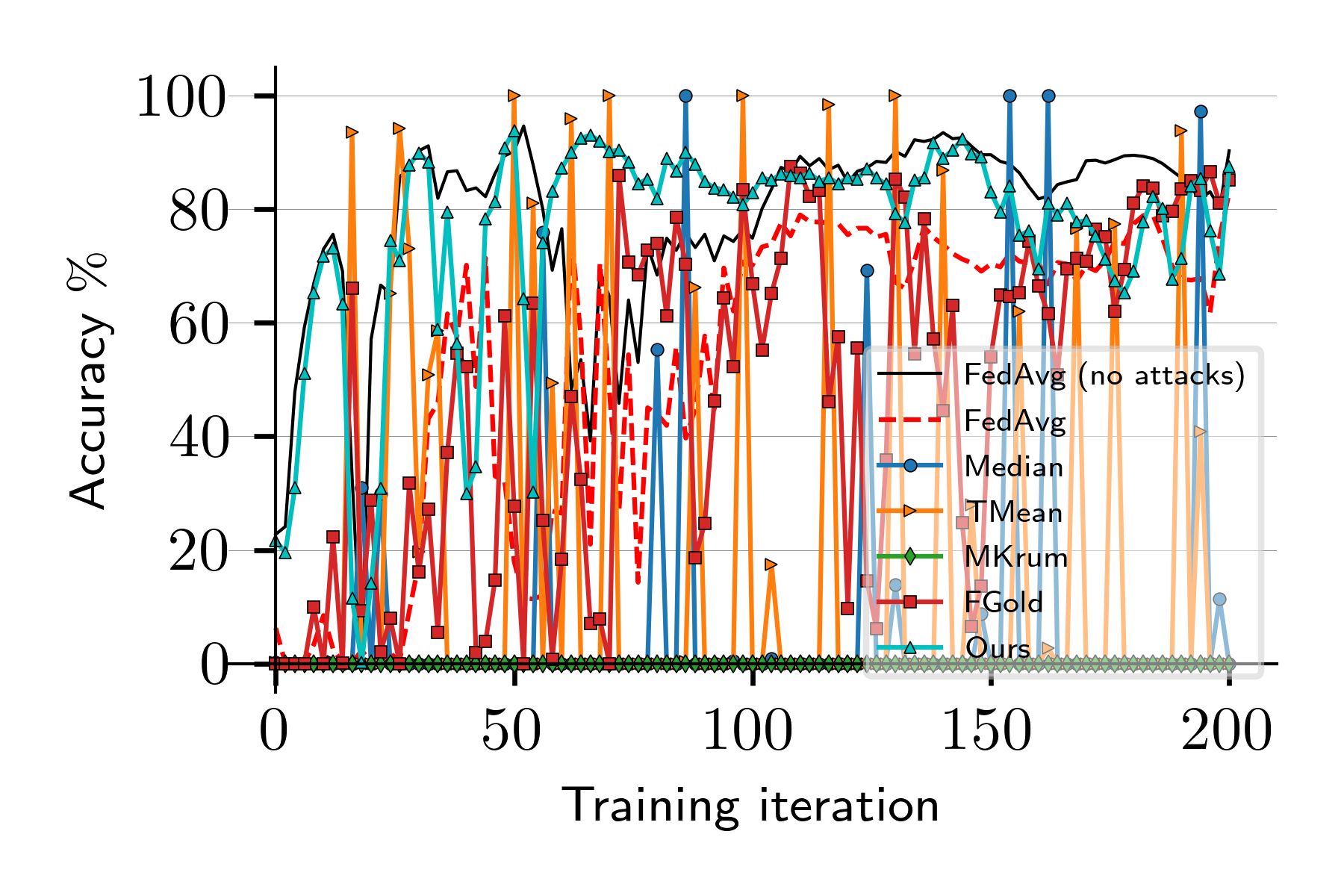}\vspace{-1\baselineskip}
      \caption{MNIST-Extreme}
    \end{subfigure}%
    
    \centering
    \begin{subfigure}{0.5\textwidth}
      \centering
      \includegraphics[width=1\linewidth]{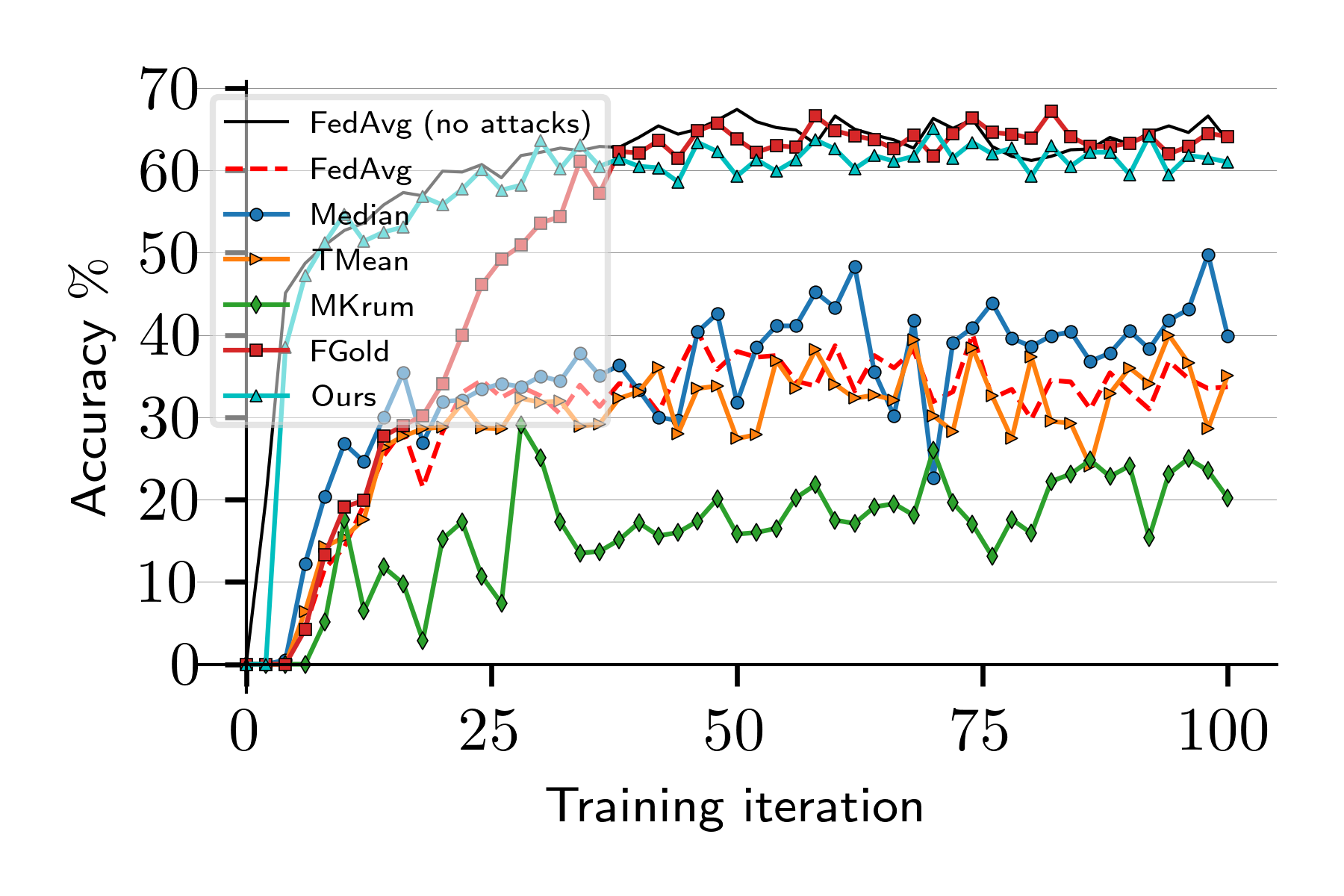}\vspace{-1\baselineskip}
      \caption{CIFAR10-iid}
    \end{subfigure}% 
    \begin{subfigure}{0.5\textwidth}
      \centering
      \includegraphics[width=1\linewidth]{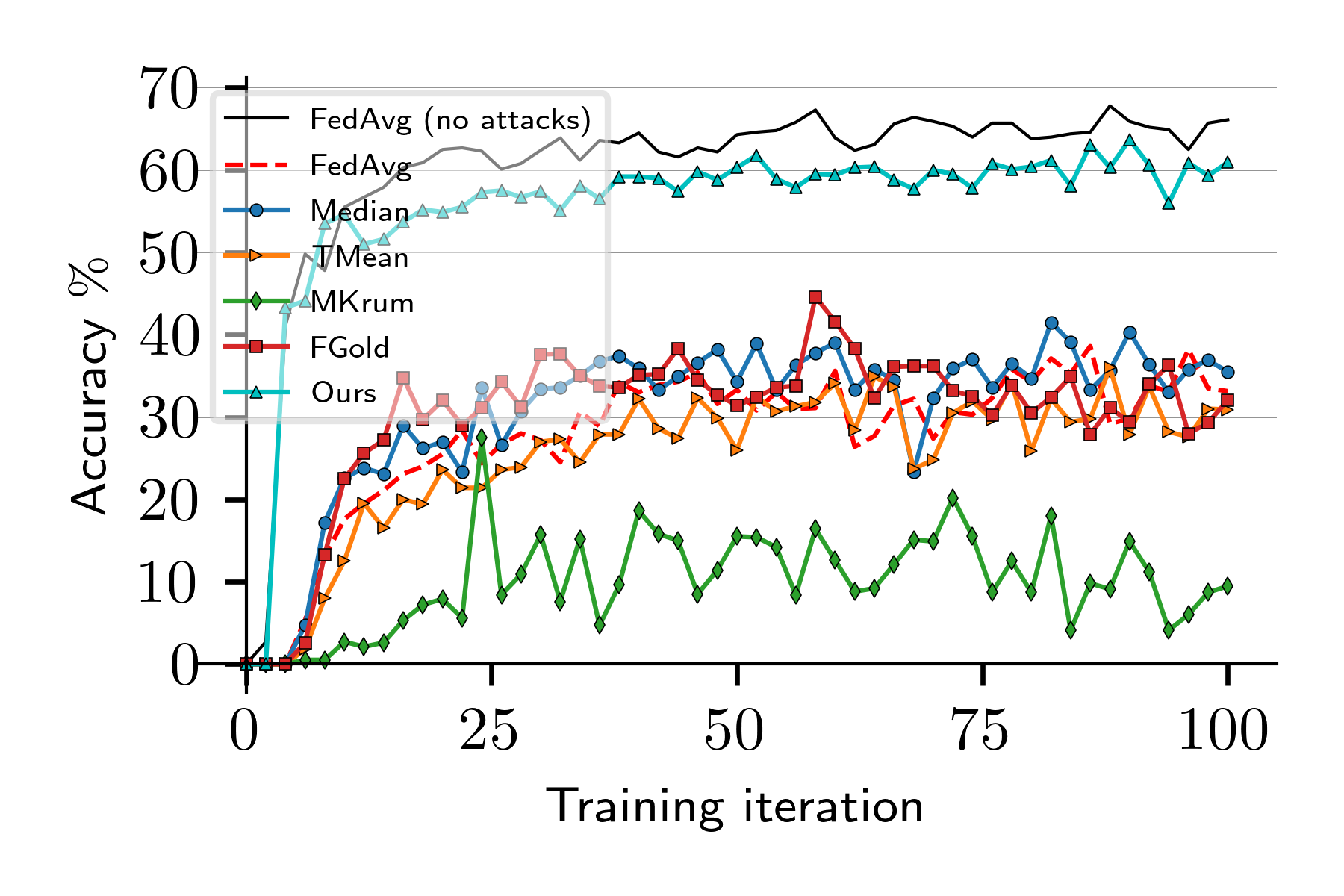}\vspace{-1\baselineskip}
      \caption{CIFAR10-Mild}
    \end{subfigure}%
\caption{Evolution of the source class accuracy with $30\%$ attackers ratio.}
\label{fig:stability}
\end{figure*}

To provide a clearer picture of the effectiveness of our defense, Figure~\ref{fig:stability} shows the evolution of the accuracy of the source class as training progresses when the attacker ratio was $30\%$ in the MNISt-Extreme, 
%JOSEP2. CIFAR -> CIFAR10
CIFAR10-iid and CIFAR10-Mild benchmarks. It is clear from the figure that the accuracy achieved by our defense was the most similar to the accuracy of the FedAvg when no attacks were performed.

\begin{figure*}[!htbp]
    \centering
    \begin{subfigure}{0.9\textwidth}
      \centering
      \includegraphics[width=1\linewidth]{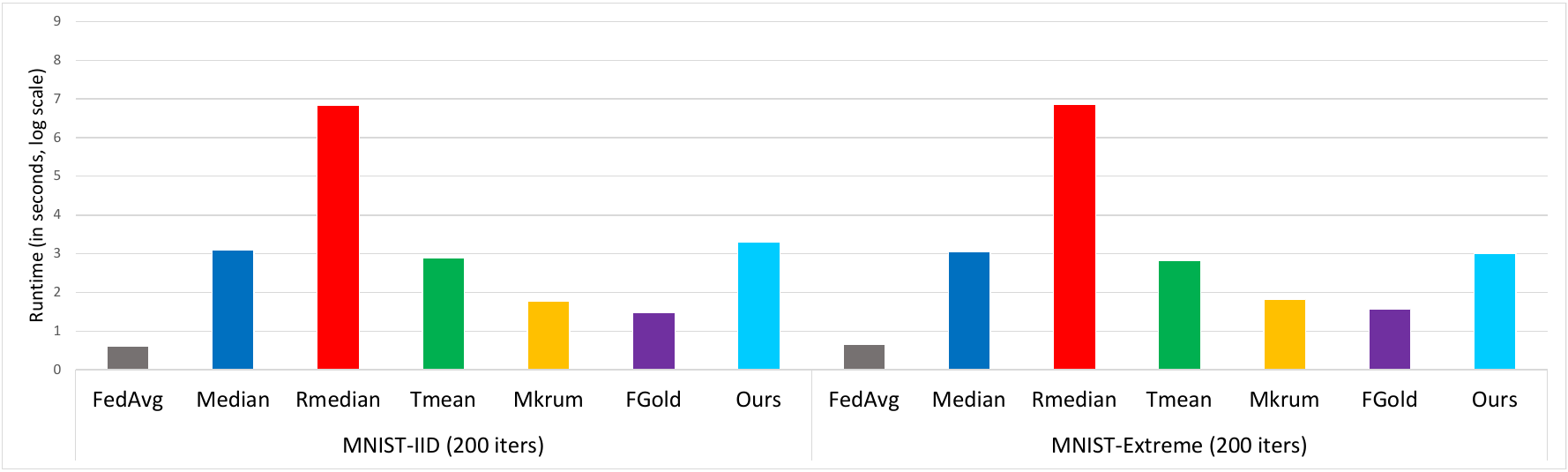}
    \end{subfigure}% 
    
    \begin{subfigure}{0.9\textwidth}
      \centering
      \includegraphics[width=1\linewidth]{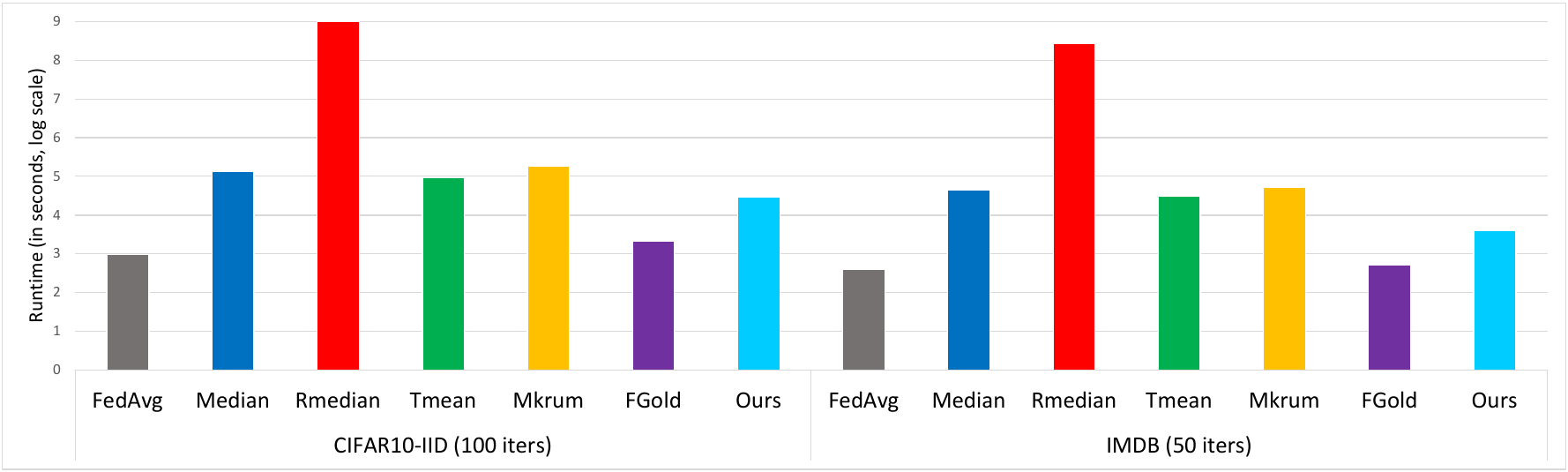}\vspace{-1\baselineskip}
    \end{subfigure}
\caption{Runtime overhead.}
\label{fig:runtime}
\end{figure*}

\textbf{Runtime overhead.}
\label{runtime}
Finally, we measured the CPU runtime of our method and compared it with the runtime of the other methods. 
Figure~\ref{fig:runtime} shows the total runtime in seconds (log scale) of each method during the whole training iterations.
The results show that FoolsGold had the smallest runtime in all cases, excluding FedAvg, which just averages updates. 
The repeated median had the highest runtime due to the regression calculations it does to estimate the median points.
%JOSEP2. Rewritten.
On the other hand, the runtime of our method was similar to that of the median and the trimmed mean when the model size was small. For the large models used 
in the CIFAR10 and the IMDB benchmarks,
our method had the second smallest runtime after FoolsGold. In fact, the runtime incurred by our method can be viewed as very tolerable, given its effectiveness at countering the LF attack.

%JOSEP2. IMPORTANT. Najeeb, this figure is very small. With the space reduction I have performed, I think you can place two figures on top and two figures below also as wide as the page. This should still keep the paper within 24 pages.
% \begin{figure}[!htbp]
%     \centering
%       \includegraphics[width=1\linewidth]{figures/runtime.pdf}
% \caption{Runtime overhead}
% \label{fig:runtime}
% \end{figure}

% \subsection{Fooling FoolsGold} 
% \label{fool_fg}
% Figure~\ref{fig:fool_fg}

% \begin{figure}[!ht]
%     \centering
%       \includegraphics[width=0.8\linewidth]{figures/fool_fg.png}
% \caption{Fooling FoolsGold.}
% \label{fig:fool_fg}
% \end{figure}

\section{Conclusions and future work}
\label{conclusion}

%JOSEP2. A bit rewritten.
In this paper, we have conducted comprehensive analyses of the label-flipping attack behavior. We have observed that the contradictory objectives of attackers and honest peers turn the parameter gradients connected to the source and target class neurons into robust discriminative features to detect the attack. Besides, we have observed that settings with different local data distributions require different strategies to defend against the attack.
Accordingly, we have presented a novel defense that uses those gradients as input features to a suitable clustering method to detect attackers. 
The empirical results we report show that our defense is very effective and performs very well simultaneously regarding test error, overall accuracy, source class accuracy, and attack success rate. In fact, our defense significantly improves on the state of the art.

As future work, we plan to test and expand our method to detect other targeted attacks such as backdoor attacks.

\begin{acks}
This research was funded by the European Commission (projects H2020-871042 ``SoBigData++'' and H2020-101006879 ``MobiDataLab''), the Government of Catalonia (ICREA Acad\`emia Prizes to J.Domingo-Ferrer and D. Sánchez, and FI grant to N. Jebreel), and MCIN/AEI /10.13039/501100011033 /FEDER, UE under project PID2021-123637NB-I00 ``CURLING''.
The authors are with the UNESCO Chair in Data Privacy, but the views in this paper are their own and are not necessarily shared by UNESCO.
\end{acks}

\bibliographystyle{ACM-Reference-Format}
\bibliography{my_bib}

\end{document}